# Two-phase Age-Structured Model of Solitarious and Gregarious Locust Population Dynamics

Vitalii V. Akimenko[1], Cyril Piou[2]

[1]*Faculty of Cybernetics, T.Shevchenko National University of Kyiv, Volodymyrska 64, 01030 Kyiv, Ukraine.*

[2]*CIRAD, UMR CBGP, F-34398 Montpellier, France & CNLAA, BP125, 86343 Inezgane, Morocco & University Ibn Zohr, Agadir, Morocco*

**Abstract**

In this paper we study a nonlinear age-structured model of locust population dynamics with variable time of egg incubation that describes the phase polyphenism and behaviour of desert locust, *Schistocerca gregaria.* The model is based on the competitive system of linear transport Eqs. with nonlinear density-dependent fertility rates with variable time delay in boundary conditions. It describes the dynamics of the density of two subpopulations or two phases of locust's polyphenism – solitarious and gregarious. The obtained system is studied both theoretically and numerically. The analysis of asymptotical stability of trivial and nontrivial equilibriums of autonomous system allows us to derive the conditions and understand the particularities of bidirectional phase transitions between solitarious and gregarious. Simulation of locust population dynamics with different conditions of phase transitions for autonomous and non-autonomous dynamical systems of our two-phase age-structured competitive model with time delay exhibits the features of behaviour of documented outbreak dynamics of *Schistocerca gregaria.*

## 1. Introduction

The understanding of complex processes of locust population dynamics is a crucial point in sustainable food security for all continents. Locusts are grasshoppers able to change behaviour, life cycle, physiology and morphology in response to a density increase under a process known as phase polyphenism, a particular case of phenotypic plasticity [37]. Locusts threat agriculture since agriculture exists and no continents are spared [40]. In particular, the periodical outbreaks of desert locust *Schistocerca gregaria* [12], [13], [14], [32], [33], [37], [40], [41], [42] populations play an important role in the economics and social life of many countries of African and Eurasian continents. This locust species create huge "swarms" that cause great harm and significant damage to agricultural lands cascading on catastrophic impacts on human societies and environmental conditions. Realistic modelling of the main life processes of locust individuals in these populations such as birth, ageing, fertility and death is possible on the basis of advanced age-structured models of population dynamics. This class of models was presented first in the mathematical theory of epidemics in the pioneer works of Kermack and McKendrick [28] – [30], theory of population of Von Foerster [45] and nonlinear theory of population dynamics in later papers of Gurtin and MacCamy [22], [23]. The main feature of age-structured models is that they allow describing the fertility process in population in an explicit way using the nonlocal integral boundary condition. As the reproduction of individuals plays a crucial role in biological populations in nature the linear and nonlinear age-structured models are widely studied and used in theoretical and applied researches of biological, demographical and ecological systems [2], [4], [5], [6], [11], [17], [18], [20], [25], [26], [27], [34], [35], [46], [47]. Despite the fact that these models are well-known and applied to numerous specific systems studied in life science, they have not been used in the simulation of locust population dynamics so far. In this paper we initiate a study of the complex processes of locust polyphenism on the basis of a two phase competitive age-structured model of desert locust population dynamics with time delay.

The phase polyphenism confers to locusts the capacity to change their 'phase' from being 'solitarious' (when they behave like ordinary grasshoppers) to becoming fully 'gregarious' (swarming behaviour) and reverse [37]. The main necessary condition of phase changing from 'solitarious' to 'gregarious' locusts (known as “gregarization”) is an increase of 'solitarious' population density. As a consequence of this phase change, there is generally a significant increase of locust’s mating and reproduction intensity that leads to incredible growth and quasi-periodical regimes of locust population dynamics. The decline of populations happen when environmental conditions such as climate, vegetation cover, soil moisture, and others are inadequate for several generations and/or human controls (treatments with pesticides) are effective. This influence of environmental change leads to the decreasing of populations with reverse phase transition from 'gregarious' to 'solitarious' locusts – “solitarization”. The impact of these factors is described by means of changing the parameters of locust population dynamics model.

In quantitative population ecology these regimes are named “population outbreaks” [1], [8]. The population dynamics (outbreaks of population density) in this case is described by quasi-periodical functions of two types: pulse sequences (populations with cyclical eruption dynamics) or single pulses (population with pulse eruptive dynamics) [8]. Simulation of these types of outbreaks were studied in previous works [2], [3] on the basis of linear and nonlinear age-structured models of population dynamics. There are two main factors which might lead to population outbreaks. The first one is described by a density-dependent death rate or/and a density-dependent fertility rate that reflects the feedback of population growth on the important life process of individuals - fertility and mortality. The second factor is described by oscillating functions of death rate and fertility rate that deals with the quasi-periodical influences of environmental changes (climate, vegetation cover, human activity, etc.,) on the cycles of population dynamics. As both factors play a crucial role in population outbreaks they have to be taken into account in the realistic models of desert locust population dynamics as well.

In the present work we apply for the first time a nonlinear competitive age-structured model of population dynamics with density-dependent fertility rate with time delay and density-independent death rate to study the population dynamics of solitarious and gregarious phases of locusts. Our objective here is the development of a mathematical model for realistic simulation of all stages of two-phase population dynamics including the incubation period of locust’s eggs in the soil, birth, ageing, maturation, the repeated processes of reproduction and death. The crucial question of this model is to represent the phase polyphenism in desert locust population – *i.e.* the transition from solitarious phase (non-swarming behaviour, without outbreaks) to gregarious phase (swarming behaviour, population outbreaks) and reverse. The model of phase polyphenism is developed on the basis of competitive epidemic model of inter-cohort mixing of locusts in population studied in the works [7], [9], [24], [25], [44]. In the present paper, the competitive system of equations of initial boundary value problems for semi-linear transport with density-dependent fertility rate in integral boundary conditions with time delay is studied both theoretically and numerically.

After section 2 presenting the model, in section 3 we prove the existence and uniqueness of a travelling wave solution of the competitive system using the explicit recurrent formulae. The obtained algorithm allows deriving in an explicit form the restrictions for the model parameters (coefficients of equations and initial values) that guarantee continuity and smoothness of the travelling wave solution presented in section 4. In section 5, we carry out the analysis of conditions of existence and local asymptotical stability of trivial and nontrivial equilibriums of the autonomous system of nonlinear initial problems with time delay for the size of solitarious and gregarious subpopulations. Our motivation for theoretical study of the features of unstable dynamical regimes of both subpopulations lies in an attempt to understand further the conditions of misbalance between mortality and fertility in locust population. In particular, the linear stability analysis provided in section 6 reveals that the asymptotical stability of nontrivial equilibrium of autonomous system does not depend from the incubation period of locust's eggs in the soil (discrete time delay parameter) and depends from the sign of derivative of density-dependent fertility rate of solitarious and gregarious. We continue in section 7 the study of dynamical regimes of autonomous system in numerical experiments with different conditions of asymptotical stability and instability of solution. The numerical experiments carried out with specific functions of death and fertility rates exhibit the existence of special regimes of population dynamics which are characterized by simultaneous phase changes and outbreaks of *Schistocerca gregaria.*

Finally, in section 8, we study the features of two-phase age-structured non-autonomous system in numerical experiments comparing the numerical solutions with a dataset of observed important swarming events in the distribution range of *S. gregaria*. The dataset used in this study covers the results of observation of swarm outbreaks over the last 30 years as a time series of number of square degrees ($1^\circ \times 1^\circ$) with swarms of gregarious locusts. We consider the parameter estimation problem for the dynamical coefficients of the vital characteristics of locusts, such as the fertility rate and death rate. Using the metaheuristic method "tabu search" [21], [38] we fit the solution of non-autonomous system with the dataset for three different periods of important gregarization with a relative error of less than 26%. On one hand, the results of these simulations verify the chosen approach of mathematical modeling on the basis of two-phase age-structured competitive model of locust population dynamics. On the other hand, these simulations provide practical insight into the relative roles of the life processes and polyphenism hence allowing for improvements of the theoretical instruments of analysis and prediction of locust population dynamics.

## 2. Competitive age-structured model of solitarious and gregarious desert locust population dynamics

The considered age-structured model of desert locust population is separated into two phases - solitarious and gregarious. The mature females of locust in each phase lay eggs in their egg pods in the soil with an incubation time period $\tau$. The age of new born locust emerged from the ground is considered as its initial age $a = 0$. After birth, locusts go through some growing and maturing stages up to the mature imago stage when it is ready for mating. After mating with a male, the adult female locust can lay eggs in the soil only at some defined age and repeat this process several times within their lifespan up to the maximum age of $a_d$ (Fig.1). It will be assumed throughout the paper that index $s$ or $g$ marks a mathematical object (constant or function) connected to solitarious or gregarious locust phase respectively. Let functions $u(a,t)$ and $g(a,t)$ denote the age-specific density of solitarious and gregarious locust population and $w(a,t) = u(a,t) + g(a,t)$ the age-specific density of all locust population defined in domain $\overline{Q} = \{(a,t): 0 \le a \le a_d, \quad t \ge 0\}$. We consider here population density as a theoretical description of a panmictic population over a large contiguous area. The number of solitarious, gregarious and all locusts in the population at time $t$ are respectively defined as:

$$U(t) = \int_0^{a_d} u(a,t)da, \quad G(t) = \int_0^{a_d} g(a,t)da, \quad W(t) = \int_0^{a_d} w(a,t)da \tag{1}$$

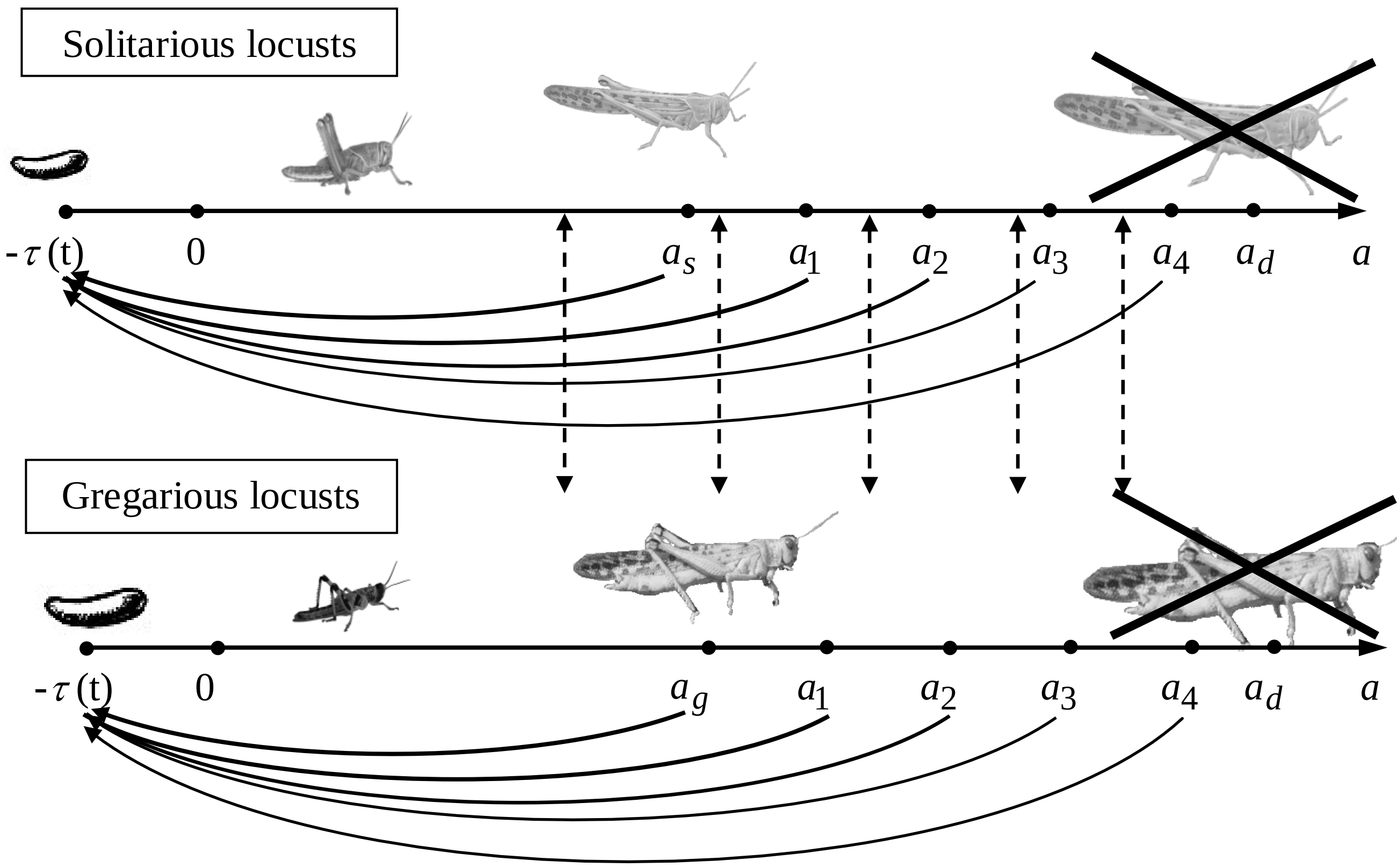

Figure 1. Schematic diagram of age-structured model of solitarious and gregarious desert locust lifespan: $\tau(t)$ is a time of egg incubation; $a = 0$ is an age of new born locust; $a_s(t)$, $a_g(t)$ are the ages of matured solitarious and gregarious females at which they lay their first pod with eggs; $a_d$ is a maximum age of locusts, ⟶ denotes the directions of reproduction, - - - ➜ denotes the direction of phase transition (photos of locusts: CIRAD).

The dynamics of $u(a,t)$ and $g(a,t)$ are governed by the competitive system of initial-boundary value problems for the semi-linear transport equations with non-local nonlinear integral boundary condition with time delay:

$$\frac{\partial u}{\partial t}=-\left(\alpha_0(a,t)+q_g(a,t,W(t))\right)u+q_s(a,t,W(t))g-\frac{\partial u}{\partial a},\ (a,t)\in Q, \tag{2}$$

$$\frac{\partial g}{\partial t}=-\left(\alpha_0(a,t)+q_s(a,t,W(t))\right)g+q_g(a,t,W(t))u-\frac{\partial g}{\partial a},\ (a,t)\in Q, \tag{3}$$

$$u(0,t)=F_s(u(a,t-\tau(t)),U_s(t-\tau(t)),t-\tau(t))=\omega_s\int_{a_s(t-\tau(t))}^{a_d} f_s(a,t-\tau(t),U_s(t-\tau(t))) \tag{4}$$

$$\times u(a,t-\tau(t))da,\ \ t>0,$$

$$g(0,t)=F_g(g(a,t-\tau(t)),G_g(t-\tau),t-\tau(t))=\omega_g\int_{a_g(t-\tau(t))}^{a_d} f_g(a,t-\tau(t),G_g(t-\tau(t))) \tag{5}$$

$$\times g(a,t-\tau(t))da,\ \ t>0,$$

$$u(a,t)=\varphi(a,t),\qquad a\in[0,a_d],\ \ t\in[-\tau(0),0]\ , \tag{6}$$

$$g(a,t)=0,\qquad a\in[0,a_d],\ t\in[-\tau(0),0]\ , \tag{7}$$

where $\alpha_0(a,t)$ is a natural *death rate* common for both phases of locusts; $q(a,t,W(t))\ge 0$ are the *phase-switching rate* which regulate the locust's changing from solitarious to gregarious phase $\left(q_s(a,t,W(t))>0\right)$ and reverse $\left(q_g(a,t,W(a))>0\right)$ (see Fig.2.); $0<\omega<1$ is a fraction of locust females of reproductive age in population; $a_d$ is a maximum age of locusts; $a_s(t)$, $a_g(t)$ are the ages of matured solitarious and gregarious females of which they lay the first pod with eggs, $0<\breve{a}_s\le a_s(t)\le\hat{a}_s<a_d$, $0<\breve{a}_g\le a_g(t)\le\hat{a}_g<a_d$, where $\breve{a}_s$, $\hat{a}_s$, $\breve{a}_g$, $\hat{a}_g$ are prescribed constants; $\varphi(a,t)$ is an initial value of density of solitarious population; $f_s(a,t,U_s(t))$, $f_g(a,t,G_g(t))$ are the (per unit density) locust's *fertility rates* which depend from the density of matured solitarious and gregarious respectively (sizes of matured locust subpopulations):

$$U_s(t)=\int_{a_s(t-\tau(t))}^{a_d} u(a,t)da,\ \ G_g(t)=\int_{a_g(t-\tau(t))}^{a_d} g(a,t)da \tag{8}$$

In the model (2) – (4) we use a variable time delay parameter $\tau(t)$ which is a incubation time for locust eggs in the soil included in the boundary conditions (4), (5). Since this parameter can depend from environmental conditions, it is considered as positive bounded function of time $0<\tau_0\le\tau(t)\le\tau_{\max}$, where $\tau_0$ and $\tau_{\max}$ are a minimum and maximum possible incubation periods.

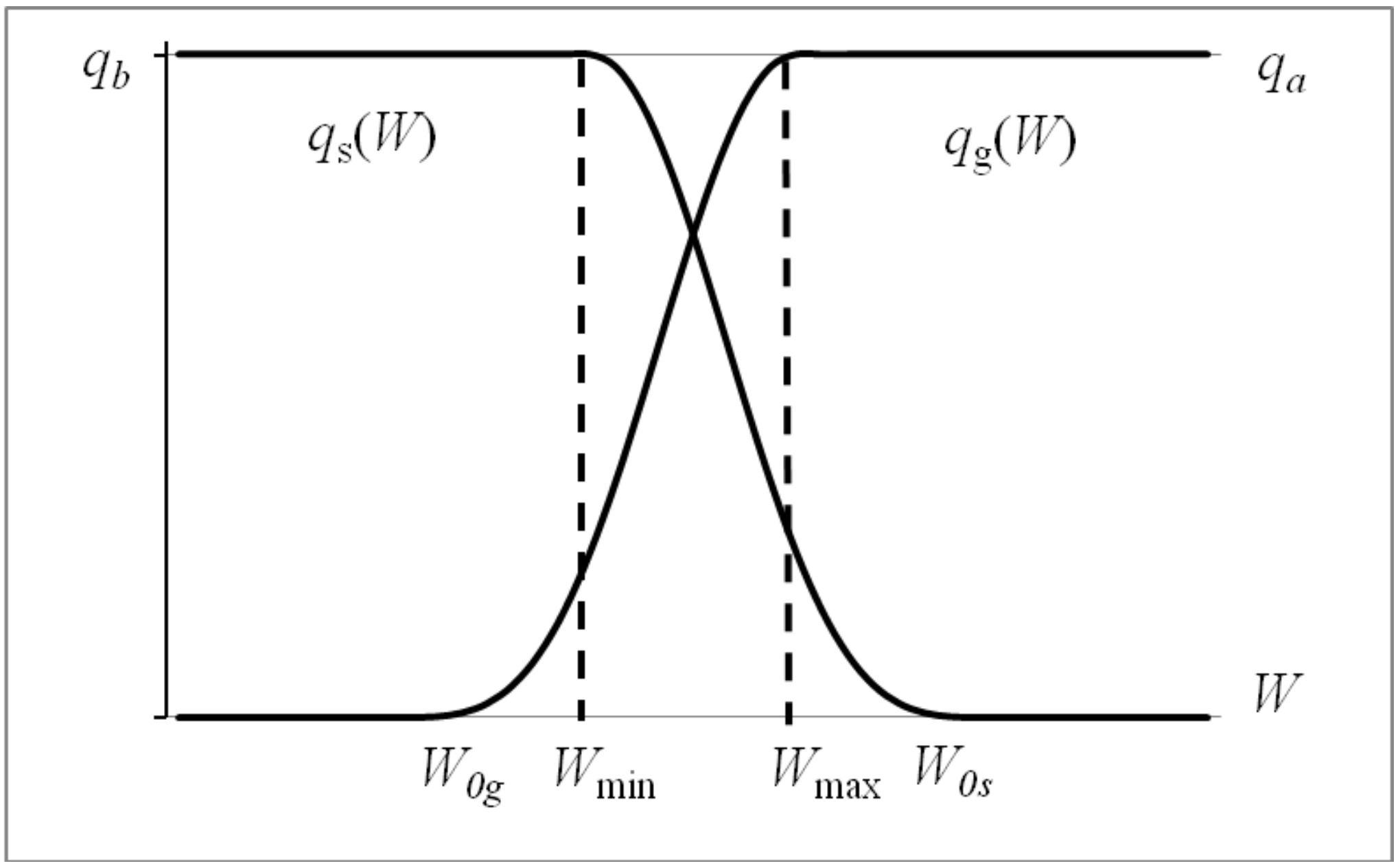

Figure 2. Graphs of *phase-switching rates*; $W_{0g}$ is a maximum number of solitarious in the bounded area when population increase before the onset of the gregarization phenomenon ($q_g(W)>0$); $W_{0s}$ is a minimum density of gregarious in the same area when population size decrease before the reverse phase change – solitarization ($q_s(W)>0$); $[W_{0s},W_{0g}]$ is an interval of locust density of bidirectional phase changing ($q_s(W)>0$, $q_g(W)>0$).

From the system (2) – (7) follows that the dynamics of the locust population density $w(a,t)$ is governed by the initial-boundary value problem for the semi-linear transport Eq. with non-local integral boundary condition with discrete time delay:

$$\frac{\partial w}{\partial t} = -\alpha_0(a,t)w - \frac{\partial w}{\partial a}, \quad (a,t)\in Q, \tag{9}$$

$$w(0,t) = F_w(u(a,t-\tau(t)), g(a,t-\tau(t)), U_s(t-\tau(t)), G_g(t-\tau(t)), t-\tau(t)) \tag{10}$$

$$= F_s(u(a,t-\tau(t)), U_s(t-\tau(t)), t-\tau(t)) + F_g(g(a,t-\tau(t)), G_g(t-\tau(t)), t-\tau(t)), \quad t>0,$$

$$w(a,t) = \varphi(a,t), \quad a\in[0,a_d], \; t\in[-\tau(0),0] \;, \tag{11}$$

Solution of problem (9) – (11) $w(a,t)$ and function $W(t)$ (1) are used in the coefficients of Eqs. (2), (3) and boundary conditions (4), (5).

### 3. "Travelling wave" solution of the problem (2) - (11)

The real semi-line $[0,\infty)$ is covered by the infinite union of disjoint intervals $[t_{k-1},t_k)$, $(t_k = ka_d,\; k\in N,\; t_0=0)$ which defines the following sets:

$$Q_k^{(1)} = \{(a,t) | \, t \in [(k-1)a_d, (a+(k-1)a_d)), \, a \in [0, a_d]\}, \tag{12}$$

$$Q_k^{(2)} = \{(a,t) | t \in [(a+(k-1)a_d), k\,a_d), \, a \in [0, a_d]\}, \tag{13}$$

where $\overline{Q} = \bigcup_{k=1}^{\infty} \left( Q_k^{(1)} \cup Q_k^{(2)} \right)$. Using the method of characteristics we can write Eq. (9) in the new characteristic variables $v = a - t$, $t' = t$ and reduce the problem (9) – (11) to the Cauchy problem for the linear ODE of first order:

$$\frac{\partial w}{\partial t'} = -\alpha_0(v + t', t')w, \tag{14}$$

$$w(v,0) = \varphi(v,0), \tag{15}$$

Boundary value (10) completes the problem (14) – (15). Solution of this problem is considered in the first couple of subsets $Q_1^{(1)}$, $Q_1^{(2)}$ (for $k = 1$), at the interval $t \in [0, \tau_0]$. Solution of problem (14) – (15) in former variables $a$, $t$ is given by:

$$w(a,t) = \begin{cases} w^{(1)}(v,t) = w^{(1)}(a-t,t), \quad if \ (a,t) \in Q_1^{(1)}; \\ \\ w^{(2)}(v,t) = w^{(2)}(a-t,t), \quad if \ (a,t) \in Q_1^{(2)}; \end{cases} \tag{16}$$

$$w^{(1)}(v,t) = \varphi(v,0) \exp\left( -\int_0^t \alpha_0(v+\xi,\xi)d\xi \right), \tag{17}$$

$$w^{(2)}(v,t) = F_w(u(a,-v-\tau(t)), g(a,-v-\tau(t)), U_s(-v-\tau(t)), G_g(-v-\tau(t)), -v-\tau(t)) \tag{18}$$

$$\times \exp\left( - \int_{-v/l}^{t} \alpha_0(v+\xi,\xi)d\xi \right)$$

The boundary value $F_0$ in (18) depends from the known values of functions $u(a,t-\tau(t))$, $g(a,t-\tau(t))$, $U_s(-v-\tau(t))$, $G_g(-v-\tau(t))$ taken at the previous moment of time from the initial conditions (6), (7), (11). Obtained solution $w(a,t)$ is substituted to the Eq. (2) and boundary condition (4). The problem (2) – (7) in new characteristic variables is reduced to the given Cauchy problem for the linear ODE:

$$\frac{\partial u}{\partial t'} = -\big(\alpha_0(v+t',t') + q_s(v+t',t',w(v+t',t')) + q_g(v+t',t',w(v+t',t'))\big)u \tag{19}$$

$$+ q_s(v+t',t',w(v+t',t'))w,$$

$$\frac{\partial g}{\partial t'} = -\big(\alpha_0(v+t',t') + q_s(v+t',t',w(v+t',t')) + q_g(v+t',t',w(v+t',t'))\big)g \tag{20}$$

$$+ q_g(v+t',t',w(v+t',t'))w,$$

$$u(v,0) = \varphi(v,0), \tag{21}$$

$$g(a,0) = 0, \tag{22}$$

Boundary values (4), (5) complete the problem (19) - (22). Solution of system (19), (20) at the current step of algorithm is considered also in the first couple of subsets $Q_1^{(1)}$, $Q_1^{(2)}$ at the interval $t \in [0, \tau_0]$ and may be written in the former variables $a$, $t$ in the following form:

$$u(a,t) = \begin{cases} u^{(1)}(v,t) = u^{(1)}(a-t,t), \quad if\ (a,t) \in Q_1^{(1)}; \\ \\ u^{(2)}(v,t) = u^{(2)}(a-t,t), \quad if\ (a,t) \in Q_1^{(2)}; \end{cases} \tag{23}$$

$$u^{(1)}(v,t) = \varphi(v,0)\exp\left(-\int_0^t \alpha(v+\xi,\xi)d\xi\right) + \int_0^t \exp\left(-\int_\xi^t \alpha(v+\eta,\eta)d\eta\right) \tag{24}$$

$$\times q_s(v+\xi,\xi,W(\xi))w^{(1)}(v+\xi,\xi)d\xi,$$

$$u^{(2)}(v,t) = F_s(u(a,-v-\tau(t)),U_s(-v-\tau(t)),-v-\tau(t))\exp\left(-\int_{-v}^t \alpha(v+\xi,\xi)d\xi\right) \tag{25}$$

$$+ \int_{-v}^t \exp\left(-\int_\xi^t \alpha(v+\eta,\eta)d\eta\right) q_s(v+\xi,\xi,W(\xi))w^{(2)}(v+\xi,\xi)d\xi,$$

$$g(a,t)=\begin{cases} g^{(1)}(v,t)=g^{(1)}(a-t,t), & if\ (a,t)\in Q_1^{(1)};\\ \\ g^{(2)}(v,t)=g^{(2)}(a-t,t), & if\ (a,t)\in Q_1^{(2)};\end{cases} \tag{26}$$

$$g^{(1)}(v,t)=\int_0^t \exp\left(-\int_\xi^t \alpha(v+\eta,\eta)d\eta\right) q_g(v+\xi,\xi,W(\xi))w^{(1)}(v+\xi,\xi)d\xi\,, \tag{27}$$

$$g^{(2)}(v,t)=F_g(g(a,-v-\tau(t)),G_g(-v-\tau(t)),-v-\tau(t))\exp\left(-\int_{-v}^t \alpha(v+\xi,\xi)d\xi\right) \tag{28}$$

$$+\int_{-v}^t \exp\left(-\int_\xi^t \alpha(v+\eta,\eta)d\eta\right) q_g(v+\xi,\xi,W(\xi))w^{(2)}(v+\xi,\xi)d\xi\,,$$

$$\alpha(a,t)=\alpha_0(a,t)+q_s(a,t,W(t))+q_g(a,t,W(t)). \tag{29}$$

where $\alpha(a,t)=\alpha_0(a,t)+q_s(a,t,w(a,t))+q_g(a,t,w(a,t))$. We repeat this procedure considering the next time intervals $t\in[\tau_0,2\tau_0]$, $t\in[2\tau_0,3\tau_0]$, etc., up to the boundary of subset $Q_1^{(2)}$ where $t=a_d$. Solution of system (2) - (11) in domain $Q_1^{(1)}\cup Q_1^{(2)}$ $(k=1)$ is given by recurrent formulae (16) – (18), (23) – (29). Then we repeat this procedure considering the next subsets $Q_k^{(1)}$, $Q_k^{(2)}$. For the arbitrary value of $k>1$ we obtain the travelling wave solution of system (2) – (11) in the following form:

$$w(a,t)=\begin{cases} w^{(k)}(v,t)=w^{(k)}(a-t,t), & if\ (a,t)\in Q_k^{(1)};\\ \\ w^{(k+1)}(v,t)=w^{(k+1)}(a-t,t), & if\ (a,t)\in Q_k^{(2)};\end{cases} \tag{30}$$

$$w^{(k+1)}(v,t)=F_w(u(a,-v-\tau(t)),g(a,-v-\tau(t)),U_s(-v-\tau(t)),G_g(-v-\tau(t)),-v-\tau(t)) \tag{31}$$

$$\times\exp\left(-\int_{-v}^t \alpha_0(v+\xi,\xi)d\xi\right)$$

$$u(a,t)=\begin{cases} u^{(k)}(v,t)=u^{(k)}(a-t,t), \quad if\ (a,t)\in Q_k^{(1)}; \\ \\ u^{(k+1)}(v,t)=u^{(k+1)}(a-t,t), \quad if\ (a,t)\in Q_k^{(2)}; \end{cases} \tag{32}$$

$$u^{(k+1)}(v,t)=F_S(u(a,-v-\tau(t)),U_S(-v-\tau(t)),-v-\tau(t))\exp\left(-\int_{-v}^{t}\alpha(v+\xi,\xi)d\xi\right) \tag{33}$$

$$+\int_{-v}^{t}\exp\left(-\int_{\xi}^{t}\alpha(v+\eta,\eta)d\eta\right)q_S(v+\xi,\xi,W(\xi))w^{(k+1)}(v+\xi,\xi)d\xi\,,$$

$$g(a,t)=\begin{cases} g^{(k)}(v,t)=g^{(k)}(a-t,t), \quad if\ (a,t)\in Q_k^{(1)}; \\ \\ g^{(k+1)}(v,t)=g^{(k+1)}(a-t,t), \quad if\ (a,t)\in Q_k^{(2)}; \end{cases} \tag{34}$$

$$g^{(k+1)}(v,t)=F_g(g(a,-v-\tau(t)),G_g(-v-\tau(t)),-v-\tau(t))\exp\left(-\int_{-v}^{t}\alpha(v+\xi,\xi)d\xi\right) \tag{35}$$

$$+\int_{-v}^{t}\exp\left(-\int_{\xi}^{t}\alpha(v+\eta,\eta)d\eta\right)q_g(v+\xi,\xi,W(\xi))w^{(k+1)}(v+\xi,\xi)d\xi\,,$$

Combining Eqs. (16) – (18), (23) - (29) with (30) – (35) we obtain the solution of the system (2) – (11) in all domain $\overline{Q}$. Although the density of gregarious phase can be found through already known solutions $g(a,t)=w(a,t)-u(a,t)$, we use the explicit form of problem (20), (22) and solution (26) – (28), (34), (35) for the analysis of continuity and smoothness of solution and equilibrium states of system in the following sections.

## 4. Compatibility (continuity and smoothness) conditions for travelling wave solution

For the two parts of travelling wave solution in Eqs. (16), (23), (26), (30), (32), (34) we have to consider the compatibility (continuity and smoothness) conditions. The continuity conditions $u(a,t)\in C\left(\overline{Q}\right)$, $g(a,t)\in C\left(\overline{Q}\right)$, $w(a,t)\in C\left(\overline{Q}\right)$ mean that the travelling wave solution at the points of characteristic $a=lt$ in directions $a=const$, $t=const$ must satisfy equation:

$$\lim_{a\to t^-}u(a,t)=\lim_{a\to t^+}u(a,t)\,,\quad \lim_{t\to a^-}u(a,t)=\lim_{t\to a^+}u(a,t)\,, \tag{36}$$

$$\lim_{a\to t^-} g(a,t) = \lim_{a\to t^+} g(a,t), \quad \lim_{t\to a^-} g(a,t) = \lim_{t\to a^+} g(a,t), \tag{37}$$

In order to fulfill the conditions (36), (37) we have to impose the continuity condition on the initial value $\varphi(a,t) \in C([0,a_d]\times[-\tau(0),0])$. From Eqs. (16) – (18), (23) - (29) emanates the continuity condition for the solutions $u(a,t)$, $g(a,t)$, $w(a,t)$, $(a,t)\in Q_1^{(1)} \cup Q_1^{(2)}$ in the following form:

$$u^{(1)}(0,t) = u^{(2)}(0,t), \quad g^{(1)}(0,t) = g^{(2)}(0,t). \tag{38}$$

From the condition (7) it follows that for each point $(a,t)\in Q_1^{(1)} \cup Q_1^{(2)}$ we have

$$F_g(0,0,t-\tau(t)) = 0, \tag{39}$$

$$F_w(\varphi(a,t-\tau(t)),0,U_s(t-\tau(t)),0,t-\tau(t)) = F_s(\varphi(a,t-\tau(t)),U_s(t-\tau(t)),t-\tau(t)). \tag{40}$$

Substituting Eqs. (17), (18), (24), (25), (27), (28) in (38) and using Eqs. (39), (40) yields the sufficient “continuity restriction” for the initial function $\varphi(a,t)$ and fertility rate from the boundary conditions (4), (5), (10):

$$\varphi(0,0) = F_s(\varphi(a,-\tau(0)),U_s(-\tau(0)),-\tau(0)) = \omega_s \int_{a_s(-\tau(0))}^{a_d} f_s(a,-\tau(0),U_s(-\tau(0)))\varphi(a,-\tau(0))da \tag{41}$$

Eq. (41) has a simple interpretation: the initial distribution of locust population have to satisfy the reproduction boundary condition at the initial moment of time. A second consequence of this Eq. is that the fertility rate of gregarious locusts is equal to null at the initial moment of time. So, if initial value $\varphi(a,t)$ and fertility rate $\sigma_1(a,t,W)$, $\sigma_2(a,t,W)$ fulfill the compatibility conditions (38) - (40), the solution is continuous: $u(a,t)\in C\left(Q_1^{(1)} \cup Q_1^{(2)}\right)$, $g(a,t)\in C\left(Q_1^{(1)} \cup Q_1^{(2)}\right)$, $w(a,t)\in C\left(Q_1^{(1)} \cup Q_1^{(2)}\right)$. From the boundary conditions (4), (5) it follows that if conditions (39) - (41) are fulfilled the solution remains continuous in the next domains $Q_k^{(1)} \cup Q_k^{(2)}$ for $k>1$, and therefore in all considered domain $u(a,t)\in C(\overline{Q})$, $g(a,t)\in C(\overline{Q})$, $w(a,t)\in C(\overline{Q})$.

The smoothness conditions $u(a,t)\in C^{(1)}(\overline{Q})$, $g(a,t)\in C^{(1)}(\overline{Q})$, $w(a,t)\in C^{(1)}(\overline{Q})$ at the points of characteristic $a=t$ in directions $a=const$, $t=const$ are given by:

$$\lim_{a\to t^-} \frac{\partial u}{\partial a} = \lim_{a\to t^-} \frac{\partial u}{\partial a}, \quad \lim_{t\to a^-} \frac{\partial u}{\partial t} = \lim_{t\to a^+} \frac{\partial u}{\partial t}, \tag{42}$$

$$\lim_{a\to t^-}\frac{\partial g}{\partial a}=\lim_{a\to t^+}\frac{\partial g}{\partial a},\quad \lim_{t\to a^-}\frac{\partial g}{\partial t}=\lim_{t\to a^+}\frac{\partial g}{\partial t},\tag{43}$$

Solutions (23), (26) allow us to write conditions (42), (43) for $k=1$ in the following form:

$$\left.\frac{\partial u^{(1)}}{\partial a}\right|_{a=0}=\left.\frac{\partial u^{(2)}}{\partial a}\right|_{a=0},\quad \left.\frac{\partial u^{(1)}}{\partial t}\right|_{a=0}=\left.\frac{\partial u^{(2)}}{\partial t}\right|_{a=0},\tag{44}$$

$$\left.\frac{\partial g^{(1)}}{\partial a}\right|_{a=0}=\left.\frac{\partial g^{(2)}}{\partial a}\right|_{a=0},\quad \left.\frac{\partial g^{(1)}}{\partial t}\right|_{a=0}=\left.\frac{\partial g^{(2)}}{\partial t}\right|_{a=0}.\tag{45}$$

The existence of partial derivatives in (44), (45) imposes the restrictions on the coefficients of Eqs. (2), (3), (9) and fertility rate in boundary conditions (4), (5), (10) which were formulated in [10], [26], [47]:

$$\varphi(a,t)\in C^{(1)}([0,a_d]\times[-\tau(0),0]),\ \ \alpha_0(a,t)\in C^{(1)}(Q),\ a_s(t),a_g(t)\in C^{(1)}([-\tau(0),\infty)),\tag{46}$$

$q_s(a,t,F)$, $q_g(a,t,F)$, $f_s(a,t,F)$, $f_g(a,t,F)$ $\in C^{(1)}\left(Q,R^+\right)$ ($i=\overline{1,2}$) are locally Lipchitz continuous functions with respect to $F$ uniformly for $(a,t)\in Q$ (47)

Substituting Eqs. (23) – (29), (32) – (35) in (44), (45) yields the first "smoothness" conditions for the initial function $\varphi(a,t)$, fertility rate $\sigma_i(a,t,W)$ and coefficients $\alpha_0(a,t)$, $q_s(a,t,W)$, $q_g(a,t,W)$ of system (2) – (11):

$$\left.\frac{\partial\varphi}{\partial a}\right|_{\substack{t=0\\a=0}}-\left.\frac{\partial F_s(\varphi(a,-v-\tau(0)),U_s(-v-\tau(0)),-v-\tau(0))}{\partial a}\right|_{\substack{v=0\\a=0}}\tag{48}$$

$$+\varphi(0,0)\alpha_0(0,0)=0,$$

$$q_g(a,0,W(0))=0.\tag{49}$$

In Eq. (49) we assumed that initial distribution $\varphi(a,t)$ satisfies the condition (41). The second "smoothness" condition for the initial function $\varphi(a,t)$ follows from the compatibility condition for the solution $w(a,t)$, $u(a,t)$ at the initial moment of time given in [19]:

$$\lim_{t\to 0^+}\frac{\partial u}{\partial t}=\lim_{t\to 0^-}\frac{\partial\varphi}{\partial t},\tag{50}$$

$$\lim_{t\to 0^+} \frac{\partial g}{\partial t} = 0 \tag{51}$$

Substituting Eq. (19) in Eq. (50) and using (7) yields the following restriction for the initial distribution $\varphi(a,t)$

$$\left.\frac{\partial \varphi}{\partial t}\right|_{t=0} = -\alpha_0(a,0)\varphi(a,0) - \left.\frac{\partial \varphi}{\partial a}\right|_{t=0},\ a \in [0, a_d] \tag{52}$$

The validity of condition (52) follows from Eqs. (7) and (50). From the boundary conditions (4), (5) follows that if conditions (48), (49), (52) are fulfilled the solution of system (2) – (11) remains continuously differentiable in the next domains $Q_k^{(1)} \cup Q_k^{(2)}$ for $k > 1$, and therefore in all domain $w(a,t) \in C^{(1)}(\overline{Q})$, $u(a,t) \in C^{(1)}(\overline{Q})$, $g(a,t) \in C^{(1)}(\overline{Q})$. We impose the biologically correct additional restrictions on the model parameters:

$$\alpha_0(a,t) > 0,\ \varphi(a,t) \ge 0,\ \omega > 0,\ a_s(t) \in [\breve{a}_s, \hat{a}_s],\ a_g(t) \in [\breve{a}_g, \hat{a}_g],\ \tau(t) \in [\tau_0, \tau_{\max}], \tag{53}$$

$$q(a,t,W(t)) \ge 0,\ f(a,t,W(t)) > 0,\ \forall W(t) \ge 0, \tag{54}$$

From all results from above we can formulate the following theorem.

**Theorem 1.** Let model parameters of the system (1) – (10) $\alpha_0$, $\varphi$, $q_s$, $q_g$, $f_s$, $f_g$, $\omega_s$, $\omega_g$, $a_s$, $a_g$ satisfy sufficient smoothness conditions (48), (49) and conditions (41), (50) - (54) then there exist a unique smooth travelling wave solution of problem (2) - (11) $u(a,t) \in C^{(1)}(\overline{Q})$, $g(a,t) \in C^{(1)}(\overline{Q})$, $w(a,t) \in C^{(1)}(\overline{Q})$ which can be obtained by explicit recurrent formulas (16) - (18), (22) - (37).

## 5. Existence of equilibriums of autonomous system for adult subpopulations

We first consider the autonomous system (2) – (11) for the constant coefficients:

$$\alpha_0 > 0,\ \tau > 0,\ 0 < a_s < a_d,\ 0 < a_g < a_d. \tag{55}$$

The phase-switching rates and fertility rates do not depend directly from time:

$$q_s(W(t)) \ge 0,\ q_g(W(t)) \ge 0, \tag{56}$$

$$f_s(a, U_s(t)) = m_s \mu_s \beta_s(a) h_s(U_s(t)), \tag{57}$$

$$\beta_s(a) = \chi(a - a_s)\chi(a_d - a)(a_d - a_s)^{-1}, \tag{58}$$

$$f_g(a, G_g(t)) = m_g \mu_g \beta_g(a) h_g(G_g(t)), \tag{59}$$

$$\beta_g(a) = \chi(a - a_g)\chi(a_d - a)(a_d - a_g)^{-1}, \tag{60}$$

where $\beta(a)$ is a dimensionless *maturation function* which is approximated by the continuous uniform distribution by age at the intervals $[a_s, a_d]$ and $[a_g, a_d]$ - “reproductive windows”*;* $\mu > 0$ is a number of new born locusts from one pod; $m > 0$ is a *number of eggpods* locust female lays over her lifetime (dimensionless constants); $\chi(x) = \begin{cases} 1, \ if \ x \geq 0 \\ 0, \ if \ x < 0 \end{cases}$ is a dimensionless Heaviside function; $h_s(U_s(t))$, $h_g(G_g(t)) \in [0,1]$ are dimensionless density-dependent *fertility response functions* (or *fecundity response function* [15], [16], survival rate of eggs to mortality by predation [7]). The fertility rates of solitarious (57) and gregarious (59) contain the constants $\mu$ and $m$ which are the assessments of maximum values of number of new born locusts from one pod and number of eggpods that a locust female can lay over her lifetime. In general case these parameters depend from the density of solitarious and gregarious. We assume here that the density-dependent fertility response functions simulate also the dependencies of $\mu$ and $m$ from population density.

The asymptotically stable states of locust population is studied theoretically in approximation that one subpopulation (solitarious or gregarious) reaches some steady state when phase switching processes in both directions has been already ended. Thus, the “gregarization” or “solitarization” occurs only in the intermediate period when the locust population evolves to some asymptotically stable state. The stability analysis is carried out for the dynamics of a number of matured locusts which are able to reproduce new generation, for solitarious at age $a \geq a_s$ and gregarious at age $a \geq a_g$. We consider two particular cases of equilibriums in locust population.

*1. Asymptotically stable solitarious dynamics without gregarization* $W \leq W_{0s}$, $q_g(W) = 0$, $q_s(W) > 0$, $W(t) = U(t)$, $G(t) = 0$ *(see Fig.2).*

The nontrivial equilibrium points $W_j^* > 0$, $U_j^* = W_j^*$, $G^* = 0$, $j \in N^* \subseteq N$ are obtained from the system (2) - (11). Integrating Eq. (9) with respect to age from $a_s$ to $a_d$ and using Eq. (31), boundary conditions (4), (5) with Eqs. (56) - (60), yield the initial problem for nonlinear autonomous ODE for dynamics of the numbers of adult locusts $W_s(t)$ in solitarious phase:

$$W_s^{'}(t) = -\alpha_0 W_s(t) + \omega_s \mu_s m_s (a_d - a_s)^{-1} \Big( h_s(W_s(t-\tau_s)) W_s(t-\tau_s) \qquad (61)$$

$$\times \exp\left(-\alpha_0 a_s\right) - h_s(W_s(t-\tau_d)) W_s(t-\tau_d) \exp\left(-\alpha_0 a_d\right) \Big),$$

$$W_s(t) = \int_{a_s}^{a_d} w(a,t) da \,, \quad t \in \left[-\tau, a_d\right] \,, \qquad (62)$$

where $\tau_s = \tau + a_s$, $\tau_d = \tau + a_d$. Initial problem (61), (62) always has a trivial equilibrium point $W_s^* = 0$, $U_s^* = 0$, $G_s^* = 0$ and may have countable set of nontrivial equilibrium points $W^*_{s_j} > 0$, $U^*_{s_j} = W^*_{s_j}$, $G^*_{s_j} = 0$, $j \in N^* \subseteq N$, at the time moment $t^*$, which satisfy the nonlinear algebraic Eqs., obtained from Eqs. (61):

$$q_s(W) = 0,\ \ q_g(W) > 0,\ W^*_{s_j} > 0,\ U^*_{s_j} = W^*_{s_j},\ \ G^*_s = 0, \tag{63}$$

$$-\alpha_0 + \omega_s \mu_s m_s (a_d - a_s)^{-1} h_s(W^*_{s_j})\big(\exp(-\alpha_0 a_s) - \exp(-\alpha_0 a_d)\big) = 0. \tag{64}$$

*2. Asymptotically stable gregarious dynamics without reverse phase transition to solitarious phase* $W^*_j \ge W_{0g}$, $q_g(W) > 0$, $q_s(W) = 0$, $W(t) = G(t)$, $U(t) = 0$ *(see Fig.2).*

Integrating Eq. (9) by age from $a_g$ to $a_d$, and using the condition $w(a_d, t) = 0$, Eq. (31), boundary conditions (4), (5) with Eqs. (56) - (60), yield the initial problem for nonlinear autonomous ODE for dynamics of the numbers of adult locusts $W_g(t)$ in gregarious phase:

$$W'_g(t) = -\alpha_0 W_g(t) + \omega_g \mu_g m_g (a_d - a_g)^{-1}\big(h_g(W_g(t-\tau_g))W_g(t-\tau_g) \tag{65}$$

$$\times \exp(-\alpha_0 a_g) - h_g(W_g(t-\tau_d))W_g(t-\tau_d)\exp(-\alpha_0 a_d)\big),$$

$$W_g(t) = \int_{a_g}^{a_d} w(a,t)\,da,\ \ t \in [-\tau, a_d]\ , \tag{66}$$

where $\tau_g = \tau + a_g$, $\tau_d = \tau + a_d$. We do not consider a trivial equilibrium of problem (65), (66) because the gregarious phase follows the solitarious phase and cannot appear immediately from the zero value of locust population. Thus, initial problem (65), (66) may have countable set of nontrivial equilibriums $W^*_{g_j} > 0$, $G^*_{g_j} = W^*_{g_j}$, $U^*_{g_j} = 0$, $j \in N^* \subseteq N$, at the time moment $t^*$, which satisfy the nonlinear algebraic Eqs., obtained from Eqs. (65):

$$q_g(W) > 0,\ \ q_s(W) = 0,\ W^*_{g_j} > 0,\ G^*_{g_j} = W^*_{g_j},\ U^*_g = 0, \tag{67}$$

$$-\alpha_0 + \omega_g \mu_g m_g (a_d - a_g)^{-1} h_g(W^*_{g_j})\big(\exp(-\alpha_0 a_g) - \exp(-\alpha_0 a_d)\big) = 0 \tag{68}$$

From Eqs. (64), (68) it follows that the necessary conditions of existence of nontrivial equilibriums for solitarious and gregarious populations respectively have a form:

$$h^*_s = (\Lambda_1 - \Lambda_2)^{-1} \le 1, \tag{69}$$

$$\Lambda_1 = \alpha_0^{-1}(a_d - a_s)^{-1}\omega_s \mu_s m_s \exp(-\alpha_0 a_s), \tag{70}$$

$$\Lambda_2 = \alpha_0^{-1}(a_d - a_s)^{-1}\omega_s \mu_s m_s \exp(-\alpha_0 a_d), \tag{71}$$

$$h_g^* = (\Lambda_3 - \Lambda_4)^{-1} \le 1, \tag{72}$$

$$\Lambda_3 = \alpha_0^{-1}(a_d - a_g)^{-1}\omega_g \mu_g m_g \exp\left(-\alpha_0 a_g\right), \tag{73}$$

$$\Lambda_4 = \alpha_0^{-1}(a_d - a_g)^{-1}\omega_g \mu_g m_g \exp\left(-\alpha_0 a_d\right), \tag{74}$$

where $h_s^*$, $h_g^*$ and $\Lambda_i$ $(i = 1,...,4)$ are dimensionless positive constants which indicate the existence of nontrivial system equilibriums. If condition (69) does not hold, the derivative $\dot{W}_s(t^*) < 0$ in Eq. (61) and the problem (61), (62) possesses only trivial equilibrium. If condition (72) does not hold, the derivative $\dot{W}_g(t^*) < 0$ and solution of the problem (65), (66) evolves also to the trivial equilibrium. We can collect all obtained results from this section in the following Theorem.

**Theorem 2.** Equations (61) and (65) can possess nontrivial (positive) equilibriums $W_{s_j}^* > 0$, $U_{s_j}^* = W_{s_j}^*$, $G_{s_j}^* = 0$, and $W_{g_j}^* > 0$, $G_{g_j}^* = W_{g_j}^*$, $U_{g_j}^* = 0$, respectively. The nontrivial equilibrium of solitarious phase (first equilibrium) exists if the dimensionless parameters $\Lambda_1 > 0$ (70) and $\Lambda_2 > 0$ (71) satisfy conditions (69) and fertility response function of solitarious satisfies Eq. $h_s(U_{s_j}^*) = h_s^*$. The nontrivial equilibrium of gregarious phase (second equilibrium) exists if the dimensionless parameters $\Lambda_3 > 0$ (73) and $\Lambda_4 > 0$ (74) satisfy conditions (72) and fertility response function of gregarious satisfies Eq. $h_g(G_{g_j}^*) = h_g^*$. If condition (69) or (72) do not hold Eq. (61) or (65) possesses only trivial equilibrium.

It is worth to note that from biological point of view Theorem 2 indicates that nontrivial equilibrium only exist at given situations of length of reproductive age ("reproductive window") for given parameters of other life processes.

## 6. Stability analysis of autonomous system for adult subpopulations

By analogy with the Theorem 1 from [7] formulated for a nonlinear age-structured model with density-dependent fertility response function, conditions (69) and (72) show that nontrivial equilibriums of Eqs. (61) and (65) are the results of balance between the mortality and fertility of population which influence the probability of locust birth. Obviously, if conditions (69) and (72) are not valid, then expressions (64) and (68) are negative and, as a consequence, $W_s^{'}(t) < 0$ in Eq. (61), $W_g^{'}(t) < 0$ in Eq. (65), and the nontrivial equilibriums do not exist. To identify trivial and nontrivial equilibriums we carry out the analysis of local asymptotical stability.

To begin with, we follow the procedure of stability analysis from [31] and arrive to the linearized Eqs. (61) at trivial equilibrium $W_s^* = 0$, $U_s^* = 0$, $G_s^* = 0$ for continuously differentiable function $V(t)$:

$$\dot{V}(t) = -\alpha_0 V(t) + \alpha_0 h_s(0)\left(\Lambda_1 V(t-\tau_s) - \Lambda_2 V(t-\tau_d)\right), \quad (75)$$

The corresponding characteristic Eq. reads as:

$$z(\lambda) = \lambda + \alpha_0 - \alpha_0 Y\left(\Lambda_1 \exp(-\lambda\tau_s) - \Lambda_2 \exp(-\lambda\tau_d)\right) = 0, \quad (76)$$

were we introduce the auxiliary parameter $Y = h_s(0)$. If $Y = h_s(0) = 0$ Eq. (76) has only one real negative root $\lambda = -\alpha_0$ and trivial equilibrium is unconditionally stable. For $h_s(0) \in (0,1]$ function $z(\lambda)$ possesses the following properties:

$$z(0) = \alpha_0\left(1 - Y(\Lambda_1 - \Lambda_2)\right), \quad (77)$$

$$\lim_{\lambda\to\infty} z(\lambda) = \infty. \quad (78)$$

Thus, characteristic Eq. (76) always has at least one real positive or trivial root if the density-dependent fertility response function of solitarious satisfies the following restrictions:

$$(\Lambda_1 - \Lambda_2)^{-1} = h_s^* \le Y. \quad (79)$$

In this case $z(0) \le 0$ and function $z(\lambda)$ intersects a nonnegative real semi-line $R^+$ at least once at a fixed point $\lambda \ge 0$. Trivial equilibrium is not stable for any given values of time delay $\tau > 0$ and maturing parameter $a_s \in (0, a_d)$. In other case, if condition (79) does not hold the trivial equilibrium exists, and instead of condition (79) we obtain two inequalities:

$$0 < Y < h_s^* = (\Lambda_1 - \Lambda_2)^{-1}. \quad (80)$$

In such case, function $z(0) > 0$ (Eq. (77)) and derivative $z'(\lambda)$ is positive for all $\lambda \ge 0$:

$$z'(\lambda) > 1 + \alpha_0(1 - \Lambda_1^{-1}\Lambda_2)^{-1}\tau_s \exp(-\lambda\tau_s) - \alpha_0(\Lambda_2^{-1}\Lambda_1 - 1)^{-1}\tau_d \exp(-\lambda\tau_d) \quad (81)$$

$$= \left(\exp(\alpha_0(\tau_d - \tau_s) - 1\right)^{-1}\alpha_0\tau_d \exp(-\lambda\tau_d)\Big((\alpha_0\tau_d)^{-1}\exp(\lambda\tau_d)(\exp(\alpha_0(\tau_d - \tau_s)) - 1)$$

$$+\sigma_s \exp((\alpha_0 + \lambda)(\tau_d - \tau_s)) - 1\Big) = \left(\exp(\alpha_0(\tau_d - \tau_s) - 1\right)^{-1}\alpha_0\tau_d \exp(-\lambda\tau_d)$$

$$\times\left((1 + \lambda\tau_d + R_{11})(1 - \sigma_s + R_{12}) - (1 - \sigma_s) + \sigma_s((\alpha_0 + \lambda)(\tau_d - \tau_s) + R_{13})\right) > 0.$$

where $\sigma_s = \tau_s\tau_d^{-1}$ is a dimensionless constant, $\sigma_s \in (0,1)$; $R_{1i} > 0$ are positive remainder terms in the Taylor series expansion, $(i = 1,\ldots,3)$. In this case, the characteristic Eq. (76) does not have real positive roots, but it can have complex root $\lambda = \varsigma + i\zeta$ with nonnegative real part $\varsigma > 0$ and nontrivial imaginary part $\zeta \ne 0$. Substituting complex $\lambda$ in Eq.

(76) we arrive to the two equations for the real $z^{(1)}(\varsigma,\zeta)$ and imaginary $z^{(2)}(\varsigma,\zeta)$ parts of $z(\lambda)$:

$$z^{(1)}(\varsigma,\zeta) = \varsigma + \alpha_0 - \alpha_0 Y\left(\Lambda_1 \exp(-\varsigma\tau_s)\cos(\zeta\tau_s) - \Lambda_2 \exp(-\varsigma\tau_d)\cos(\zeta\tau_d)\right) = 0\,, \tag{82}$$

$$z^{(2)}(\varsigma,\zeta) = \zeta + \alpha_0 Y\left(\Lambda_1 \exp(-\varsigma\tau_s)\sin(\zeta\tau_s) - \Lambda_2 \exp(-\varsigma\tau_d)\sin(\zeta\tau_d)\right) = 0\,. \tag{83}$$

For $\varsigma \geq 0$ functions $z^{(1)}(\varsigma,\zeta)$ and $z^{(2)}(\varsigma,\zeta)$ possess the properties:

$$z^{(1)}(\varsigma,0) > 0\,, \quad \left(z^{(1)}\right)'_\zeta \Big|_{\zeta=0} > 0\,, \tag{84}$$

$$z^{(2)}(\varsigma,0) = 0\,, \quad \left(z^{(2)}\right)'_\zeta \Big|_{\zeta=0} > 0\,, \tag{85}$$

$$\min_{\substack{\varsigma\geq 0,\\ \zeta\neq 0}} \left(z^{(1)}(\varsigma,\zeta)\right) = \min_{\zeta\neq 0}\left(z^{(1)}(0,\zeta)\right) \geq \alpha_0\left(1 - Y\left(\Lambda_1 + \Lambda_2\right)\right), \tag{86}$$

$$\min_{\substack{\varsigma\geq 0,\\ \zeta\neq 0}} \left(z^{(2)}(\varsigma,\zeta)\right) = \min_{\zeta\neq 0}\left(z^{(2)}(0,\zeta)\right) \geq -|\zeta| - \alpha_0 Y\left(\Lambda_1 + \Lambda_2\right). \tag{87}$$

From Eqs. (84) - (87) it follows that Eqs. (82), (83) can have a root $\varsigma^* \geq 0$, $\zeta^* \neq 0$ if functions $z^{(1)}(\varsigma,\zeta)$ and $z^{(2)}(\varsigma,\zeta)$ take trivial or negative values at $\varsigma = 0$ (neutral stability). Otherwise, $\min\limits_{\zeta\neq 0}\left(z^{(1)}(0,\zeta)\right) > 0$ and Eq. (82) does not have a root $\varsigma^* \geq 0$, $\zeta^* \neq 0$. Substituting the neutral stability value $\varsigma = 0$ into Eqs. (82) and (83) we have

$$Y\left(\Lambda_1 \cos(\zeta\tau_s) - \Lambda_2 \cos(\zeta\tau_d)\right) = 1, \tag{88}$$

$$Y\left(\Lambda_1 \sin(\zeta\tau_s) - \Lambda_2 \sin(\zeta\tau_d)\right) = -\zeta\alpha_0^{-1}. \tag{89}$$

Squaring and summing Eqs. (88) and (89) yields the transcendent equation for real parameter $\zeta \neq 0$:

$$\zeta^2 = \alpha_0^2\left(Y^2\left(\Lambda_1^2 + \Lambda_2^2 - 2\Lambda_1\Lambda_2\cos(\zeta(a_d - a_s))\right) - 1\right). \tag{90}$$

**Lemma 1.** If conditions (80) hold Eq. (90) does not have nontrivial real root $\zeta \neq 0$. (The proof of Lemma 1 is given in Appendix A).

Overall, we can conclude that if condition (80) holds, characteristic equation (76) does not have trivial, real positive roots or complex roots with nonnegative real part. The results of stability analysis of trivial equilibriums obtained in this section are collected in Theorem 3.

**Theorem 3.** The trivial equilibrium $W_s^* = 0$, $U_s^* = 0$, $G_s^* = 0$ of autonomous system (61), (62) and (65), (66) is not locally asymptotically stable for all $\tau > 0$, $a_s \in (0, a_d)$ if fertility response function of solitarious $h_s(0)$ satisfies conditions (79). Otherwise, if $h_s(0)$ satisfies condition (80) the trivial equilibrium is locally asymptotically stable for all $\tau > 0$ and $a_s \in (0, a_d)$.

Secondly, linearized Eqs. (61) and (65) at the nontrivial equilibrium points $W_{s_j}^*$ (solution of Eq. (64)) and $W_{g_j}^*$ (solution of Eq. (68)) together with Eqs. (64), (68) respectively, read as:

$$\dot{V}(t) = -\alpha_0 V(t) + \alpha_0 \left( (\Lambda_1 - \Lambda_2)^{-1} + h_s'(W_{s_j}^*) W_{s_j}^* \right) \left( \Lambda_1 V(t - \tau_s) - \Lambda_2 V(t - \tau_d) \right), \qquad (91)$$

$$\dot{Y}(t) = -\alpha_0 Y(t) + \alpha_0 \left( (\Lambda_3 - \Lambda_4)^{-1} + h_g'(W_{g_j}^*) W_{g_j}^* \right) \left( \Lambda_3 V(t - \tau_g) - \Lambda_4 V(t - \tau_d) \right). \qquad (92)$$

The corresponding characteristic equations read:

$$z_1(\lambda_1) = \lambda_1 + \alpha_0 - \alpha_0 Y_1 \left( \Lambda_1 \exp(-\lambda_1 \tau_s) - \Lambda_2 \exp(-\lambda_1 \tau_d) \right) = 0, \qquad (93)$$

$$z_2(\lambda_2) = \lambda_2 + \alpha_0 - \alpha_0 Y_2 \left( \Lambda_3 \exp(-\lambda_2 \tau_g) - \Lambda_4 \exp(-\lambda_2 \tau_d) \right) = 0, \qquad (94)$$

where $Y_1 = (\Lambda_1 - \Lambda_2)^{-1} + h_s'(W_{s_j}^*) W_{s_j}^*$, $Y_2 = (\Lambda_3 - \Lambda_4)^{-1} + h_g'(W_{g_j}^*) W_{g_j}^*$. The nontrivial equilibriums $W_{s_j}^* > 0$, $U_{s_j}^* = W_{s_j}^*$, $G_{s_j}^* = 0$ and $W_{g_j}^* > 0$, $G_{g_j}^* = W_{g_j}^*$, $U_{g_j}^* = 0$, are not stable if Eqs. (93) and (94) have nonnegative roots. By analogy with Eqs.(77), (78), functions $z_1(\lambda_1)$ and $z_2(\lambda_2)$ possess the same properties for $Y_1, Y_2 \neq 0$:

$$z_1(0) = \alpha_0 \left( 1 - Y_1 (\Lambda_1 - \Lambda_2) \right), \qquad (95)$$

$$\lim_{\lambda_1 \to \infty} z_1(\lambda_1) = \infty, \qquad (96)$$

$$z_2(0) = \alpha_0 \left( 1 - Y_2 (\Lambda_3 - \Lambda_4) \right), \qquad (97)$$

$$\lim_{\lambda_2 \to \infty} z_2(\lambda_2) = \infty, \qquad (98)$$

By analogy with Eq. (79), from Eqs. (95), (97) it follows that characteristic Eqs. (93), (94) always have at least one real positive or trivial root if coefficients of system satisfy the following restrictions:

$$0 < (\Lambda_1 - \Lambda_2)^{-1} = h_s^* \leq Y_1, \qquad (99)$$

$$0 < (\Lambda_3 - \Lambda_4)^{-1} = h_g^* \leq Y_2. \qquad (100)$$

which can be given in the more compact form:

$$h_s^{'}(W_{s_j}^{*}) = \frac{dh_s}{dU_s}\bigg|_{U_s = W_{s_j}^{*}} \geq 0, \tag{101}$$

$$h_g^{'}(W_{g_j}^{*}) = \frac{dh_g}{dG_g}\bigg|_{G_g = W_{g_j}^{*}} \geq 0 \tag{102}$$

Indeed, conditions (101), (102) provide the change of sign of functions $z_1(\lambda_1)$ and $z_2(\lambda_2)$ at least once at some fixed points of nonnegative real semi-line $\lambda_1 \geq 0$, $\lambda_2 \geq 0$ respectively. In this case the nontrivial equilibriums $W_{s_j}^{*} > 0$, $U_{s_j}^{*} = W_{s_j}^{*}$, $G_s^{*} = 0$ and $W_{g_j}^{*} > 0$, $G_{g_j}^{*} = W_{g_j}^{*}$, $U_g^{*} = 0$ are not stable for any given values of time delay $\tau > 0$ and maturing parameters $a_s \in (0, a_d)$, $a_g \in (0, a_d)$. In the next case the coefficients of system satisfy the following restrictions:

$$Y_1 < h_s^{*} = (\Lambda_1 - \Lambda_2)^{-1}, \tag{103}$$

$$Y_2 < h_g^{*} = (\Lambda_3 - \Lambda_4)^{-1}. \tag{104}$$

which can be given in the more compact form:

$$h_s^{'}(W_{s_j}^{*}) = \frac{dh_s}{dU_s}\bigg|_{U_s = W_{s_j}^{*}} < 0, \tag{105}$$

$$h_g^{'}(W_{g_j}^{*}) = \frac{dh_g}{dG_g}\bigg|_{G_g = W_{g_j}^{*}} < 0. \tag{106}$$

In this case we consider two groups of conditions for $Y_1 > 0$, $Y_2 > 0$ and $Y_1 < 0$, $Y_2 < 0$. The first one is given in the form of two inequalities:

$$-(\Lambda_1 - \Lambda_2)^{-1} < h_s^{'}(W_{s_j}^{*})W_{s_j}^{*} < 0, \tag{107}$$

$$-(\Lambda_3 - \Lambda_4)^{-1} < h_g^{'}(W_{g_j}^{*})W_{g_j}^{*} < 0. \tag{108}$$

The characteristic Eqs. (93), (94) in this case have the same form as Eq. (76) and we can apply the results of stability analysis obtained for the trivial equilibrium that satisfied Eq. (80) to the solution of Eqs. (93), (94). Thus, if conditions (97), (98) hold, characteristic equations (93), (94) do not have trivial, real positive roots or complex roots with nonnegative real part.

The second group of conditions for $Y_1 < 0$, $Y_2 < 0$ is given in the form of two inequalities:

$$h_s^{'}(W_{s_j}^*)W_{s_j}^* < -(\Lambda_1 - \Lambda_2)^{-1} < 0, \tag{109}$$

$$h_g^{'}(W_{g_j}^*)W_{g_j}^* < -(\Lambda_3 - \Lambda_4)^{-1} < 0. \tag{110}$$

Using Eqs. (69), (72) we obtain the following auxiliary inequalities:

$$\alpha_0 + \alpha_0\left(\left|h_s^{'}(W_{s_j}^*)\right|W_{s_j}^* - (\Lambda_1 - \Lambda_2)^{-1}\right)\left(\Lambda_1 \exp(-\lambda_1\tau_s) - \Lambda_2 \exp(-\lambda_1\tau_d)\right) > 0,\ \lambda_1 \geq 0, \tag{111}$$

$$\alpha_0 + \alpha_0\left(\left|h_g^{'}(W_{g_j}^*)\right|W_{g_j}^* - (\Lambda_3 - \Lambda_4)^{-1}\right)\left(\Lambda_3 \exp(-\lambda_2\tau_g) - \Lambda_4 \exp(-\lambda_2\tau_d)\right) > 0,\ \lambda_2 \geq 0, \tag{112}$$

From inequalities (111), (112) it follows that characteristic Eqs. (93), (94) do not have nonnegative real roots. On the other hand they can have complex roots $\lambda_1 = \varsigma_1 + i\zeta_1$, $\lambda_2 = \varsigma_2 + i\zeta_2$ with nonnegative real parts $\varsigma_1 \geq 0$, $\varsigma_2 \geq 0$ and nontrivial imaginary parts $\zeta_1 \neq 0$, $\zeta_2 \neq 0$. Substituting complex $\lambda_1$ in Eq. (88) and complex $\lambda_2$ in Eq. (89) we arrive to the four equations for the real $z_1^{(1)}(\varsigma_1, \zeta_1)$, $z_2^{(1)}(\varsigma_2, \zeta_2)$ and imaginary $z_1^{(2)}(\varsigma_1, \zeta_1)$, $z_1^{(2)}(\varsigma_2, \zeta_2)$ parts of $z_1(\lambda_1)$, $z_2(\lambda_2)$, respectively, of the same type as Eqs. (82), (83). Using the same approach as in the case of trivial equilibrium, we arrive to two transcendent equations for the imaginary parts $\zeta_1 \neq 0$, $\zeta_2 \neq 0$ which have the same form as Eq. (90):

$$\zeta_1^2 = \alpha_0^2\left(Y_1^2\left(\Lambda_1^2 + \Lambda_2^2 - 2\Lambda_1\Lambda_2\cos(\zeta_1(a_d - a_s))\right) - 1\right), \tag{113}$$

$$\zeta_2^2 = \alpha_0^2\left(Y_2^2\left(\Lambda_3^2 + \Lambda_4^2 - 2\Lambda_3\Lambda_4\cos(\zeta_2(a_d - a_g))\right) - 1\right). \tag{114}$$

**Lemma 2.** If conditions (103), (104), (109), (110) hold, Eqs. (113), (114) do not have nontrivial real roots $\zeta_1 \neq 0$, $\zeta_2 \neq 0$. (The proof of Lemma 2 is given in Appendix B).

Overall, if conditions (103), (104) hold, characteristic Eqs. (93), (94) do not have real nonnegative roots or complex roots with nonnegative real part. We can collect all obtained conditions of asymptotical stability of nontrivial equilibriums of the initial problems (61), (62) and (65), (66) in the form of the following Theorem 4.

**Theorem 4.** The nontrivial equilibriums $W_{s_j}^* > 0$, $U_{s_j}^* = W_{s_j}^*$, $G_s^* = 0$ and $W_{g_j}^* > 0$, $G_{g_j}^* = W_{g_j}^*$, $U_g^* = 0$, $j \in N^* \subseteq N$ of autonomous system (61), (62) and (65), (66) are not locally asymptotically stable for all $\tau > 0$, $a_s \in (0, a_d)$, $a_g \in (0, a_d)$ if the derivatives of density-dependent fertility response functions $h_s^{'}(W_{s_j}^*)$ and $h_g(W_{g_j}^*)$ satisfy conditions (101), (102) respectively. Otherwise, if $h_s(W_{s_j}^*)$ and $h_g(W_{g_j}^*)$ satisfy conditions (105), (106) nontrivial equilibriums are locally asymptotically stable for all $\tau > 0$, $a_s \in (0, a_d)$ and $a_g \in (0, a_d)$.

The result of theoretical analysis of trivial and nontrivial equilibriums and asymptotically stable regimes of initial problems with autonomous Eqs. (61), (62) and (65), (66) can be applied to the system (2) – (11): if the conditions of Theorems 2 - 4 hold, system (2) – (11) with autonomous equations has an asymptotically stable distributions $u_j(a,t^*)$, $g_j(a,t^*)$, $w_j(a,t^*)$ which correspond to the equilibriums of adult locust subpopulations $U^*_{s_j}$, $G^*_{s_j}$, $W^*_{s_j}$ or $U^*_{g_j}$, $G^*_{g_j}$, $W^*_{g_j}$ of initial problems (61), (62) and (65), (66) respectively.

## 7. Simulation of locust population dynamics for autonomous system

A first series of experiments is carried out for the autonomous system (2) – (10) with coefficients which satisfy the conditions of Theorems 1 - 4. Our strategy in these experiments is to study the different dynamical regimes for the tree types of equilibriums – trivial and nontrivial (positive) with various fertility response functions $h_s(U_s(t))$, $h_g(G_g(t))$ and coefficients of the system. For the simulation of phase polyphenism we use the model of age and time-dependent force of infection from the age-structured epidemic model of population dynamics [7], [9], [24], [25], [44]. The set of algebraic functions for the phase-switching rates and fertility response functions is given in the following form:

$$q_g(W(t)) = q_a\chi(W(t)-W_{\max}) + \chi(W(t)-W_{0g})\chi(W_{\max}-W(t)) \tag{115}$$

$$\times 0.5q_a\left(1+\sin\left((W(t)-W_{\max})(W_{\max}-W_{0g})^{-1}\pi - 0.5\pi\right)\right),$$

$$q_s(W(t)) = q_b\chi(W_{\min}-W(t)) + \chi(W_{0s}-W(t))\chi(W(t)-W_{\min}) \tag{116}$$

$$\times 0.5q_b\left(1+\sin\left((W(t)-W_{\min})(W_{0s}-W_{\min})^{-1}\pi + 0.5\pi\right)\right)$$

$$h_s(U_s) = \sigma_b + (\sigma_a-\sigma_b)\left(0.5 \pm arctg(\gamma(U_s - U_{s_{\min}}))/\pi\right), \tag{117}$$

$$h_s^{'}(U_s) = \pm(\sigma_a-\sigma_b)\pi^{-1}\gamma\left(1+\gamma^2(U_s-U_{s_{\min}})^2\right)^{-1}, \tag{118}$$

$$h_g(G_g) = \sigma_d + (\sigma_c-\sigma_d)\left(0.5 \pm arctg(\gamma(G_g - G_{g_{\max}}))/\pi\right), \tag{119}$$

$$h_g^{'}(G_g) = \pm(\sigma_c-\sigma_d)\pi^{-1}\gamma\left(1+\gamma^2(G_g-G_{g_{\max}})^2\right)^{-1} < 0. \tag{120}$$

where $q_a > 0$, $q_b > 0$, $\sigma_a > \sigma_b > 0$, $\sigma_c > \sigma_d > 0$, $W_{\max} > 0$, $W_{0g} > 0$, $W_{\min} > 0$, $W_{0s} > 0$, $U_{s_{\min}} > 0$, $G_{g_{\max}} > 0$ are given positive constants – parameters of the model. In all experiments we consider only monotonic functions $q_s(W(t))$, $q_g(W(t))$, $h_s(U_s(t))$, $h_g(G_g(t))$, the graphs of which are shown in the Fig.2 and Fig.3. We assume that the conditions $h_s^* < 1$ (69) and $h_g^* < 1$ (72) hold. Experiments are performed for the all types of algebraic functions of the fertility response function which correspond to different statements of Theorems 3 and 4 using the expected values of biological constants given in Table 2 (Appendix C).

*7.1 Experiment I*

In the first group of experiments (Experiment I), we study the stability of trivial equilibrium of solitarious subpopulation with the various fertility response functions which satisfy the conditions of instability (79) and stability (80) according to the statement of Theorem 3. The dynamics of the number of adult solitarious and number of all solitarious calculated in the Experiment I are shown on the Fig.4a, b. We do not consider here the direct phase changing "solitarious-gregarious" and study only the possibility and conditions of such transition. All simulations are carried out for the same initial values of population density and different fertility response functions of the type (117) ($h_s^{'}(U_s) > 0$) with various constants $\sigma_a$ and $\sigma_b$. We obtain two dynamical regimes of population: 1) when $h_s(0) < h_s^*$, the population density evolves from the neighborhood of trivial equilibrium to null (the graphs below the dashed line in the Fig.4a, b); 2) when $h_s(0) > h_s^*$ solitarious subpopulation density grows and can evolve to the stable or unstable solitarious or gregarious nontrivial equilibrium.

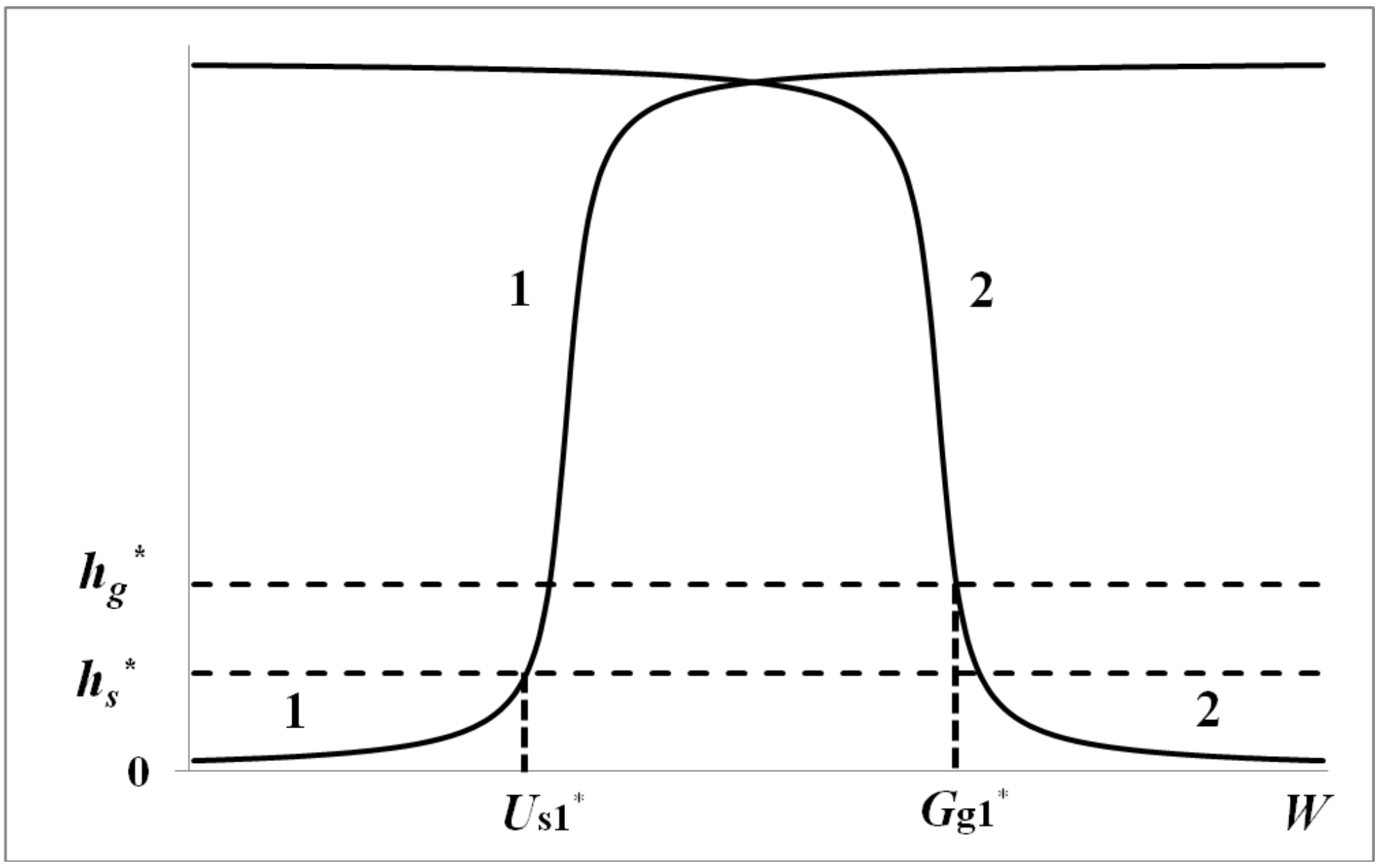

Figure 3. Graphs of fertility response functions when $h_s(U_s)$ is monotonically increasing $h_s^{'}(U_s) > 0$ (line 1) and $h_g(G_g)$ is monotonically decreasing $h_g^{'}(G_g) < 0$ (line 2).

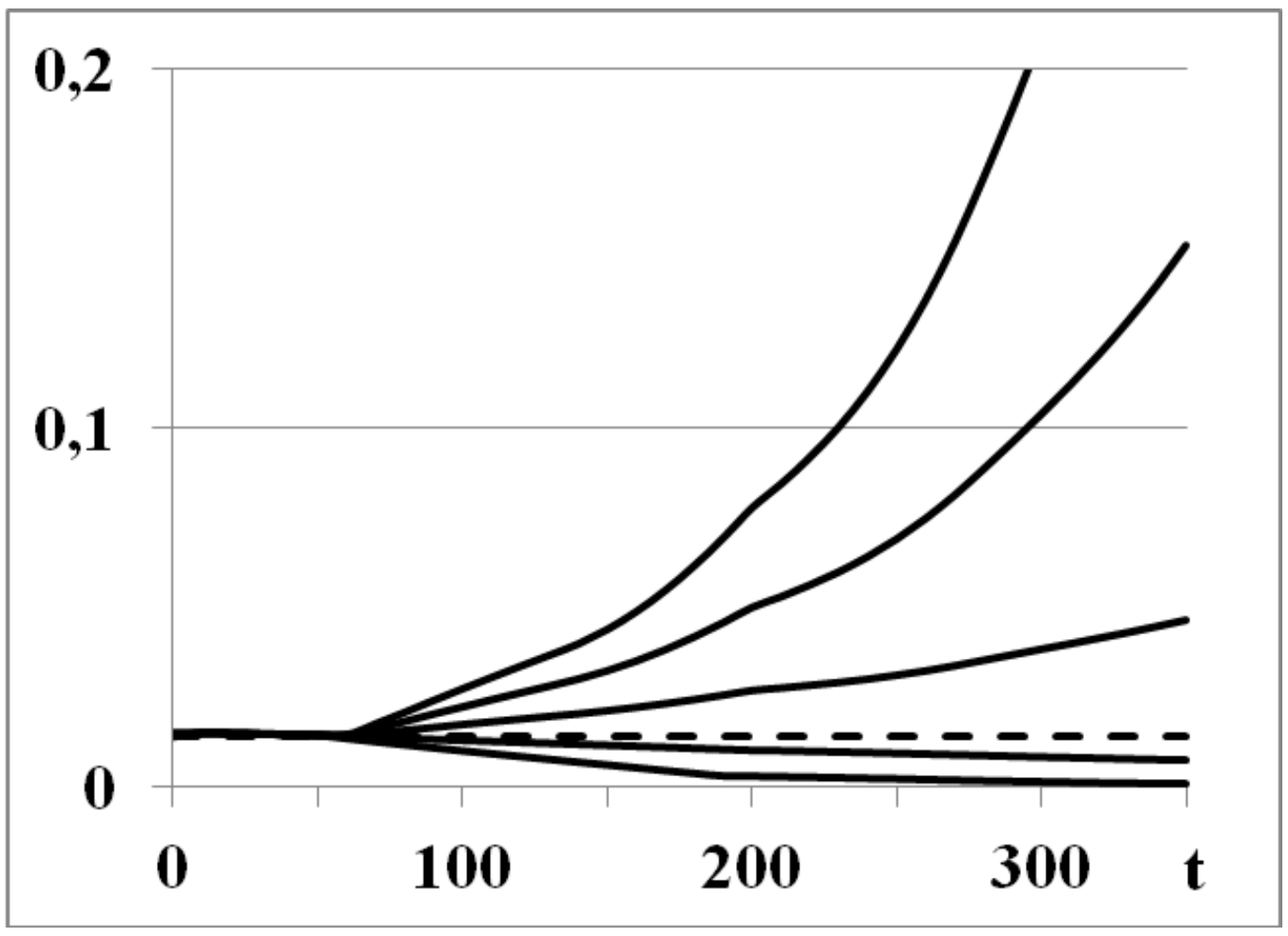

Figure 4a. Graphs of $U_S(t)$ in the neighborhood of trivial equilibrium in the Experiment I, $h_S^{'}(U_S) > 0$.

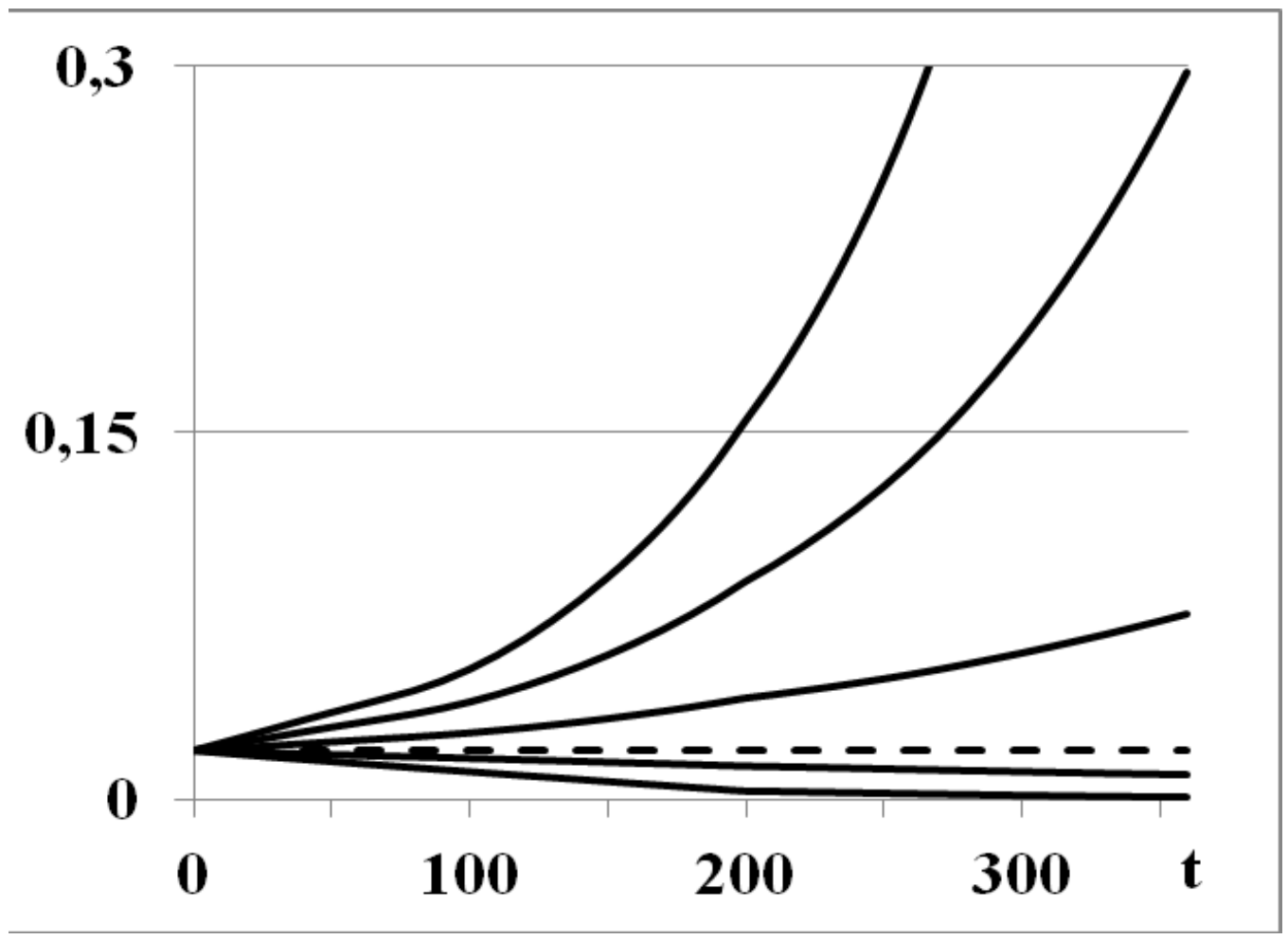

Figure 4b. Graphs of $W(t)$ in the neighborhood of trivial equilibrium in the Experiment I, $h_S^{'}(U_S) > 0$.

*7.2 Experiment II*

In the second group of experiments (Experiment II), we study the dynamical regimes of solitarious subpopulation in the neighborhood of nontrivial (positive) equilibrium. We consider various initial values of population density from the neighborhood of equilibrium $U_1^*$ and two types of fertility response function of solitarious (118) with $h_S^{'}(U_S) > 0$ (Fig. 5) and $h_S^{'}(U_S) < 0$ (Fig. 6). For the increasing fertility response function of first type we obtain the asymptotically stable dynamical regimes for $h_S(0) < h_S^*$ and asymptotically unstable dynamical regimes for $h_S(0) > h_S^*$. In the first case the population dynamics is the same as in the Experiment I when population density evolves to the trivial equilibrium (Fig.5a, b). For the second condition of this first case (initial condition just above nontrivial equilibrium of Fig.5a, b), we observe an increase of density of solitarious subpopulation. Obviously in this case, Eq. $h_S(U_{S_j}^*) = h_S^*$ does not have a root and nontrivial (positive) equilibrium of solitarious subpopulation does not exist. In fact, the condition of instability of trivial equilibrium together with increasing solitarious fertility response function always causes the gregarization of solitarious subpopulation and transit of their subpopulation to the gregarious phase.

In this group of experiments, the second case is the dynamical regimes of solitarious subpopulation with decreasing solitarious fertility response function $h_S^{'}(U_S) < 0$ for the condition $h_S(0) > h_S^*$ (Fig. 6) without gregarization $U_S^* < W_{0g}$ (see Fig.2). By virtue of Theorem 4 this type of $h_S^{'}(U_S)$ corresponds to the stable nontrivial (positive) equilibrium of solitarious subpopulation. In this case, Eq. $h_S(U_{S_j}^*) = h_S^*$ has a unique solution – asymptotically stable nontrivial equilibrium $U_{S1}^*$. We obtain the stable dynamics of solitarious subpopulation without gregarization in the form of quasi-periodical os-cillations in the neighborhood of positive equilibrium. The results of two different simulations for the smaller and larger than equilibrium initial values are shown in the Figs.6a, b. This is the unique nontrivial asymptotically stable dynamical regime of solitarious subpopulation obtained in all experiments.

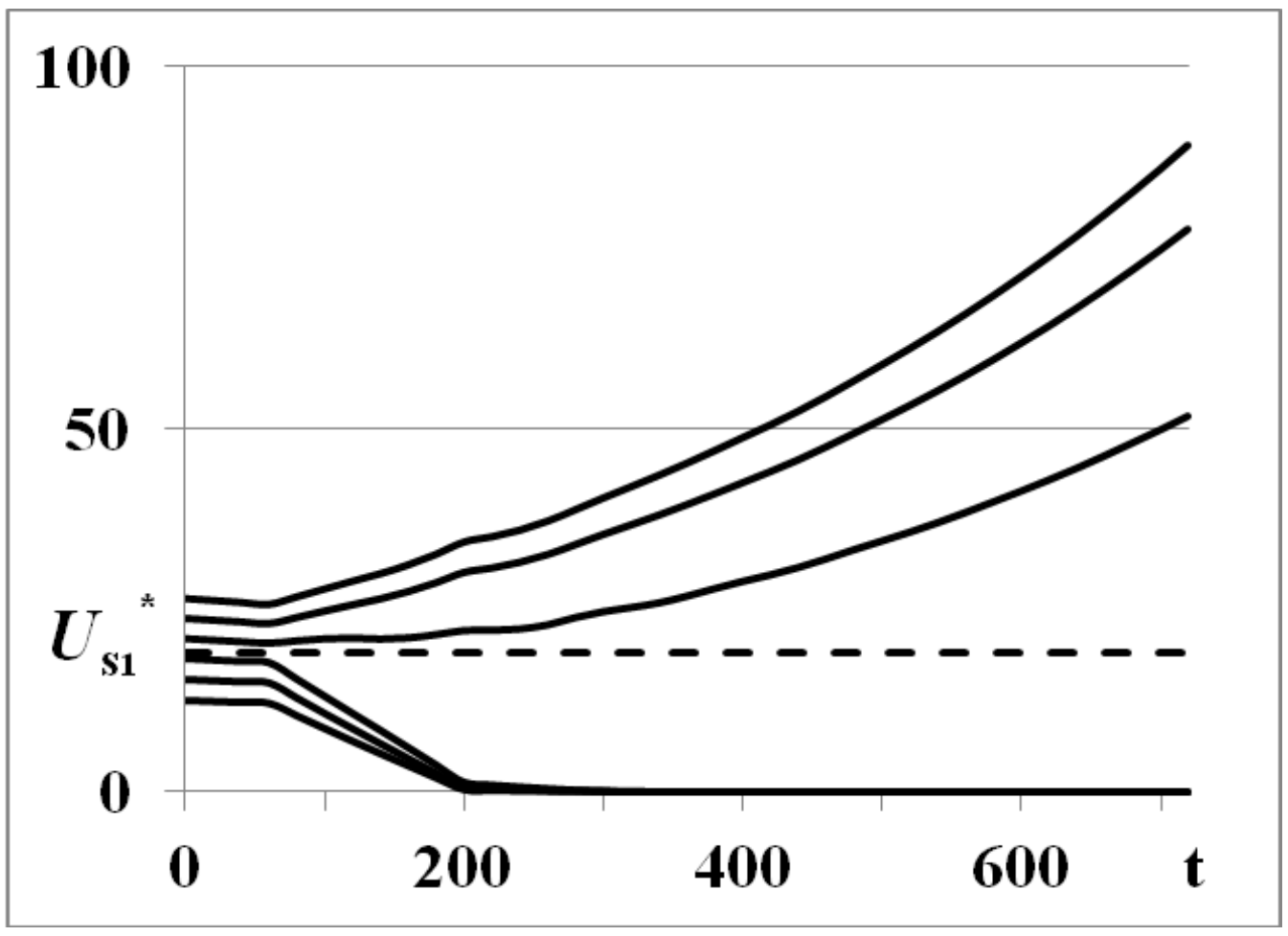

Figure 5a. Graphs of $U_s(t)$ in the neighborhood of nontrivial (positive) equilibrium $U_{s1}^*$ in the Experiment II, $h_s^{'}(U_s) > 0$.

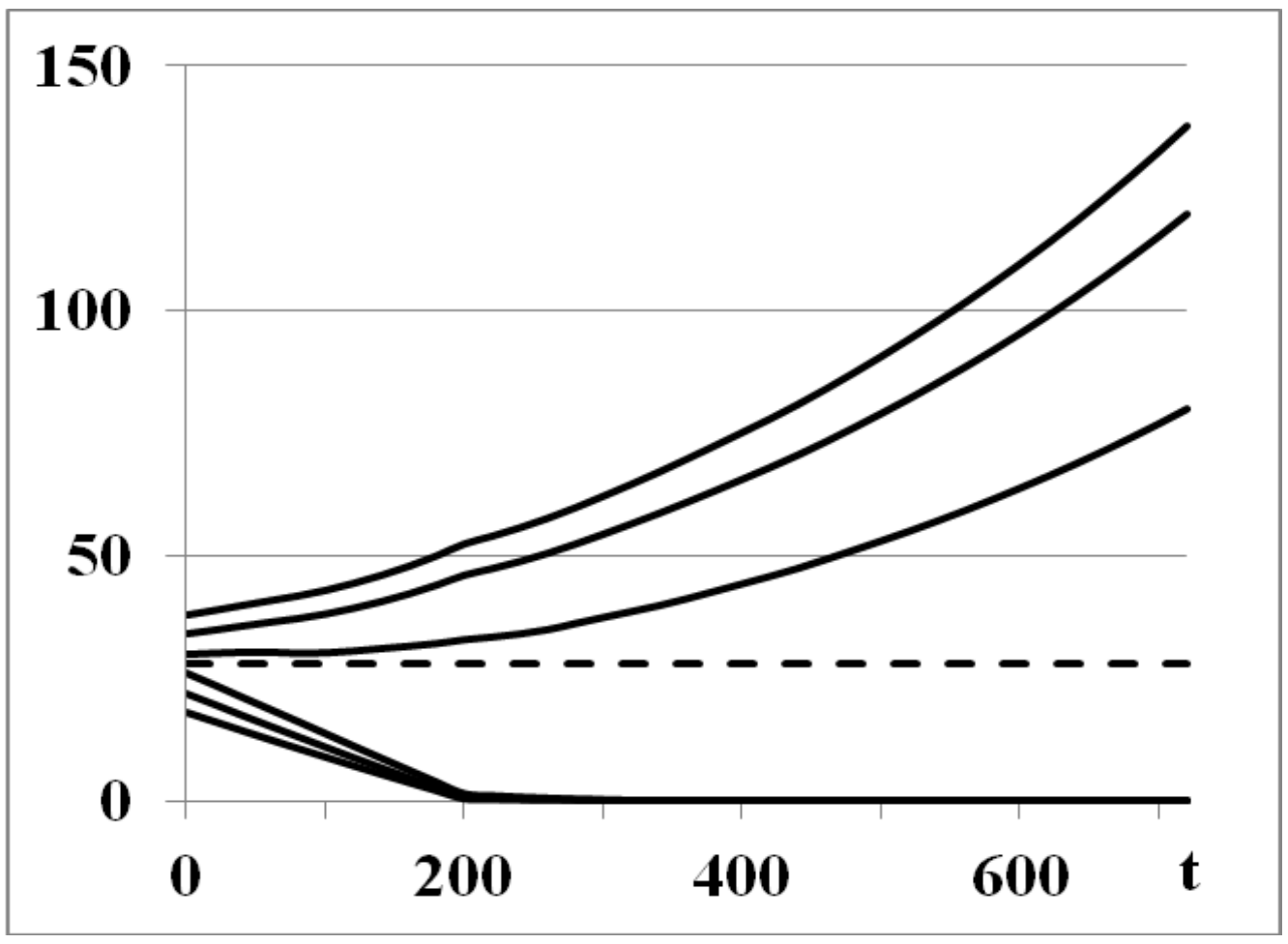

Figure 5b. Graphs of $W(t)$ in the neighborhood of nontrivial (positive) equilibrium $U_{s1}^*$ in the Experiment II, $h_s^{'}(U_s) > 0$.

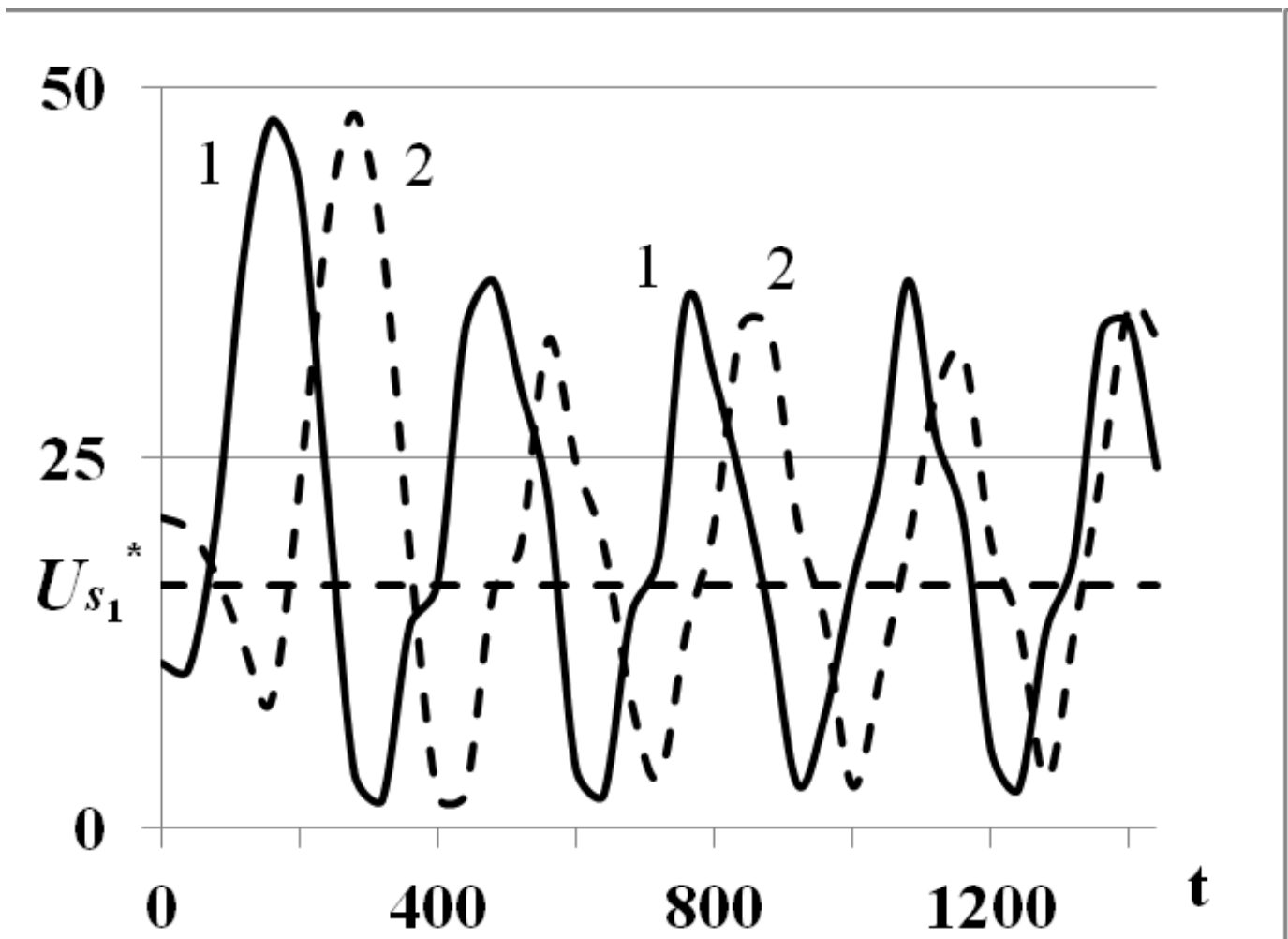

Figure 6a. Graphs of $U_s(t)$ in the neighborhood of nontrivial equilibrium $U_{s1}^*$ in the Experiment II, $h_s^{'}(U_s) < 0$, $U_s^* < W_{0g}$, for smaller (1) and larger (2) initial values of $U_s(0)$.

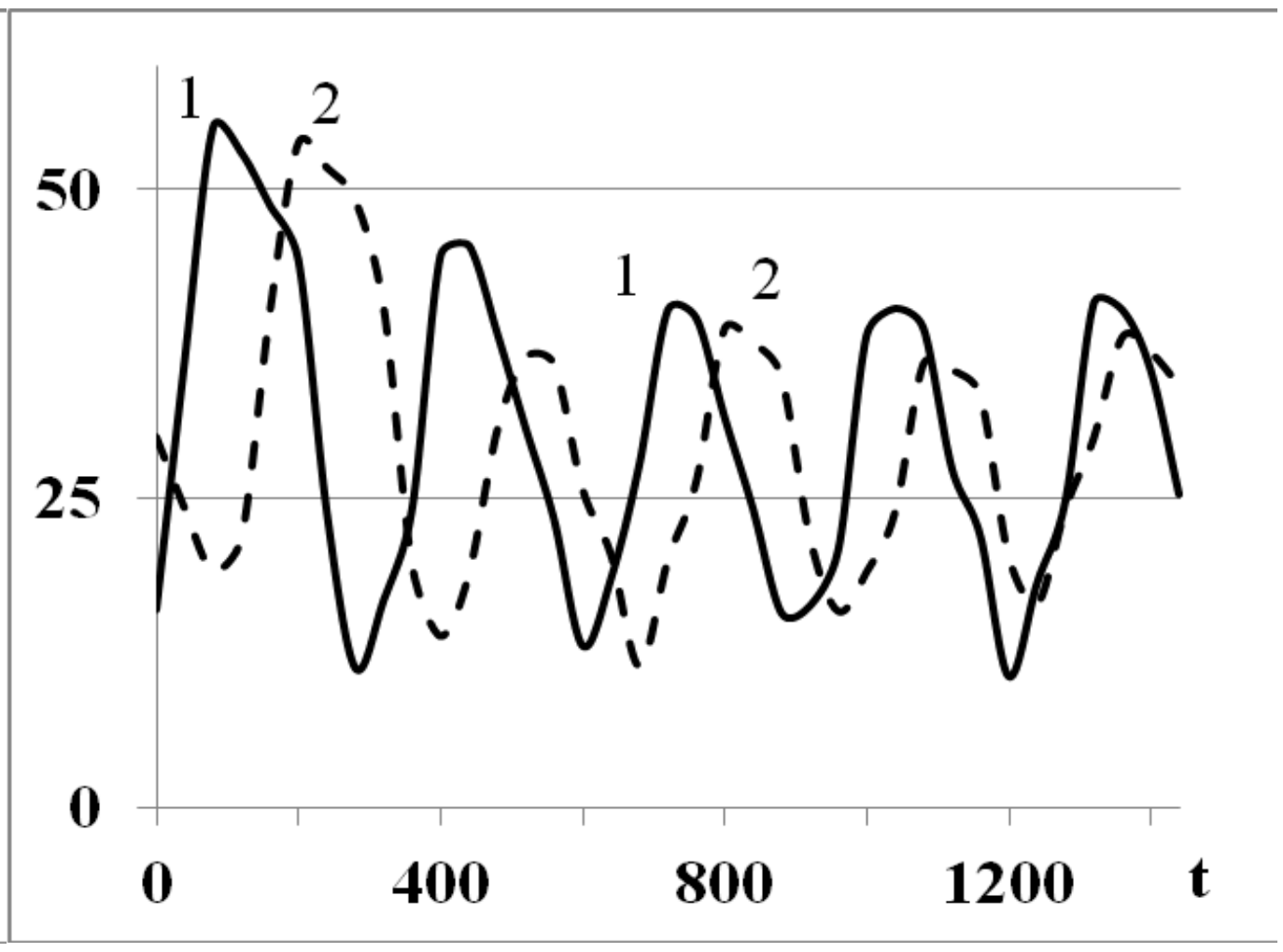

Figure 6b. Graphs of $W(t)$ in the neighborhood of nontrivial equilibrium in the Experiment II, $h_s^{'}(U_s) < 0$, $U_s^* < W_{0g}$, for smaller (1) and larger (2) initial values of $W(0)$.

### *7.3 Experiment III*

In the third group of experiments (Experiment III), we study the dynamical regimes of population in the neighborhood of nontrivial (positive) equilibrium of gregarious subpopulation $G_{g1}^*$ with parameters $h_g(0) < h_g^*$, $h_g^{'}(G_g) < 0$. In these experiments we consider the unstable dynamical regime of solitarious subpopulation with characteristics $h_s(0) > h_s^*$, $h_s^{'}(U_s) > 0$ (Figs. 5a, 5b) providing the phase transition from the solitarious phase to the gregarious one. By virtue of Theorem 2 the nontrivial solitarious and gregarious equilibriums do not exist if their fertility response functions satisfy conditions $h_s(0) > h_s^*$, $h_s^{'}(U_s) > 0$ and $h_g(0) < h_g^*$, $h_g^{'}(G_g) < 0$, respectively. The absence of nontrivial equilibrium is a result of our simplified analysis of asymptotically stable regimes of two-phase population dynamics when we studied the asymptotical regime of each subpopulation separately reducing the age-structured system (2) – (11) to the two initial problems (61), (62), and (65), (66). On the other hand, the smooth form of overlapping phase-switching rates (Fig. 2) accepts the coexistence of the two phases and, as a consequence, the existence at the same time of nontrivial equilibriums of solitarious and gregarious phases. Theorems 2 - 4 provide us the possibility of a qualitative analysis and the prediction of simultaneous coexistence of such nontrivial equilibriums as a result of balance between asymptotically unstable dynamical regime of solitarious subpopulation (conditions $h_s(0) > h_s^*$, $h_s^{'}(U_s) > 0$) and asymptotically stable dynamical regime of gregarious subpopulation (conditions $h_g(0) < h_g^*$, $h_g^{'}(G_g) < 0$). The numerical experiments allow computing and studying the equilibriums of both locust phases for these conditions. The results of simulations shown in the Figs.7a, b, c, exhibit the stable quasi-periodical oscillations of subpopulation densities in the neighborhood of some positive stationary point, which may be identified as a nontrivial positive equilibrium. Thus, autonomous system (2) – (11) may possess at the same time the solitarious and gregarious nontrivial equilibriums for the model of phase-switching rates given in the Fig.2.

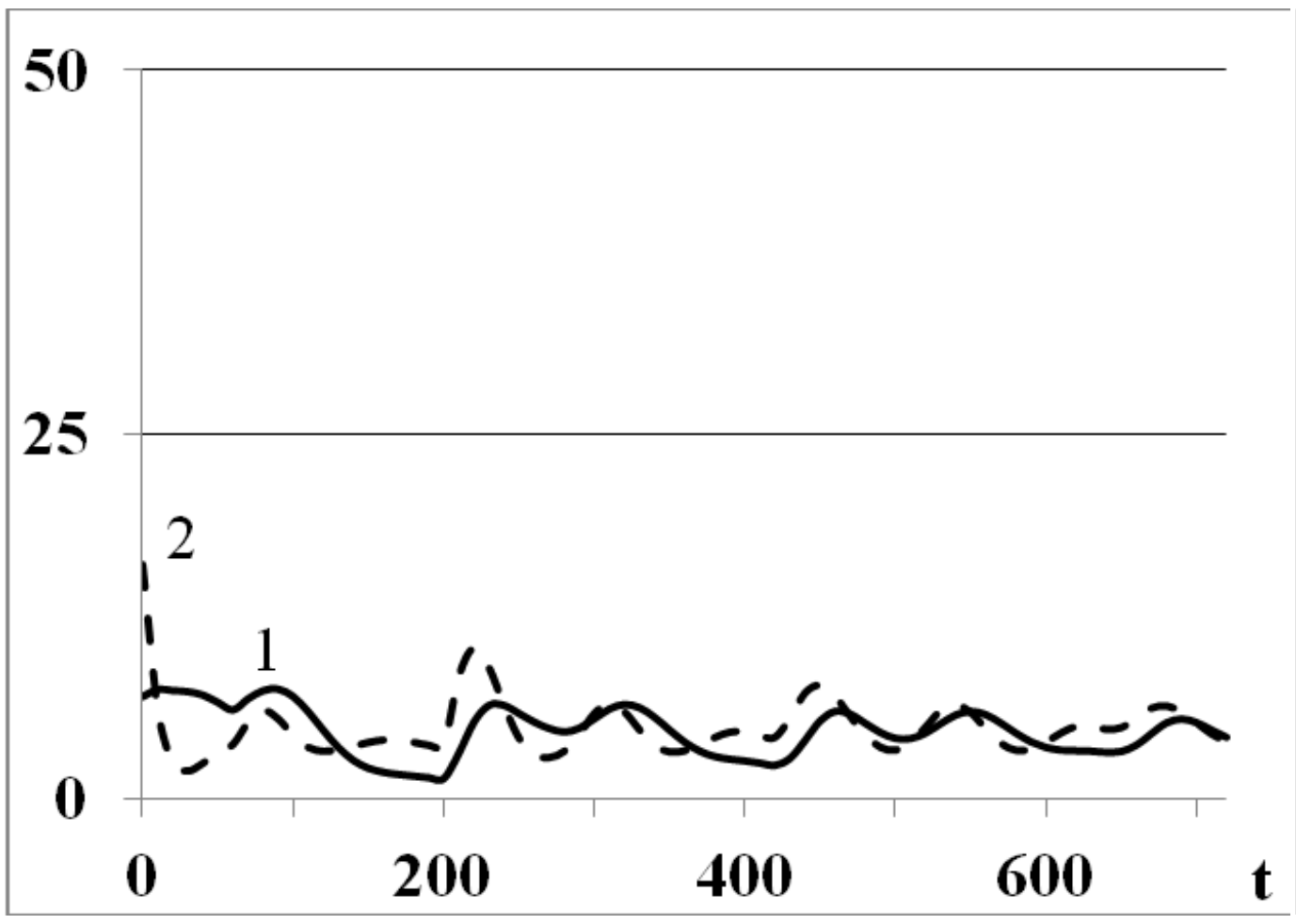

Figure 7a. Graphs of $U_s(t)$ for smaller (1) and larger (2) initial values of $U_s(0)$.

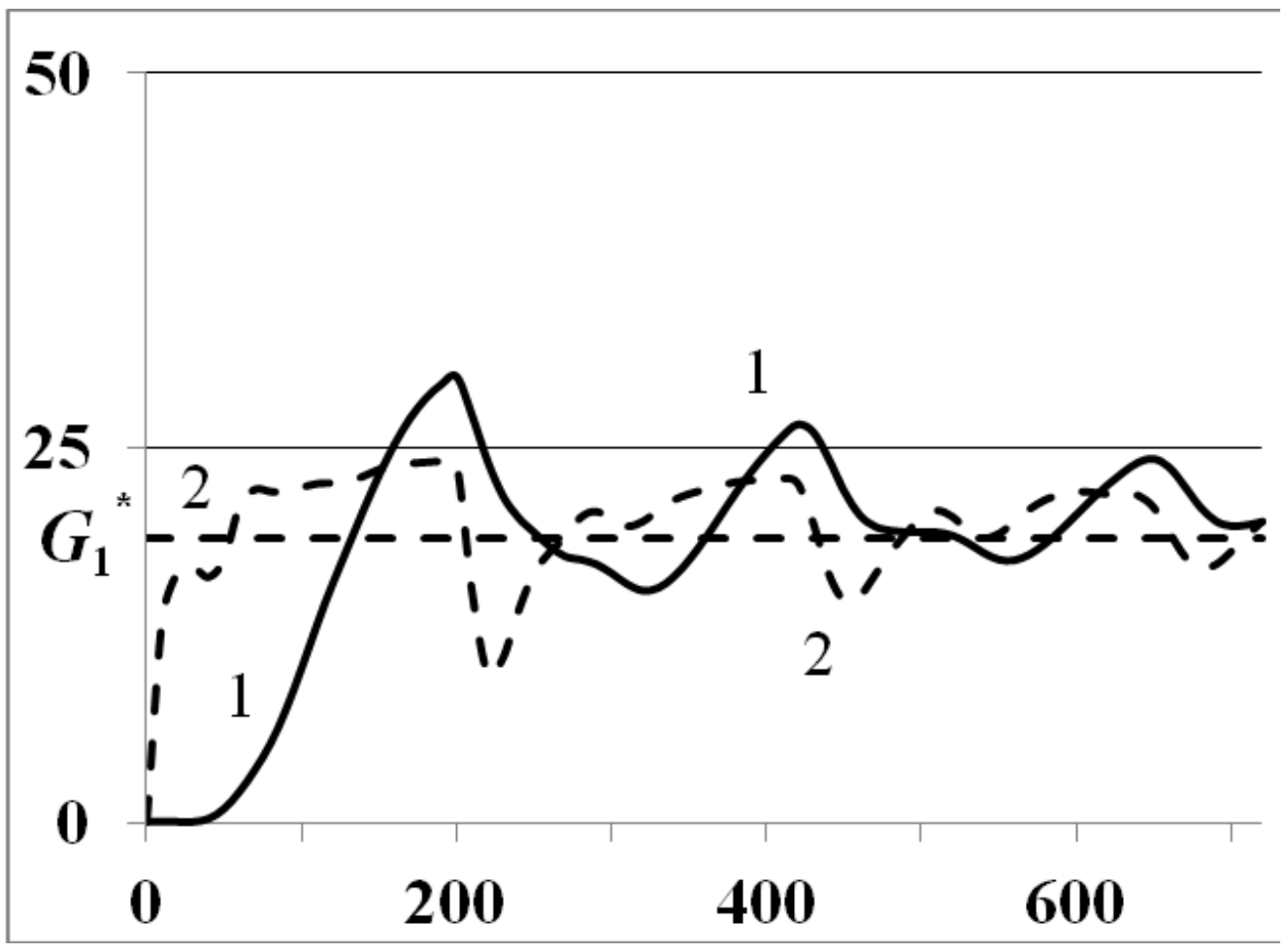

Figure 7b. Graphs of $G_g(t)$, for smaller (1) and larger (2) initial values of $U_s(0)$

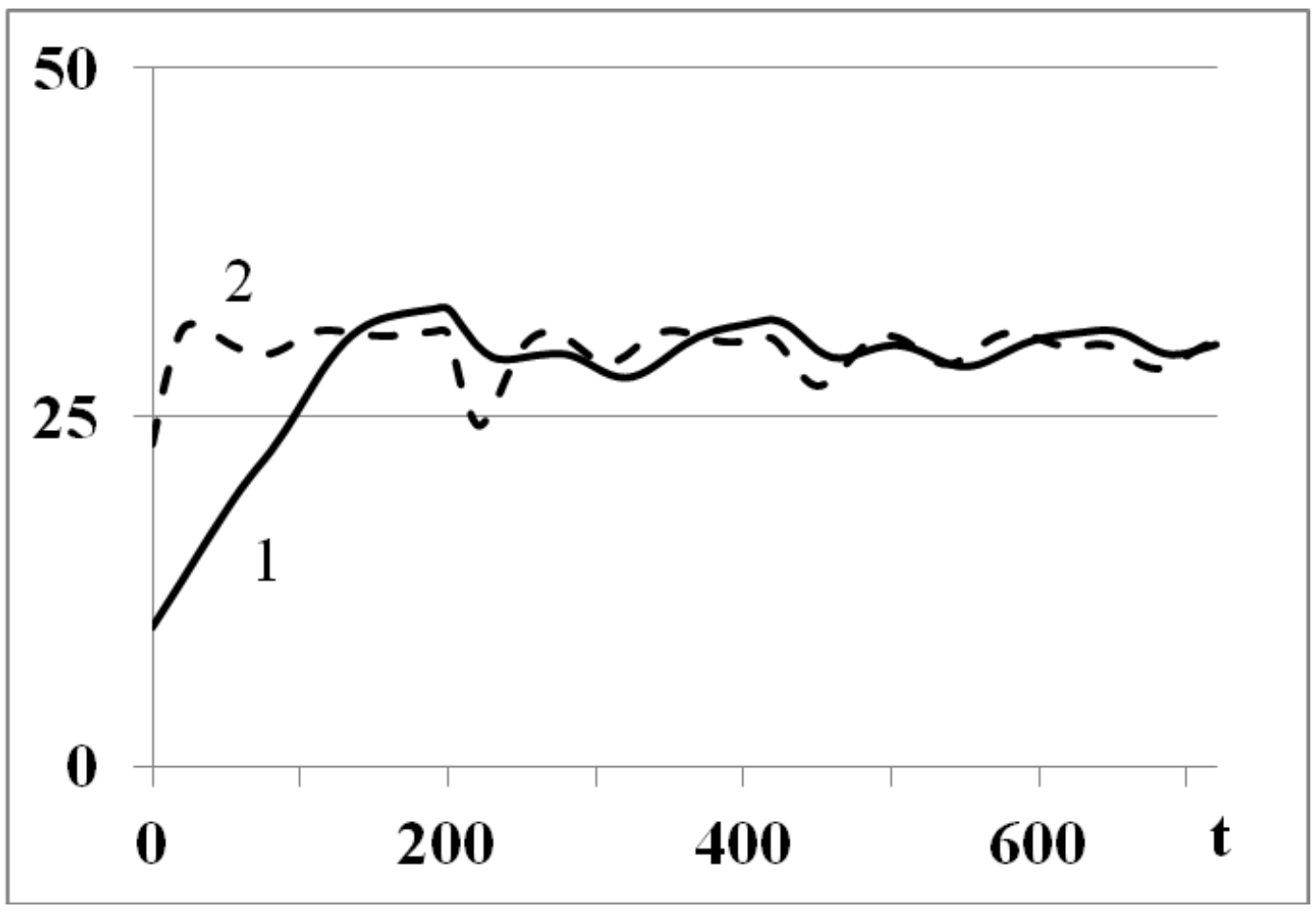

Figure 7c. Graphs of $W(t)$ in the neighborhood of asymptotically stable nontrivial equilibrium in the Experiment III for smaller (1) and larger (2) initial values of $W(0)$.

### *7.4 Experiment IV*

In the fourth group of experiments (Experiment IV), we continue to study the dynamical regimes of population in the neighborhood of nontrivial (positive) equilibrium of gregarious subpopulation $G_{g1}^*$ with parameters $h_g(0) > h_g^*$, $h_g'(G_g) < 0$ for different types of dynamical regimes of solitarious subpopulation. We consider two types of gregarious fertility response functions which satisfy the following conditions:

$$\min_{G_g > 0} h_g(G_g) \le h_g^*, \tag{121}$$

$$\min_{G_g > 0} h_g(G_g) > h_g^*. \tag{122}$$

By virtue of Theorem 2 these two types of $h_g(G_g)$ for the considered parameters of numerical experiment have a principal difference. The first type of $h_g(G_g)$ (121) together with condition $h_g'(G_g) < 0$ guaranties the existence of asymptotically stable nontrivial (positive) equilibrium of gregarious subpopulation which is achievable for the growing density of gregarious. In the second case (122) the nontrivial equilibrium of gregarious subpopulation does not exist. As a consequence, the decreasing fertility response function $h_g'(G_g) < 0$ does not guarantee the stable behavior of gregarious subpopulation because the results of Theorem 4 are valid only for the existing nontrivial equilibriums.

For the first type of $h_g(G_g)$ (121) we study two dynamical regimes with different conditions of direct phase change (solitarious - gregarious). In the first case, fertility response function of solitarious is that $h_s(0) > h_s^*$, $h_s'(U_s) > 0$ for which solitarious subpopulation is asymptotically unstable (Fig.4a, 4b). The results of numerical experiments in this case with two different initial values of $U_s(0)$ are shown in the Figs.8a, b, c. The densi-ty of gregarious subpopulation oscillates in the neighborhood of a stable positive equilibrium – unique root of the Eq. $h_g(G_{g1}^*) = h_g^*$. The density of solitarious subpopulation is negligibly small in the neighborhood of the trivial equilibrium.

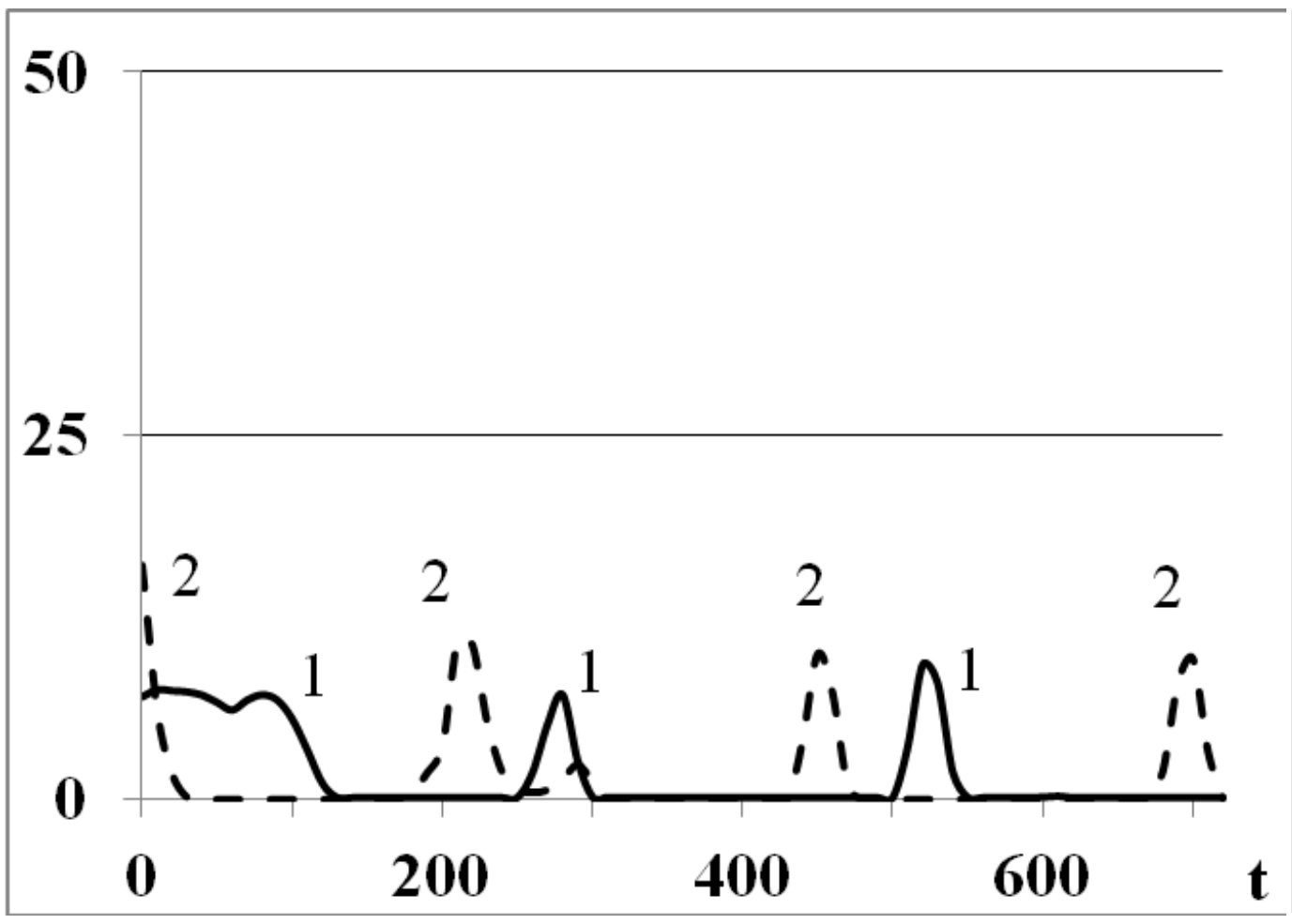

Figure 8a. Graphs of $U_s(t)$ for smaller (1) and larger (2) initial values of $U_s(0)$.

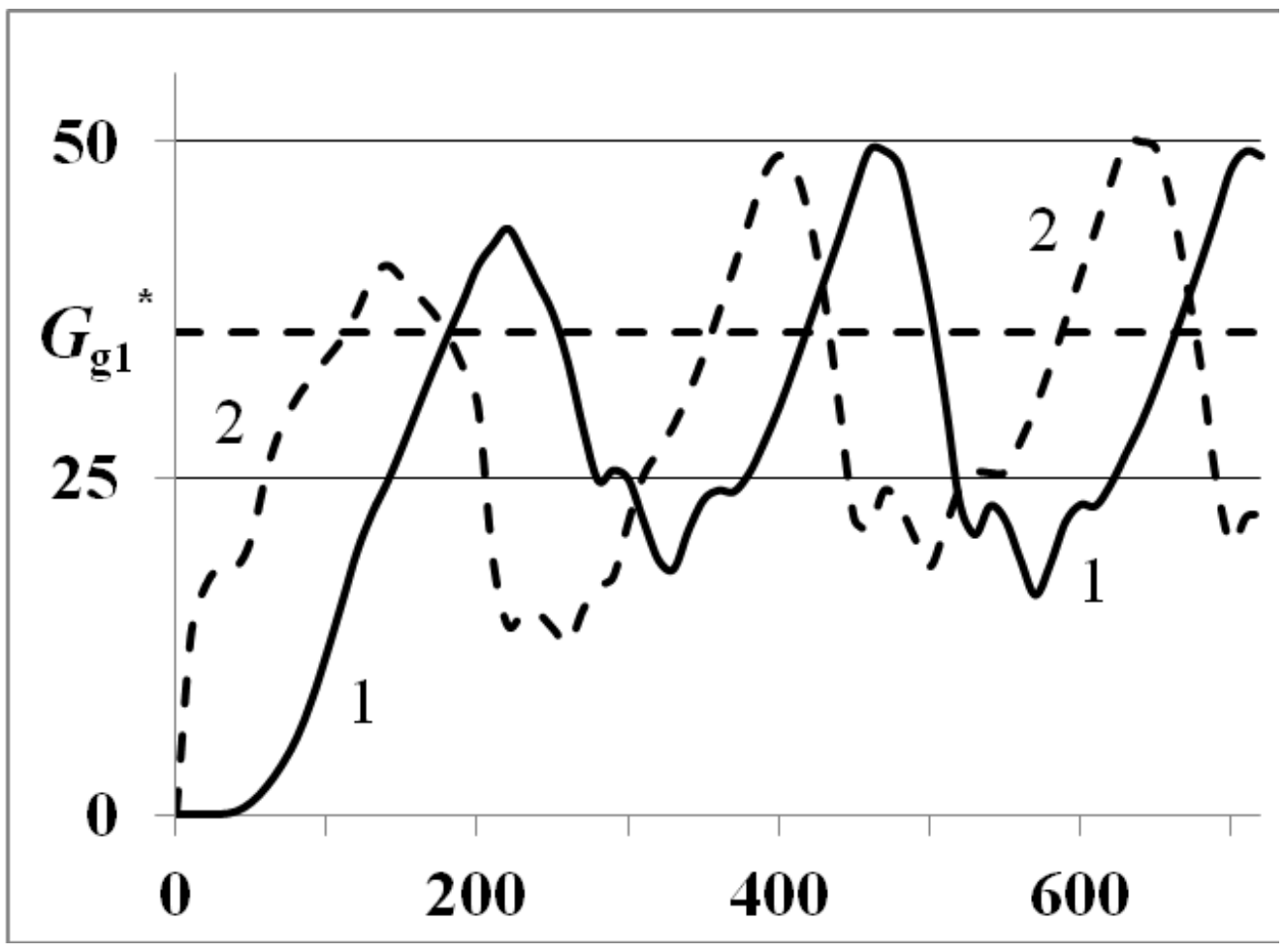

Figure 8b. Graphs of $G_g(t)$ for smaller (1) and larger (2) initial values of $U_s(0)$.

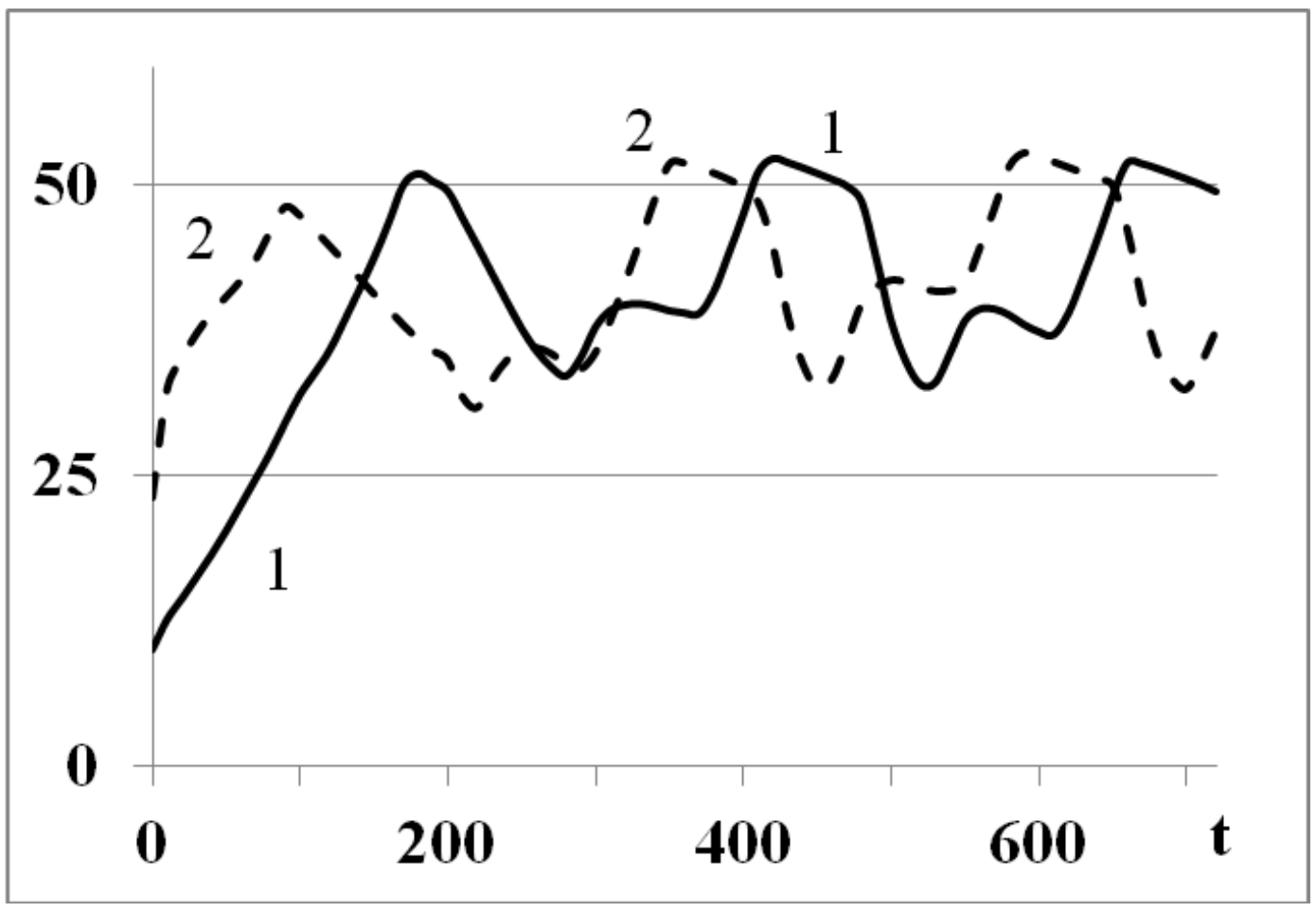

Figure 8c. Graphs of $W(t)$ in the neighborhood of asymptotically stable nontrivial equilibrium in the Experiment IV (type 1, case 1) for smaller (1) and larger (2) initial values of $W(0)$.

In the second case, we consider the dynamical regime of solitarious subpopulation with decreasing fertility response function $h_s(0) > h_s^*$, $h_s^{'}(U_s) < 0$ with $U_s^* > W_{0g}$. The particularities of this case are that 1) the value of nontrivial equilibrium of solitarious is bigger than the threshold density of gregarization; 2) both density-dependent fertility response functions of solitarious and gregarious are decreasing functions. In contrast with the second case of Experiment II (Fig.6a, 6b) condition $U_s^* > W_{0g}$ guarantees the gregarization in population. The results of numerical experiments in this case with two different initial values of $U_s(0)$ are shown in the Figs.9a, b, c. The behavior of density of all subpopulations is close to the dynamics of solitarious and gregarious densities in the previous first case. The gregarious density oscillates in the neighborhood of a stable positive equilibrium – unique root of the Eq.

$h_g(G_{g1}^*) = h_g^*$. The density of solitarious subpopulation is small in the neighborhood of the trivial equilibrium. Despite the fact that we consider two different types of fertility response functions of solitarious which cause the unstable trivial equilibrium, we obtain the same type of gregarious dynamics and overall population dynamics in both cases.

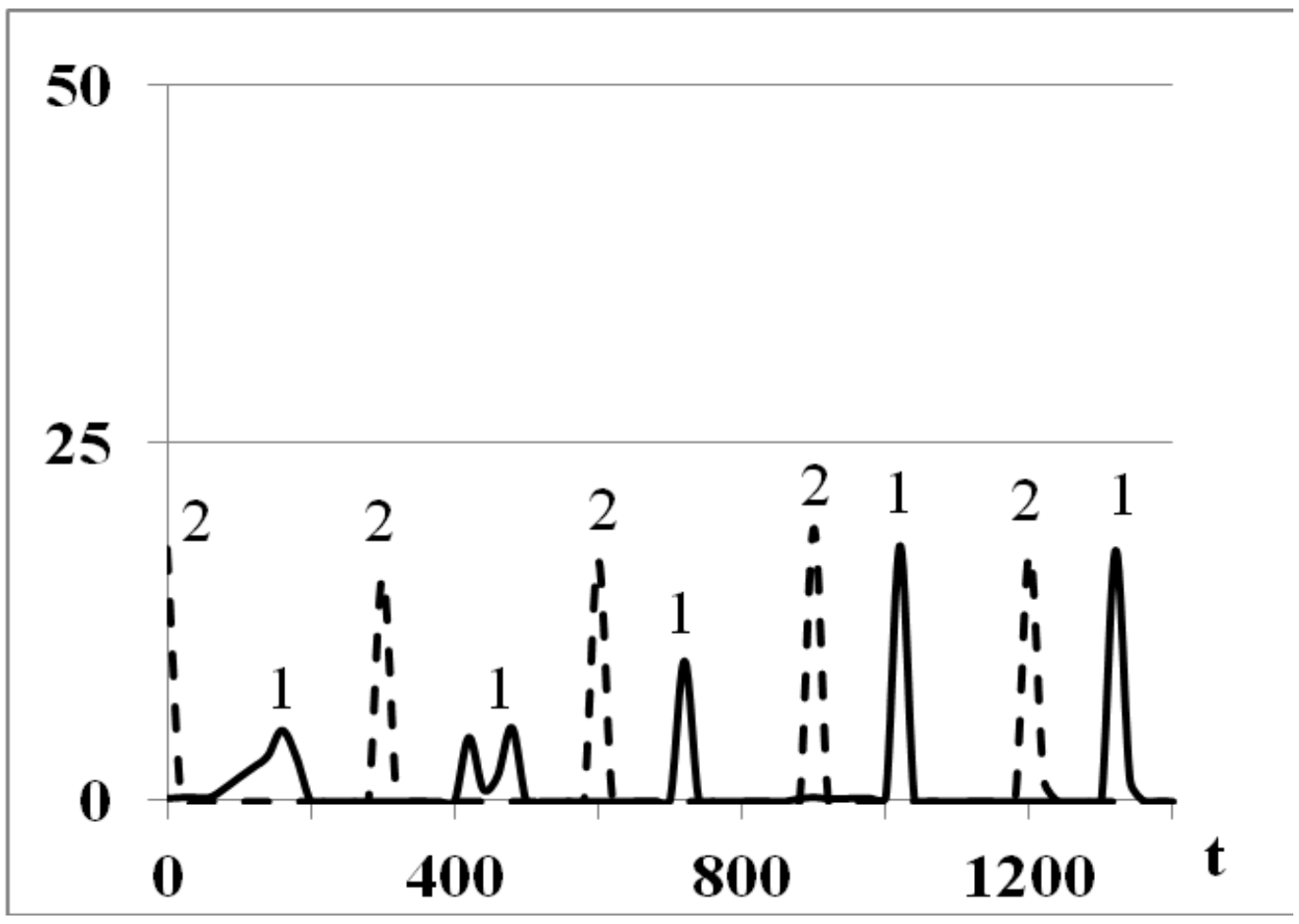

Figure 9a. Graphs of $U_s(t)$ for smaller (1) and larger (2) initial values of $U_s(0)$.

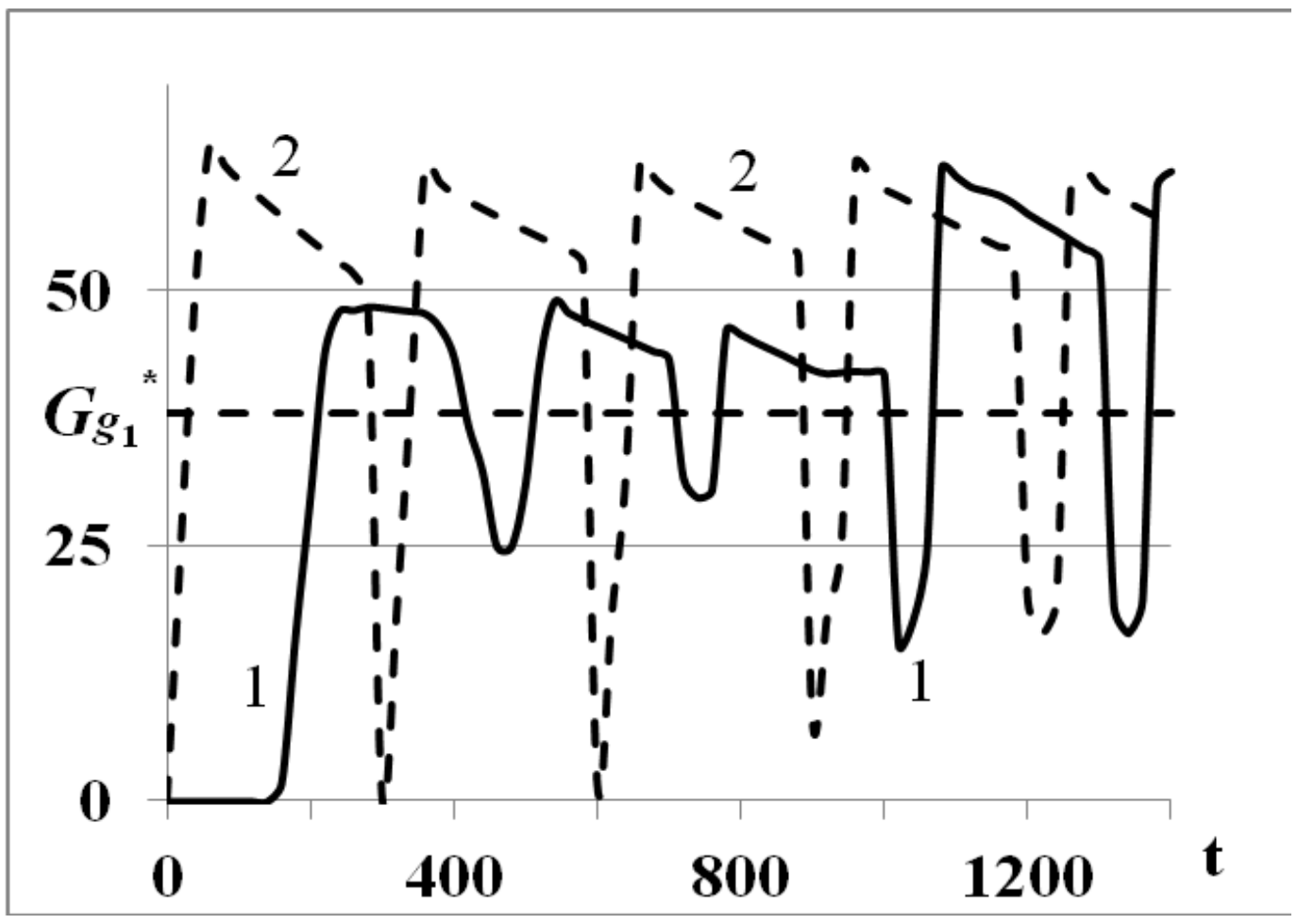

Figure 9b. Graphs of $G_g(t)$ for smaller (1) and larger (2) initial values of $U_s(0)$.

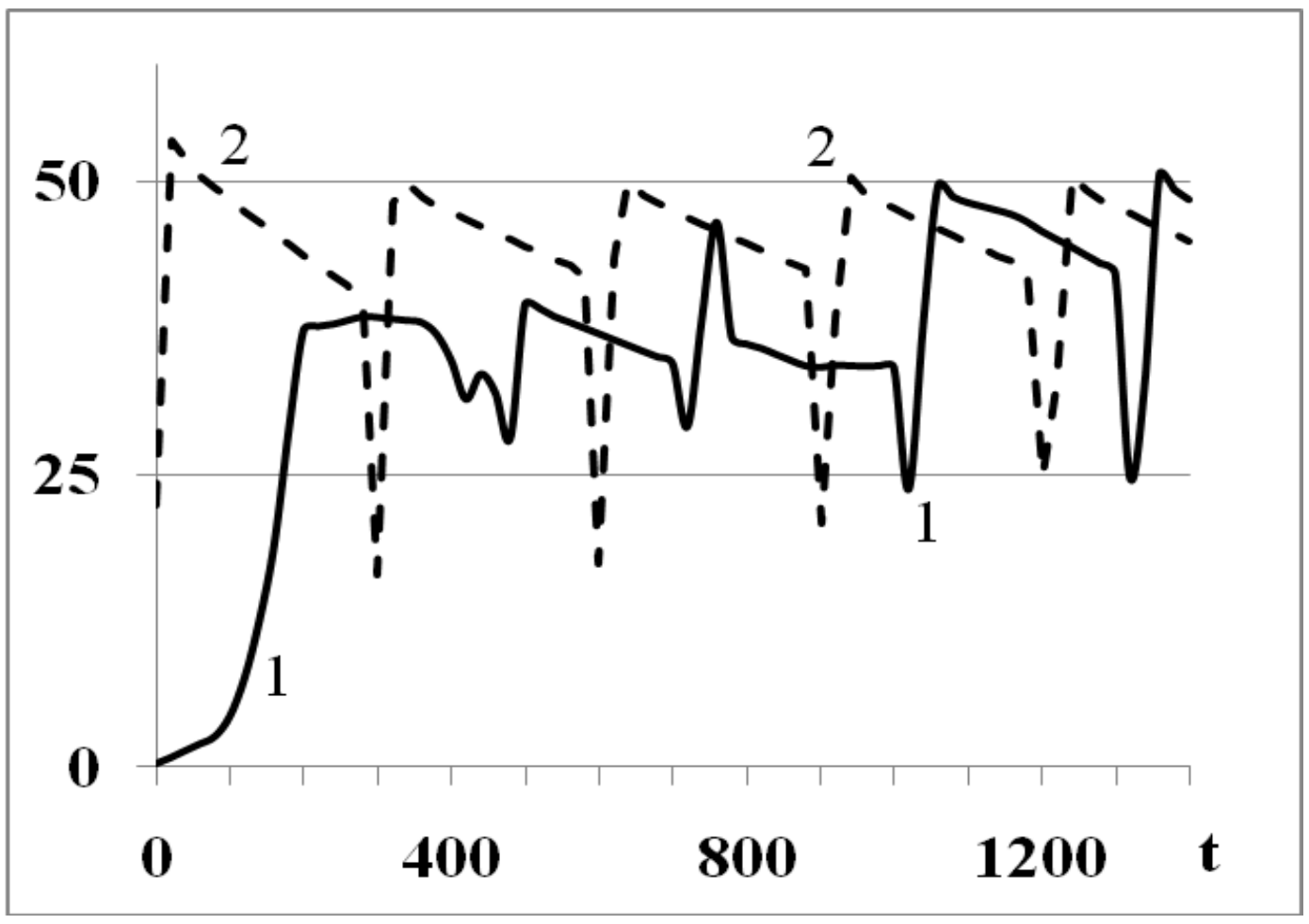

Figure 9c. Graphs of $W(t)$ in the neighborhood of asymptotically stable nontrivial equilibrium in the Experiment IV (type 1, case 2) for smaller (1) and larger (2) initial values of $W(0)$.

The results of numerical experiments for the second type of $h_g(G_g)$ (122) for unstable solitarious subpopulation with fertility response function $h_s(0) > h_s^*$, $h_s'(U_s) > 0$ are shown in the Figs.10a, b. The density of gregarious subpopulation and density of all population grow infinitely with time. The dynamics of population is not asymptotically stable even with decreasing fertility response function $h_g'(G_g) < 0$.

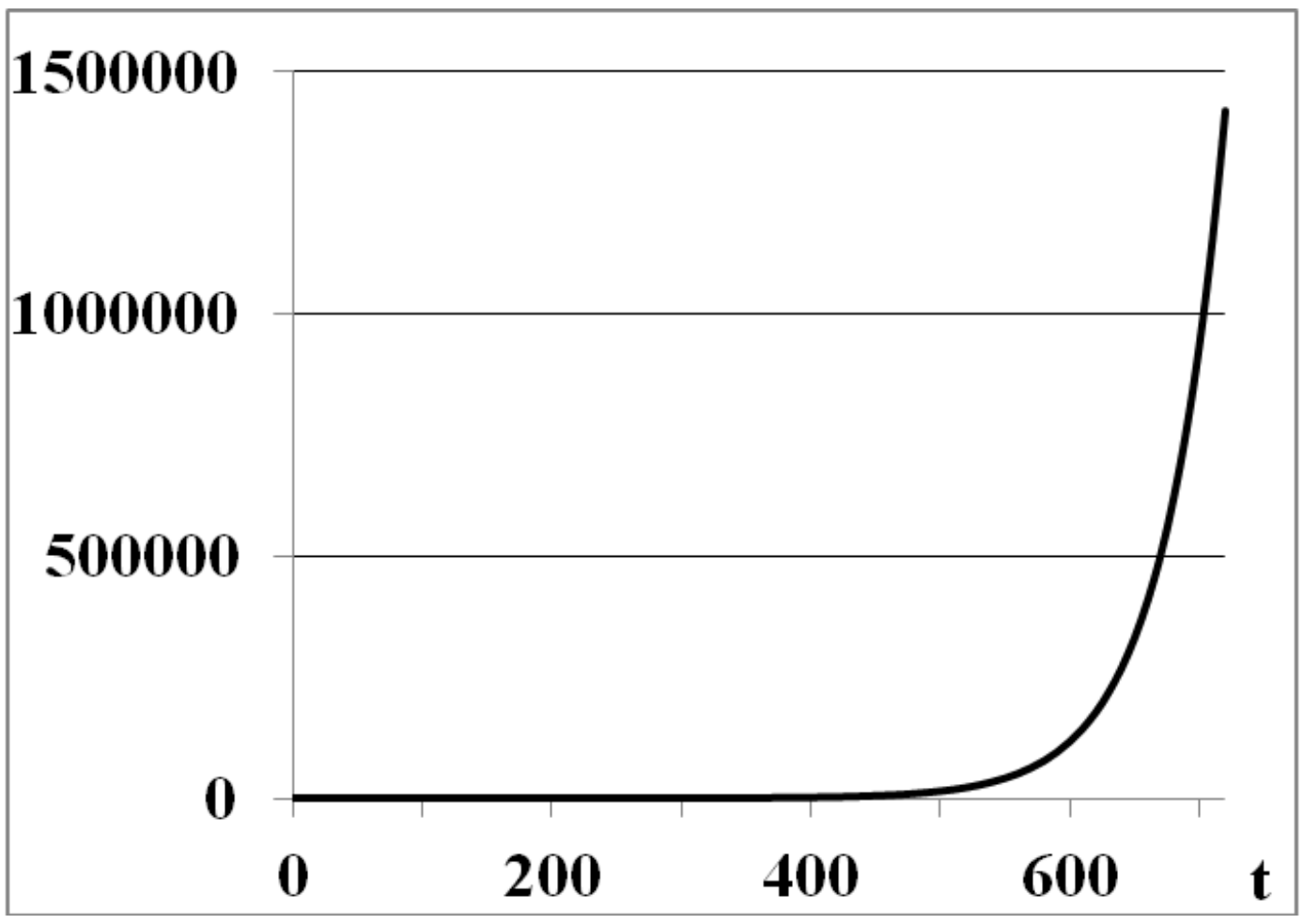

Figure 10a. Graph of $G_g(t)$ of unstable dynamics in the Experiments IV (type 2) and V.

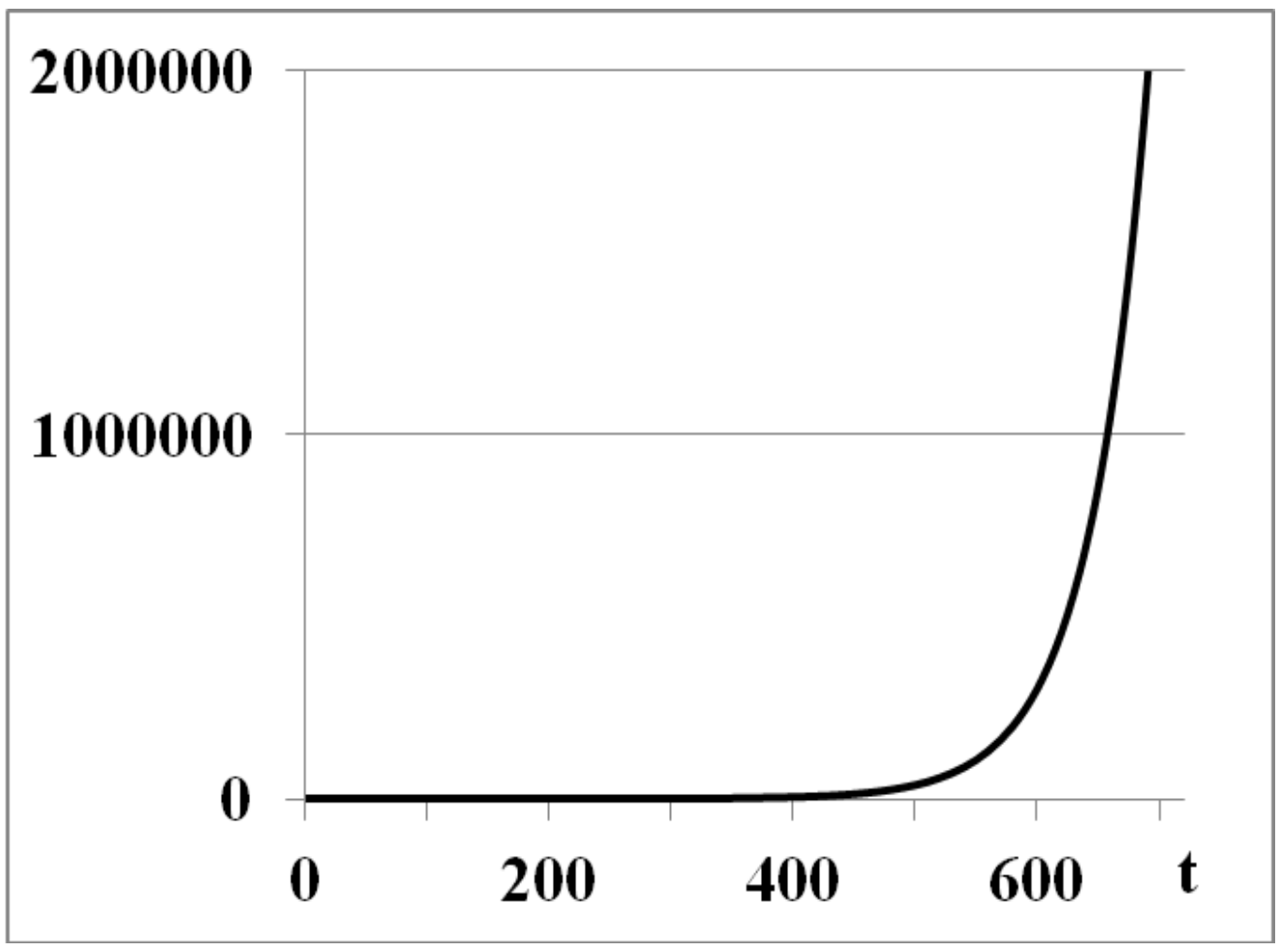

Figure 10b. Graph of $W(t)$ of unstable dynamics in the Experiments IV (type 2) and V.

*7.5 Experiment V*

The last, fifth group of numerical experiments are carried out for the two groups of gregarious fertility response function parameters: $h_g(0) < h_g^*$, $h_g^{'}(G_g) > 0$ (type 1), $h_g(0) > h_g^*$, $h_g^{'}(G_g) < 0$ (type 2) and the same solitarious fertility response function $h_s(0) > h_s^*$, $h_s^{'}(U_s) > 0$. By virtue of Theorem 2, in the first case (type 1) the nontrivial positive equilibrium of gregarious subpopulation exists. On the other hand, Theorem 4 states that this equilibrium is not asymptotically stable. In the second case (type 2) the nontrivial positive equilibrium of gregarious subpopulation does not exist and increasing fertility response function causes the unstable asymptotical dynamics of all population. The results of simulations for both types of experiments are the same as shown in the Figs. 10a, 10b. We observe a tremendous growth of gregarious subpopulation density and density of all population with time.

Overall, the performed numerical analysis of stability of trivial and nontrivial equilibriums of autonomous system (2) – (10) allow us to illustrate the main theoretical results obtained in Theorems 2 - 4, and, on the other hand, understand the main features of locust population dynamical regimes including the conditions and prerequisites of the phase changes in population.

## 8. Simulation of population outbreaks for non-autonomous dynamical system

The last group of numerical experiments is carried out for the non-autonomous system (2) – (10) to analyze the features of the two-phase age-structured model in simulation of locust population dynamics. We consider the data of number of swarm observations in the distribution area of desert locust (map in Fig.14 in Appendix C) over the last 30 years. This data are presented as a time series of number (or area) of square degrees ($1^\circ \times 1^\circ$) where at least one swarm was found within one month [41]. The time series of number of swarms for three essential periods are shown in Fig.15a – 15c (Appendix C). In this group of experiments we use phase-switching rates (56) – (60), dynamical natural death rate $\alpha_0(t)$ and dynamical coefficients $\sigma_a(t)$, $\sigma_c(t)$ for the fertility rates (115) – (120).

According to Theorems 3, 4 and the results of the numerical analysis of dynamical regimes obtained in the previous section, we choose the increasing density-dependent fertility response functions of solitarious $h_s(U_s(t))$ with $h_s^{'}(U_s(t))>0$ and decreasing density-dependent fertility response functions of gregarious $h_g(G_g(t))$ with $h_g^{'}(G_g(t))<0$ in the form of (117), (119) shown in the Fig.3. The maximum, expected and minimum values of all biological constants taken from the literature [32], [36], [37], [39], [40], [42], [43] and numerous laboratory experiments are shown in Table 2 (Appendix C). Parameters $a_s$, $a_g$, $a_d$, $\tau$ are measured in days; $\omega_s$, $\omega_g$, $\mu_s$, $\mu_g$ are dimensionless, $\alpha_0(t)$ is measured in fraction of all locust per day, $\alpha_0(t)\in[\alpha_{0\min},\alpha_{0\max}]$.

The number of matured gregarious locusts $G_g(t)$ is fitted to the observed time series of number of square degrees with swarms for three periods of important outbreaks. We consider the problem of parameter identification in the form:

$$\left(\alpha_0^*(t),\sigma_a^*(t),\sigma_c^*(t)\right)=\arg\min_{\Omega}\Phi\left(\alpha_0(t),\sigma_a(t),\sigma_c(t)\right)=\arg\min_{\Omega}\left(\sum_{i=1}^{N}\left|G_g(t_i)-G_i\right|\right), \quad (123)$$

where $\{G_i\}$ is a time-series, $G_i>0$, $N$ is a number of time-series members. The optimization problem (123) is considered on the compact set

$$\Omega=\Big\{\alpha_0(t),\sigma_a(t),\sigma_c(t)\Big|\alpha_0(t)\in C([0,T]),\alpha_0(t)\in[\alpha_{0\min},\alpha_{0\max}], \quad (124)$$

$$\left|\alpha_0^{'}(t)\right|<c_0,\ \sigma_a(t)\in C([0,T]),\ \sigma_a(t)\in[\sigma_b,\sigma_{\max}],\left|\sigma_a^{'}(t)\right|<c_0,\ \sigma_c(t)\in C([0,T]),$$

$$\sigma_c(t)\in[\sigma_d,\sigma_{\max}],\left|\sigma_c^{'}(t)\right|<c_0\Big\}.$$

For the parameter estimation problem (123) we use the adaptive metaheuristic optimization method “tabu search” which is based on the algorithm of random search of optimal solution on the compact set [5], [21], [38]. The graphs of fitted functions of total number of locusts $W(t)$ and number of matured gregarious locusts $G_g(t)$ to the data of observation for the periods of January 1988 – December 1989, November 1992 – March 1997 and October 2003 – December 2005 are shown in the Fig.11a, b, 11c, d and 11e, f, respectively.

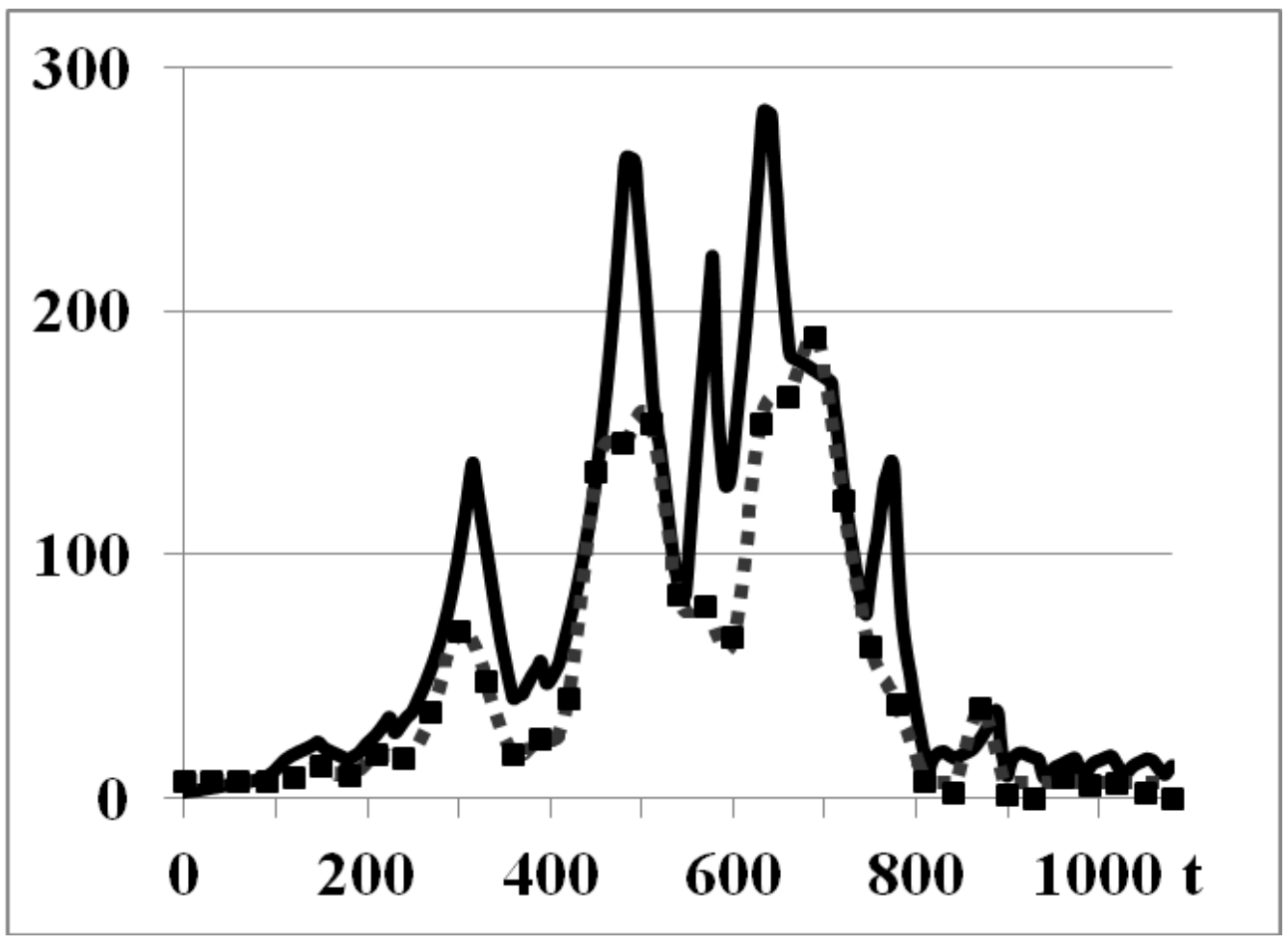

Fig.11a. January 1988 – December 1989.

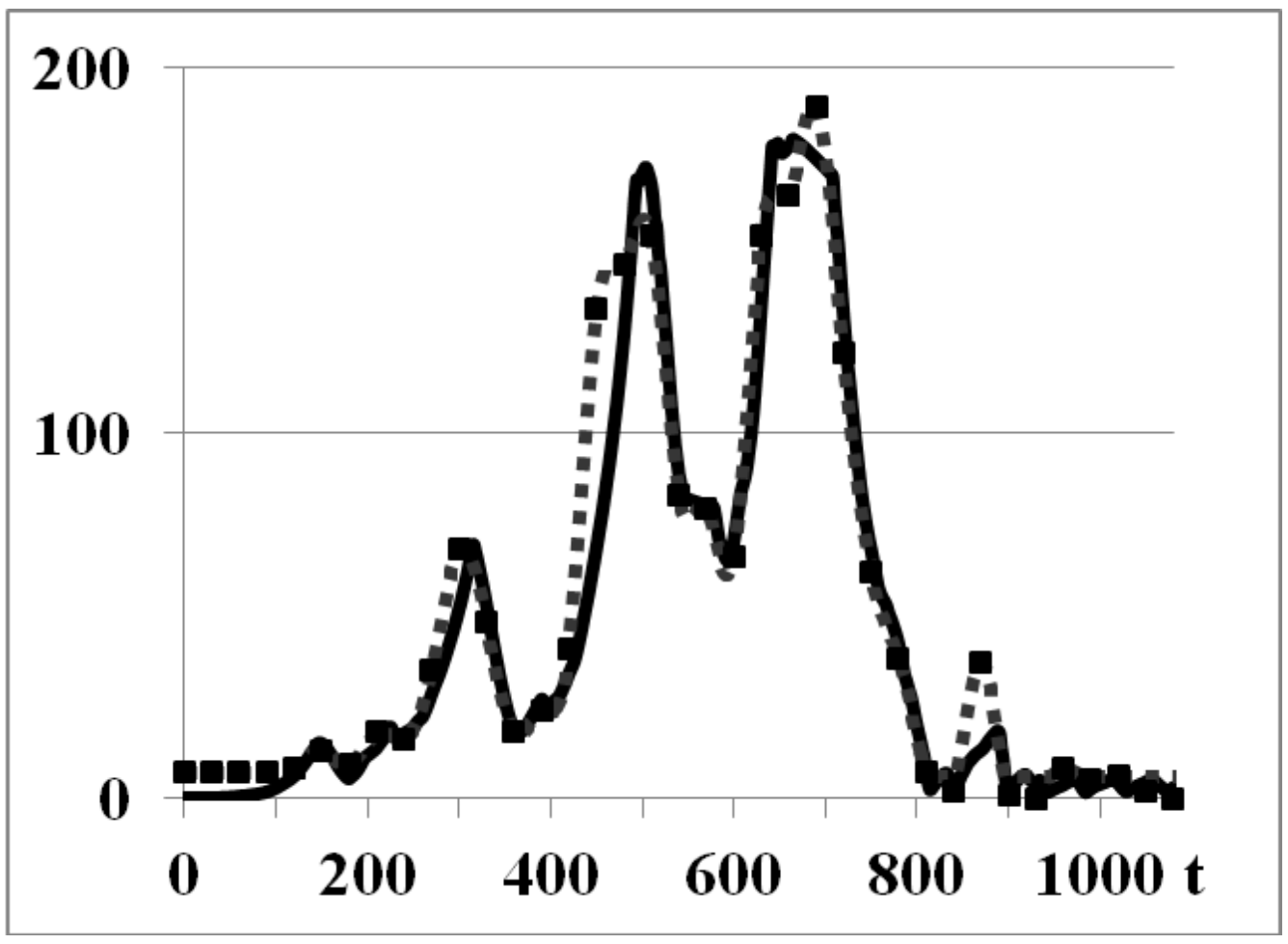

Fig.11b. January 1988 – December 1989.

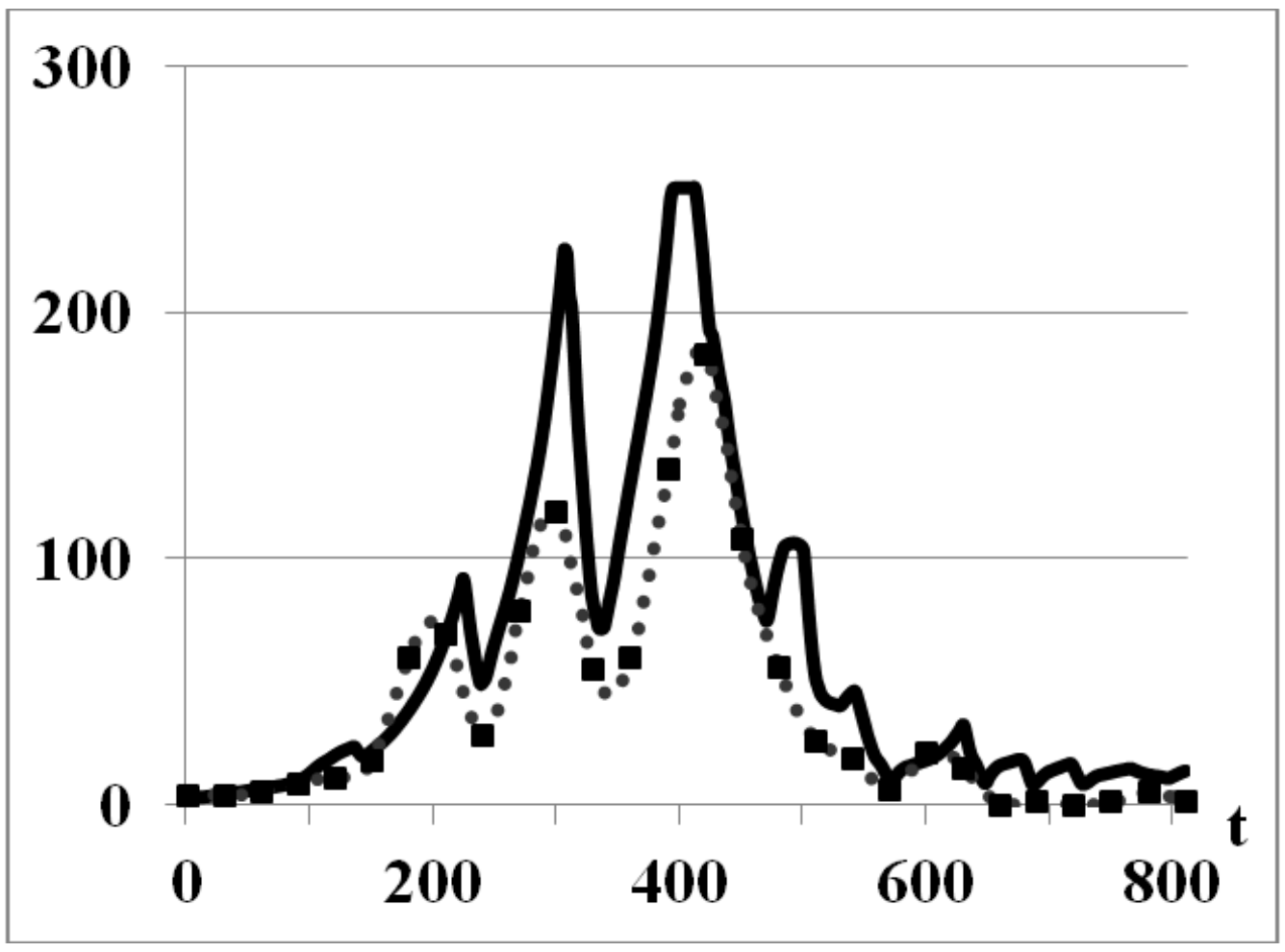

Fig.11c. November 1992 – March 1997.

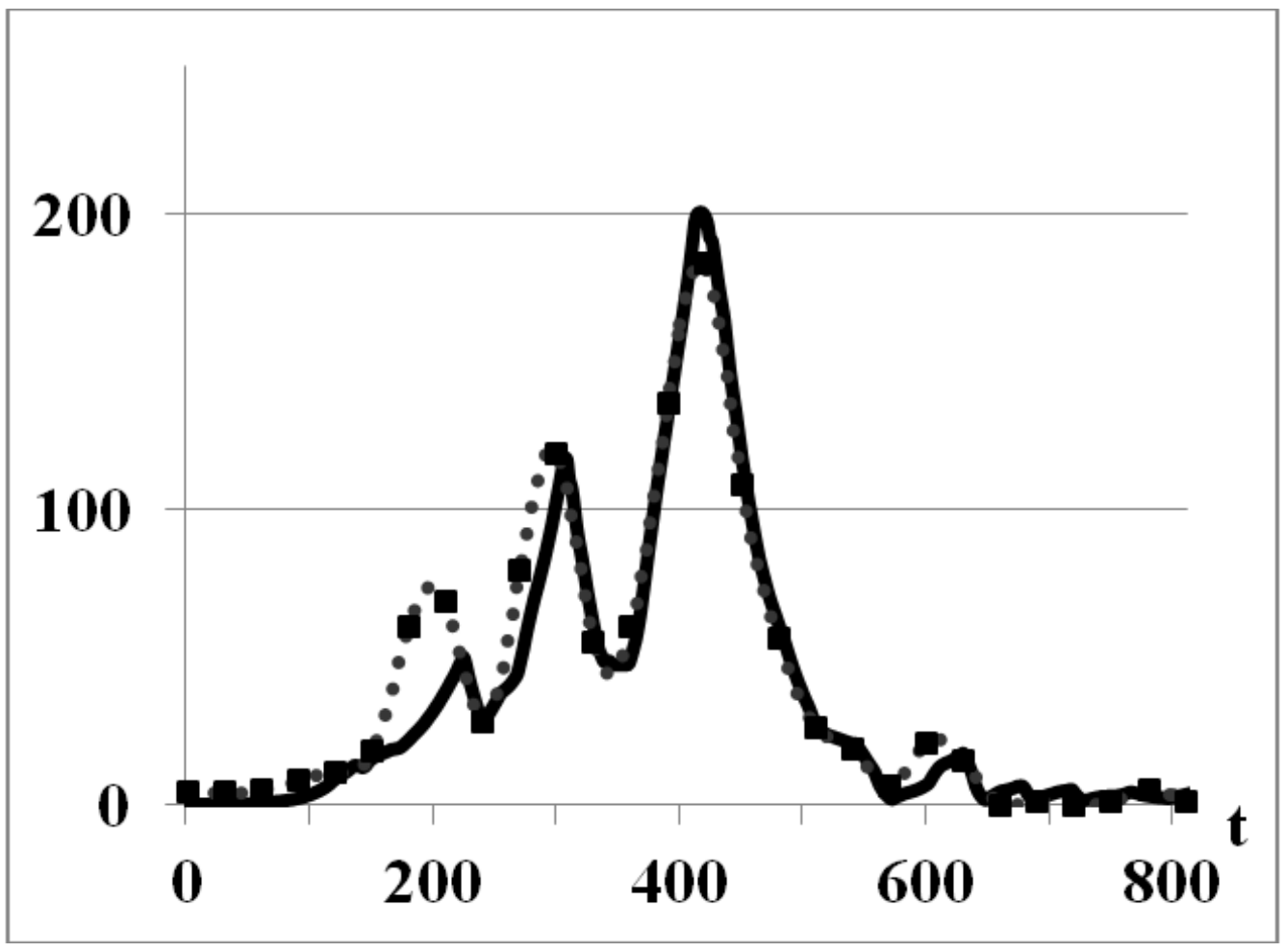

Fig.11d. November 1992 – March 1997.

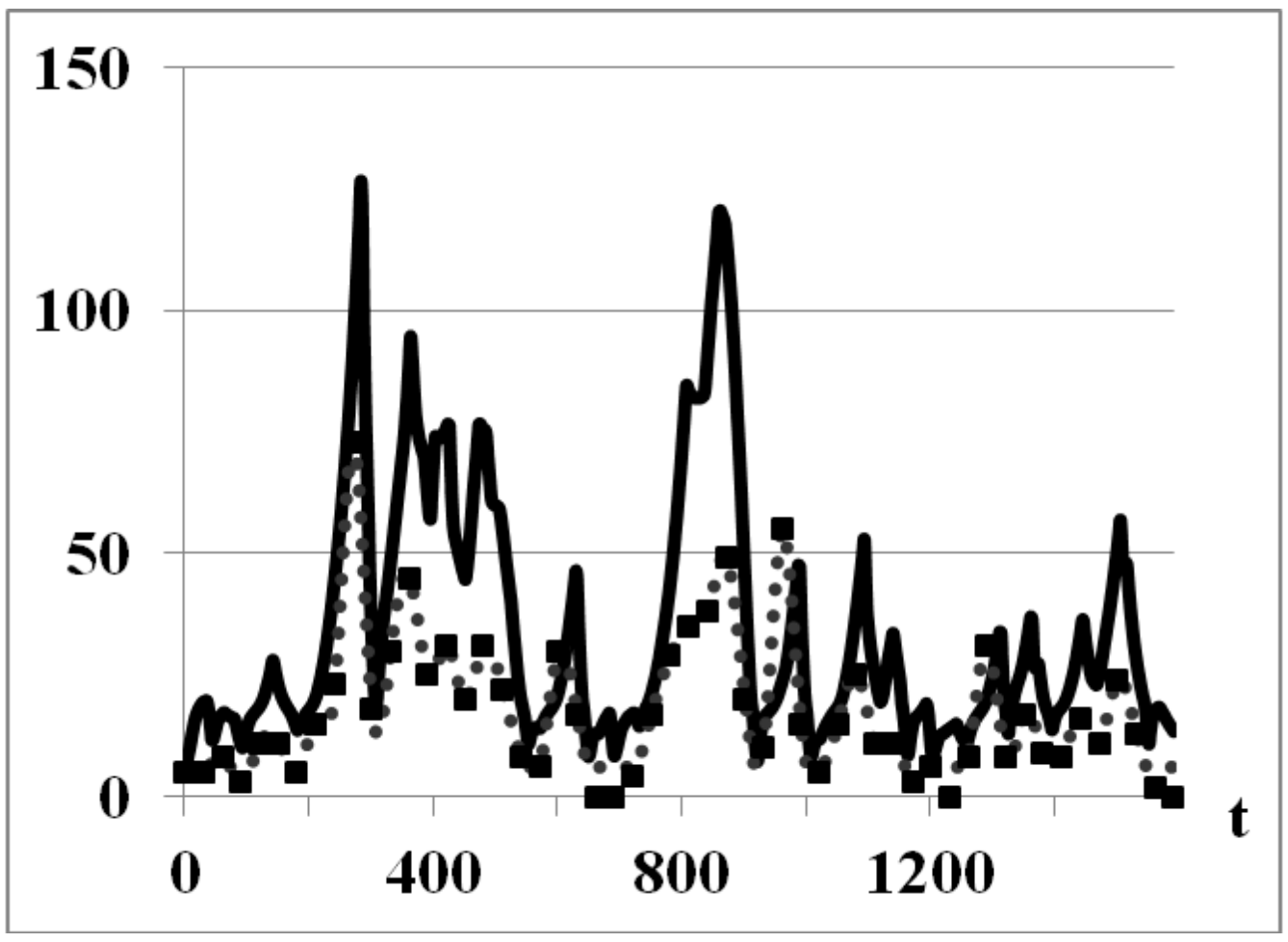

Fig.11e. October 2003 – December 2005.

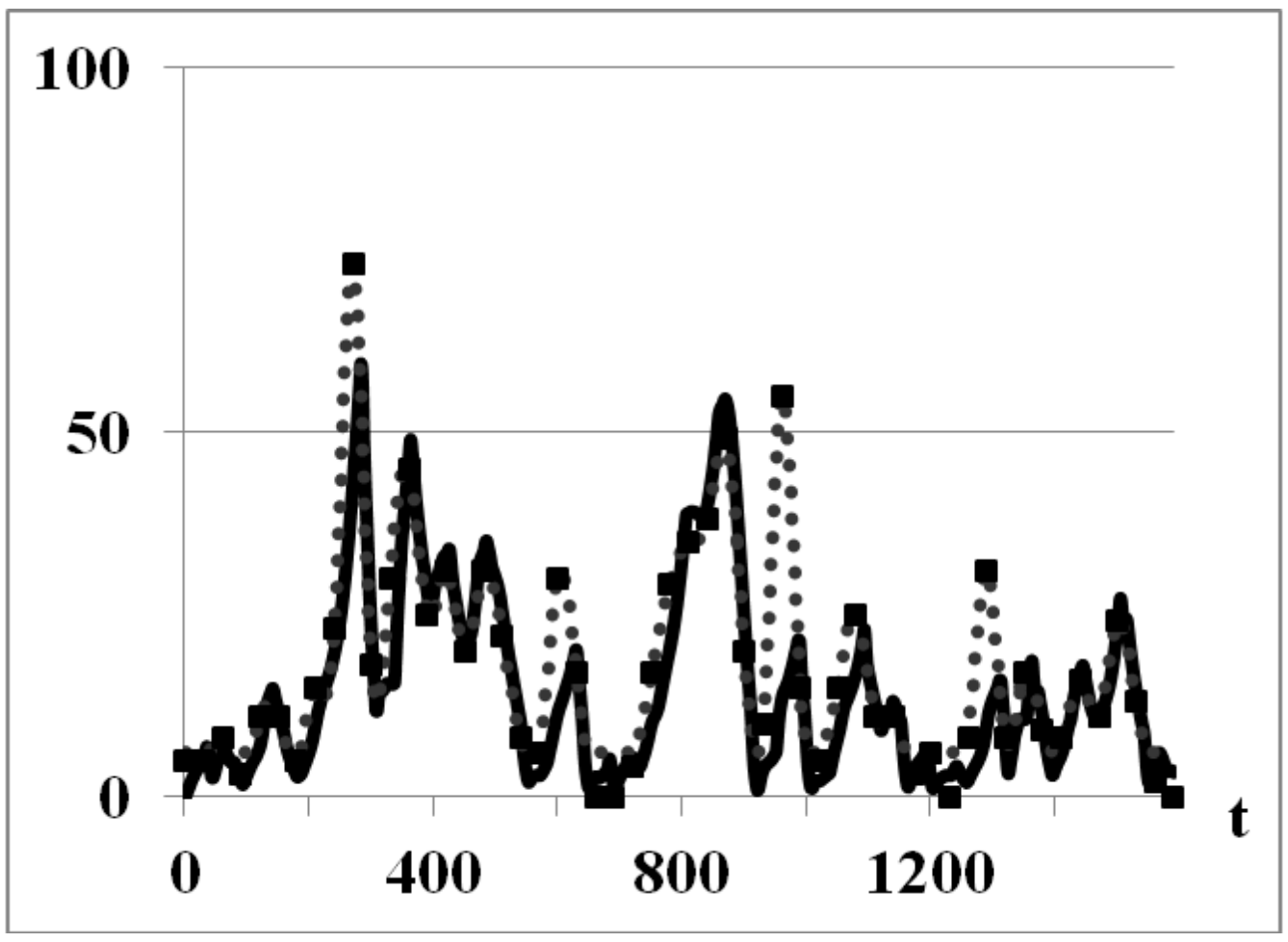

Fig.11f. October 2003 – December 2005.

Fig.11. Graphs of the time series of number of observed sq.degrees of "gregarious" desert locust swarms in the distribution area (map in Fig.14, Appendix C) at the different periods (dotted lines) and calculated total number of locusts $W(t)$ (Fig. 11a, 11c, 11e) (bold line) or number of matured gregarious locusts $G_g(t)$ (Fig. 11b, 11d, 11f) (bold line).

The size of solitarious subpopulation in all experiments was negligibly small with oscillating dynamics in the neighborhood of trivial equilibrium. The values of parameters obtained in numerical experiments allow for fitting the observed data of gregarious swarms to the theoretical curves. The relative error $\delta = 100\frac{1}{N}\sum_{i=1}^{N}\frac{\left|G_g(t_i) - G_i\right|}{G_i}$, $(G_i \neq 0)$, in all experiments ranges from 20% to 26%. The results of simulations exhibit the efficiency of mathematical modeling of locust population dynamics on the basis of two-phase age-structured competitive model which may be used as a theoretical instrument of analysis and prediction of locust behaviour.

## 9. Discussion and conclusion.

### *9.1. The model*

The two-phase competitive age-structured model with discrete time delay presented in this work is the first model of the kind considered in scientific literature for simulation of desert locust population dynamics. The main advantage of this model [15], [34], [46] is that it bridges the gap between the description in explicit and accurate form of individual biological parameters of locusts, such as a number of new born locusts from one pod, number of pods that a locust female lays over her lifetime, egg incubation period, the fraction of locust females of reproductive age in a population, the age of maturation of locust females from which they lay their first eggpod, a maximum age of locust, locust "reproductive window" and simulation of dynamics of huge swarms of locusts at the macro level using the age-specific density of locust population. As a consequence, the stability analysis of equilibriums of such dynamical system provides dimensionless indicators and corresponding conditions of direct and reverse phase transitions between solitarious and gregarious subpopulation equilibriums in terms of their individual biological parameters. Although the considered model cannot guarantee a perfectly accurate and absolutely adequate description of complex locust population dynamics, it provides a lot of useful and valuable information, improves our understanding of features and particularities of locust two-phase population dynamics.

### *9.2. The existence of equilibriums*

To study the equilibriums and stability of our two-phase competitive age-structured autonomous model we have made the assumption that the steady stationary state of locust population is possible only for solitarious or gregarious phases separately. The phase transition was assumed a temporary process which takes place before the population density reaches the final age distribution. We considered two initial problems for nonlinear ODE that describe the dynamics of solitarious and gregarious subpopulations sizes independently from each other. On the basis of linear stability analysis of these problems we found out that the trivial equilibrium of solitarious subpopulation always exists and we obtained the dimensionless parameters $\Lambda_i > 0 \quad (i = 1,\ldots,4)$ - indicators of existence of nontrivial equilibriums in solitarious $(\Lambda_1 - \Lambda_2 \geq 1)$ and gregarious $(\Lambda_3 - \Lambda_4 \geq 1)$ subpopulations (Theorem 2). These parameters depend from the individual biological parameters of locusts including the width of "reproduc-tive windows" $[a_s, a_d]$ (solitarious) and $[a_g, a_d]$ (gregarious). Also, the natural death rate $\alpha_0$ is a very important parameter of population dynamics included in the key conditions of existence of nontrivial equilibriums. The variation of this parameter depends directly from the change of environmental conditions in locust population habitat such as climate, vegetation cover, human activities, etc. Variations of $\alpha_0$ impact significantly on the values of $\Lambda_i$ and conditions of existence of nontrivial equilibriums. It is expected that for the small value of natural death rate $\alpha_0 \approx 0$ the nontrivial equilibriums of solitarious and gregarious subpopulations always exist:

$$\lim_{\alpha_0 \to 0} (\Lambda_1 - \Lambda_2) = \omega_s \mu_s m_s > 1, \quad \lim_{\alpha_0 \to 0} (\Lambda_3 - \Lambda_4) = \omega_g \mu_g m_g > 1. \tag{125}$$

On the other hand, a big value of $\alpha_0 \approx 1$ can lead to the violation of condition of existence of nontrivial equilibriums for arbitrary width of "reproductive windows" of locusts:

$$(\Lambda_1 - \Lambda_2) \approx \omega_s \mu_s m_s (a_d - a_s)^{-1} \exp(-a_s)(1 - \exp(-(a_d - a_s))) < 1, \tag{126}$$

$$(\Lambda_3 - \Lambda_4) \approx \omega_g \mu_g m_g (a_d - a_g)^{-1} \exp(-a_g)(1 - \exp(-(a_d - a_g))) < 1, \tag{127}$$

that are, the conditions of existence of only trivial equilibriums of population. Thus, the conditions of existence of nontrivial equilibrium in both subpopulations are a result of balance between mortality and fertility of locusts. The conditions of existence of nontrivial equilibriums obtained in Theorem 2 provides the opportunity to analyze and control the key ratio between biological parameters of locust to prevent the appearance of "swarming" – quasi-equilibriums or quasi-stable regimes of gregarious locust subpopulation dynamics.

*9.3. The stability analysis of equilibriums*

The density-dependent fertility response functions $h_s(U_s)$ and $h_g(G_g)$, and their derivatives $h_s^{'}(U_s)$ and $h_g^{'}(G_g)$ are among the most fundamental parameters in both the theory and practice of population dynamics [15], [16]. By virtue of Theorems 3, 4, they play a crucial role as determining factors in the asymptotically stable behavior of locust population. In work [16] authors used the model of density-dependent fertility response functions with property $h(0) = 0$ that is a natural condition of absence of fertility in empty population. In our model we considered this function with wider range of initial values: $h_s(0) \in [0,1]$. The absence of fertility of empty population is described by boundary condition of the system and does not depend from the value of $h_s(0)$. On the other hand the fertility response functions at the trivial point plays a crucial role in the stability condition of trivial equilibrium and behavior of population density in the neighborhood of trivial point. Condition $h_s(0) \in [0,1]$ also allows for the avoidance of the big value (singularity) of derivative $h_s^{'}(0)$ in the neighborhood of this trivial point. By virtue of Theorem 3 the trivial equilibrium of solitarious subpopulation is not locally asymptotically stable if and only if

$$F_s \le h_s(0) \le 1, \tag{128}$$

$$F_s = (\Lambda_1 - \Lambda_2)^{-1} = (\omega_s \mu_s m_s)^{-1} \alpha_0 (a_d - a_s)(\exp(-\alpha_0 a_s) - \exp(-\alpha_0 a_d))^{-1}. \tag{129}$$

In fact, these inequalities define the condition of growth of solitarious subpopulation from the neighborhood of trivial point to the nontrivial equilibrium of solitarious $U_s$ and they are valid only with small value of natural death rate $\alpha_0 \approx 0$ (125). The condition of asymptotical stability of trivial equilibrium (elimination of population) reads as:

$$0 \le h_s(0) < F_s. \tag{130}$$

It was not expected that the conditions of stability (130) and instability (128) of trivial equilibrium do not depend from the egg incubation period $\tau$(t) (delay function of the model) and depend from the maturing age of solitarious female $a_s$ and, as a consequence, from the width of “reproductive window” $[a_s, a_d]$. The “destabilization” of model equilibria for decreasing maturation period was studied in [15], [16] for broad class of maturation functions. In our model we used parameter $a_s$. From conditions (128), (130) it follows that the destabilization or stabilization of trivial equilibrium for the different value of $a_s$ depends on the sign of derivative $F_s^{'}(a_s)$:

$$F_s^{'}(a_s) = (\omega_s \mu_s m_s)^{-1} \alpha_0 \exp(\alpha_0 a_d)(\exp(\alpha_0 (a_d - a_s)) - 1)^{-2} \tag{131}$$

$$\times (1 - (1 - \alpha_0 (a_d - a_s)) \exp(\alpha_0 (a_d - a_s))) > 0, \quad a_s \in (0, a_d),$$

Thus, we can assess now the minimum and maximum of $F_s(a_s)$:

$$\max_{a_s \in (0, a_d)} (F_s(a_s)) < \lim_{a_s \to a_d} (F_s(a_s)) = (\omega_s \mu_s m_s)^{-1} \exp(-\alpha_0 a_d), \tag{132}$$

$$\min_{a_s \in (0, a_d)} (F_s(a_s)) > \lim_{a_s \to 0} (F_s(a_s)) = (\omega_s \mu_s m_s)^{-1} \alpha_0 a_d (1 - \exp(-\alpha_0 a_d))^{-1}. \tag{133}$$

From Eqs. (131) and (132) it follows three important conclusions. Firstly, we obtain the necessary and sufficient condition of existence of unstable trivial equilibrium independently from the maturing period in the form:

$$(\omega_s \mu_s m_s)^{-1} \exp(-\alpha_0 a_d) \le h_s(0) \le 1. \tag{134}$$

Secondly, the risk of emergence of unstable trivial equilibrium in solitarious subpopulation appears if the fertility response function of solitarious satisfies the condition:

$$(\omega_s \mu_s m_s)^{-1} \alpha_0 a_d (1-\exp(-\alpha_0 a_d))^{-1} \leq h_s(0) \leq 1 \qquad (135)$$

Finally, the decrease in maturation age $a_s \to 0$ can be a cause of destabilization of trivial equilibrium. From a rigorous biological point of view the maturation of locust females occurs earlier than they are able or have possibility to lay a pod with eggs at first. That's why in this paper we do not use the term "maturation period" for stability analysis as it was in theoretical works [15], [16] for the predator-prey age-structured model of population dynamics with density-dependent fertility rate. Owing to the fact that the age $a_s$ corresponds to the maximum value (maximum of reproductive activity) of maturation function $\beta(a)$ from [16] we can compare the conclusions obtained for the nonlinear age-structured models with density-dependent fertility response function in [16] with the results of Theorem 3. Due to the conclusion of work [16] the destabilization of equilibria occurs only for the decrease and not the increase of maturation period. In specific ecosystems with locust populations, one could observe a change of $a_s$ in a wide range from the minimum biologically motivated age to some maximum age which depends from the environmental conditions such as finding adequate conditions for egg deposition, development speed, food quality and finding mates. These changes can be a cause of destabilization of locust population dynamics and respectively the cause of their phase-changing. Indeed, taking into account Eq. (131) we can conclude that the trivial equilibrium of solitarious subpopulation can be stable for some fixed value of maturation period $\bar{a}_s$ and unstable for smaller value of $\bar{\bar{a}}_s < \bar{a}_s$ if the fertility response function of solitarious satisfies the condition:

$$0 < F_s(\bar{\bar{a}}_s) < h_s(0) < F_s(\bar{a}_s), \quad 0 < \bar{\bar{a}}_s < \bar{a}_s . \qquad (136)$$

The analysis of conditions of instability of trivial equilibrium has an important value for practice, because unstable behavior or growth of small solitarious subpopulation (initially in the neighborhood of the trivial equilibrium) can cause a switch of solitarious subpopulation to the nontrivial equilibrium or even lead to the gregarization and phase change with following move to the nontrivial equilibrium of gregarious subpopulation. In other words, the rapid change in environmental conditions that are known to trigger maturation in solitarious population may also decrease overall the maturation age and hence destabilize the solitarious subpopulation from a near zero population size towards much larger populations and hence initiate quickly good conditions for gregarization.

The conditions of local asymptotical stability and instability of nontrivial equilibriums of solitarious and gregarious subpopulations given in Theorem 4 used only the sign of derivatives of the density-dependent fertility response functions $h_s^{'}(U_s)$ and $h_g^{'}(G_g)$: the nontrivial equilibriums are locally asymptotically unstable (stable) if the functions $h_s^{'}(U_s) \geq 0$ $(< 0)$ and $h_g^{'}(G_g) \geq 0$ $(< 0)$. The instability condition of positive or nontrivial equilibria for small value of maturation period was the same in work [16] (Theorem 2, (a)). But in our work the local asymptotical stability and instability conditions of positive or nontrivial equilibriums do not depend from the egg incubation period $\tau$(t) (delay function of the model) nor from the maturing age of solitarious and gregarious female $a_s$ and $a_g$, and, as a consequence, from the width of "reproductive windows" $[a_s, a_d]$, $[a_g, a_d]$. These parameters impact however on the conditions of existence of nontrivial equilibriums.

In practice the value and sign of derivatives of fertility response function can be changed depending on different features of a locust population. For instance, we could assume that $h_s^{'}(U_s) > 0$, $h_g^{'}(G_g) > 0$ at the onset of solitarious population growth when the number of adult locust increase in a given area which consequently shortens the search time for mate, leads to more frequent contacts between locusts, increases the probability of laying egg-pods by females and hence increase birth of offspring. On the other hand, growth of population can lead to a deficit of food resource in a given area and be a cause of reduction of fertility of the locust population. Finally, it is a well-known fact that locust fertility decreases with gregarization [37]. These situations correspond to the negative sign of derivative of fertility response functions $h_s^{'}(U_s) < 0$, $h_g^{'}(G_g) < 0$ and hence to asymptotically stable non-trivial equilibrium.

*9.4. The numerical experiments*

The numerical algorithm developed in the Theorem 1 allowed carrying out numerous simulations to study the dynamical regimes of locust population dynamics for the autonomous and non-autonomous systems. The results obtained in the first part of numerical experiments (section 7) illustrated the various dynamical regimes of the locust population in a wide range of model parameters (coefficients of equations and initial values) for all types of asymptotically stable and unstable equilibriums obtained in the Theorems 2 - 4 for the autonomous system. We summarize all obtained theoretical results of this study with brief descriptions in Table 1.

| No | Type of fertility response functions | Type of dynamical regime of locust population |
|---|---|---|
| 1 | $h_s(0) < h_s^*$ | Asymptotically stable trivial equilibrium, $U_s(t) \to 0$. |
| 2 | $h_s(0) > h_s^*$, $h_s^{'}(U_s) > 0$. | Asymptotically unstable positive equilibrium $U_s^*$, $U_s(t)$ growths, gregarization. |
| 3 | $h_s(0) > h_s^*$, $h_s^{'}(U_s) < 0$, $U_s^* < W_{0g}$. | Asymptotically stable positive equilibrium $U_s^* < W_{0g}$, $U_s(t)$ oscillates in the neighborhood of $U_s^*$, without gregarization. |
| 4 | $h_s(0) > h_s^*, h_s^{'}(U_s) > 0$, $h_g(0) < h_g^*$, $h_g^{'}(G_g) < 0$. | Asymptotically stable quasi-periodical oscillations of $U_s(t)$ and $G_g(t)$ in the neighborhood of positive equilibriums $U_s^*$ and $G_g^*$, gregarization. |
| 5 | $h_s(0) > h_s^*$, $h_s^{'}(U_s) < 0$, $U_s^* > W_{0g}$, | Asymptotically stable quasi-periodical oscillations of $G_g(t)$ in the neighborhood of positive equilibrium $G_g^*$, oscillations of $U_s(t)$ in the neighborhood of triv- |
| | $h_g(0) > h_g^*$, $h_g^{'}(G_g) < 0$, $\min_{G_g>0} h_g(G_g) \le h_g^*$. | ial equilibrium, gregarization. |
| 6 | $h_s(0) > h_s^*, h_s^{'}(U_s) > 0$, $h_g(0) > h_g^*$, $h_g^{'}(G_g) < 0$, $\min_{G_g>0} h_g(G_g) \le h_g^*$. | Asymptotically stable quasi-periodical oscillations of $G_g(t)$ in the neighborhood of positive equilibrium $G_g^*$, oscillations of $U_s(t)$ in the neighborhood of trivial equilibrium, gregarization. |
| 7 | $h_s(0) > h_s^*, h_s^{'}(U_s) > 0$, $h_g(0) > h_g^*$, $h_g^{'}(G_g) < 0$, $\min_{G_g>0} h_g(G_g) > h_g^*$ | Asymptotically unstable dynamics, oscillations of $U_s(t)$ in the neighborhood of trivial equilibrium, $G_g(t)$ evolves to infinity $G_g(t) \to \infty$. |
| 8 | $h_s(0) > h_s^*$, $h_s^{'}(U_s) > 0$, $h_g(0) < h_g^*$, $h_g^{'}(G_g) > 0$. | Asymptotically unstable dynamics, oscillations of $U_s(t)$ in the neighborhood of trivial equilibrium, $G_g(t)$ evolves to infinity $G_g(t) \to \infty$. |
| 9 | $h_s(0) > h_s^*$, $h_s^{'}(U_s) > 0$, $h_g(0) > h_g^*$, $h_g^{'}(G_g) < 0$. | Asymptotically unstable dynamics, oscillations of $U_s(t)$ in the neighborhood of trivial equilibrium, $G_g(t)$ evolves to infinity $G_g(t) \to \infty$. |

Table 1. The dynamical regimes of autonomous system for monotonic fertility response functions with conditions (69) and (72).

Earlier studies of conditions of “gregarization” [13], [14] have focused on the environmental characteristics of solitarious density threshold when the phase change takes place. With our age-structured modeling approach, we shifted our attention to the study of the conditions of existence of trivial and nontrivial (positive), steady and unsteady equilibriums of solitarious subpopulation.

The dynamical regimes 1 and 3 of Table 1 correspond to the steady states of solitarious without gregarization. In the first case, solitarious do not have the stable nontrivial equilibrium. In the second one, the nontrivial equilibrium exists and solitarious density evolves to some positive value. Conditions of regime 2 provide a direct phase transition from solitarious to gregarious phase that is also found in the regimes 4 - 9. The regime 4 describes the steady dynamics of both subpopulations in the neighborhood of positive equilibriums. But, from numerous field observations, it is known that both phases of solitarious and gregarious do not coexist together in one local area. This dynamical regime may be acceptable in practice for large region with distributed important aggregations of locust.

In our opinion, the most realistic dynamical regimes of gregarization for a given local area correspond to the specific cases of results 5 and 6. Although in both cases we observed the same dynamical regime of locust population, there are some significant differences in the conditions of emergence of these regimes. The dynamical regime 5 corresponds to an unstable trivial equilibrium with absence of nontrivial steady state of solitarious when the point of their stable nontrivial equilibrium is bigger than the “gregarization” threshold. In the case of regime 6 there exists an unstable nontrivial equilibrium of solitarious subpopula-tion with direct phase change - gregarization. In both cases the density of solitarious subpopulation is negligibly small and oscillates in the neighborhood of trivial equilibrium.

From a biological point of view, as mentioned above, the conditions $h_s^{'}(U_s) < 0$ and $h_g^{'}(G_g) < 0$ (case 5) correspond to known biological processes in desert locust: an increase in locust density generate a decrease of egg number in the eggpods compensated by an increase in egg size [37]. Overall, this process decreases the fertility of females in dense populations. As we did not vary $\mu_s$ or $\mu_g$ according to density, the conditions of $h_s^{'}(U_s) < 0$ and $h_g^{'}(G_g) < 0$ simulate this process. The fact that $U_s^* > W_{0g}$ needs then to be respected to have possible and realistic oscillations among solitarious and gregarious phases, imply that evolutionarily, locusts were selected to gregarize before the asymptotically stable nontrivial equilibrium of solitarious is reached. This privileges the hypothesis that natural selection would have favored the apparition of phase polyphenism and associated gregarious behavior in very varying and erratic environmental conditions. Indeed, a whole regime of changing conditions is needed to imagine a selection process of plasticity to be expressed before the actual conditions is not adequate to the survival of the solitarious phase.

The unstable behavior of gregarious subpopulation in cases 7 – 9 when the density of locust population evolves to infinity is not valuable for practice. These cases are studied theoretically for analysis of restrictions on the fertility response functions and other parameters of the locust population model.

In the second part of the numerical experiments (section 8) the data of field observations were fitted to the solutions of non-autonomous model of locust population dynamics. We considered in these experiments the number (or area) of square degrees ($1^\circ \times 1^\circ$) with observation of locust swarms in the distribution map with the computed number of matured locusts in gregarious subpopulation. We obtained the quasi-periodical outbreaks of gregarious subpopulation density and size which correspond to the observed behavior of desert locust in the considered area. Since the relative error of approximation was less than 26%, the obtained results illustrated a good agreement of computed solutions with the observed data.

Overall, we can conclude that the results of the numerical experiments shown here sustain that the considered two-phase age-structured competitive model with discrete time delay can be successfully used in the simulation and study of locust population dynamics in the different real-world applications.

**Acknowledgments.** We would like to acknowledge Professor Glen Webb for helpful and fruitful discussions and suggestions on the manuscript topic.

**Appendix A.** The proof of Lemma 1.

Eq. (90) can be written in the compact form:

$$\zeta^2 - b = -c\cos(\zeta(a_d - a_s)), \tag{137}$$

$$b = \alpha_0^2 \left( Y^2 \left( \Lambda_1^2 + \Lambda_2^2 \right) - 1 \right), \tag{138}$$

$$c = 2\alpha_0^2 \Lambda_1 \Lambda_2 Y^2 , \tag{139}$$

If conditions of Theorem 2 and condition (80) hold the coefficients of Eq. (137) satisfy inequalities:

$$0 < c < 2\alpha_0^2 \Lambda_1 \Lambda_2 , \tag{140}$$

$$b = \alpha_0^2 \left( Y^2 (\Lambda_1 - \Lambda_2)^2 - 1 \right) + c < c , \tag{141}$$

The roots of Eq. (137) may be found as the intersection points of graphs of scalar functions of real argument $f_{11}(\zeta)$ and $f_{12}(\zeta)$:

$$f_{11}(\zeta) = \zeta^2 - b , \tag{142}$$

$$f_{12}(\zeta) = -c \cos(\zeta (a_d - a_s)) . \tag{143}$$

The illustration of three possible types of graphs of function $f_{11}(\zeta)$ and graph of function $f_{12}(\zeta)$ (in correspondence with Eqs. (140), (141)) is given in Fig.12. As all graphs have the axial symmetry we analyze only the existence of positive real roots $\zeta > 0$ of Eq. (137). The graph of function $f_{11}(\zeta)$ does not intersect the graph of $f_{12}(\zeta)$ if the derivatives of functions satisfy conditions:

$$f_{11}'(\zeta) > f_{12}'(\zeta), \ \ \zeta > 0 . \tag{144}$$

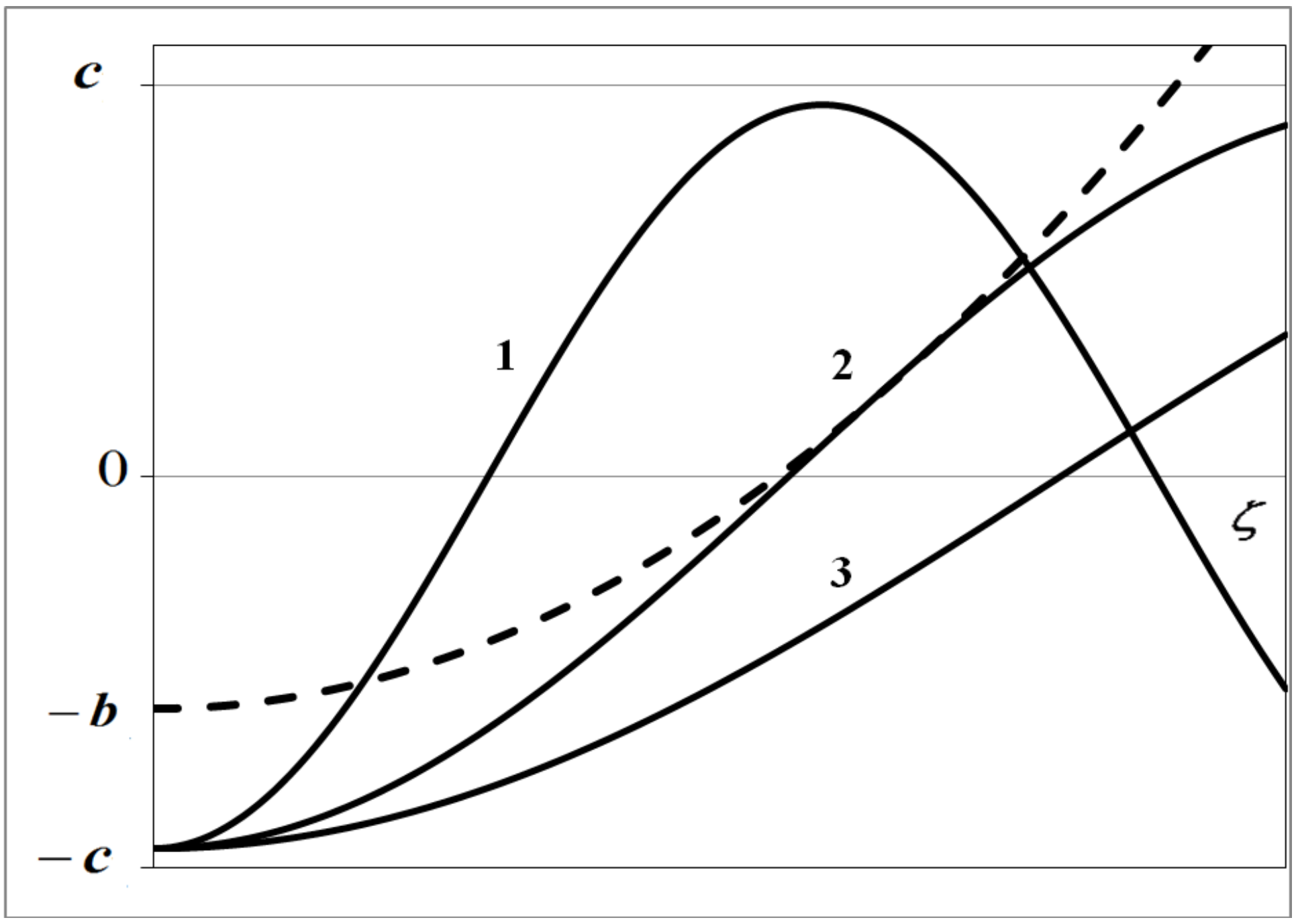

Fig.12. The graph of $f_{11}(\zeta)$ (dashed line) and graphs of $f_{12}(\zeta)$ (continuous line) (1) and (2) for $f_{21}'(0) > 1$; (3) for $f_{21}'(0) < 1$.

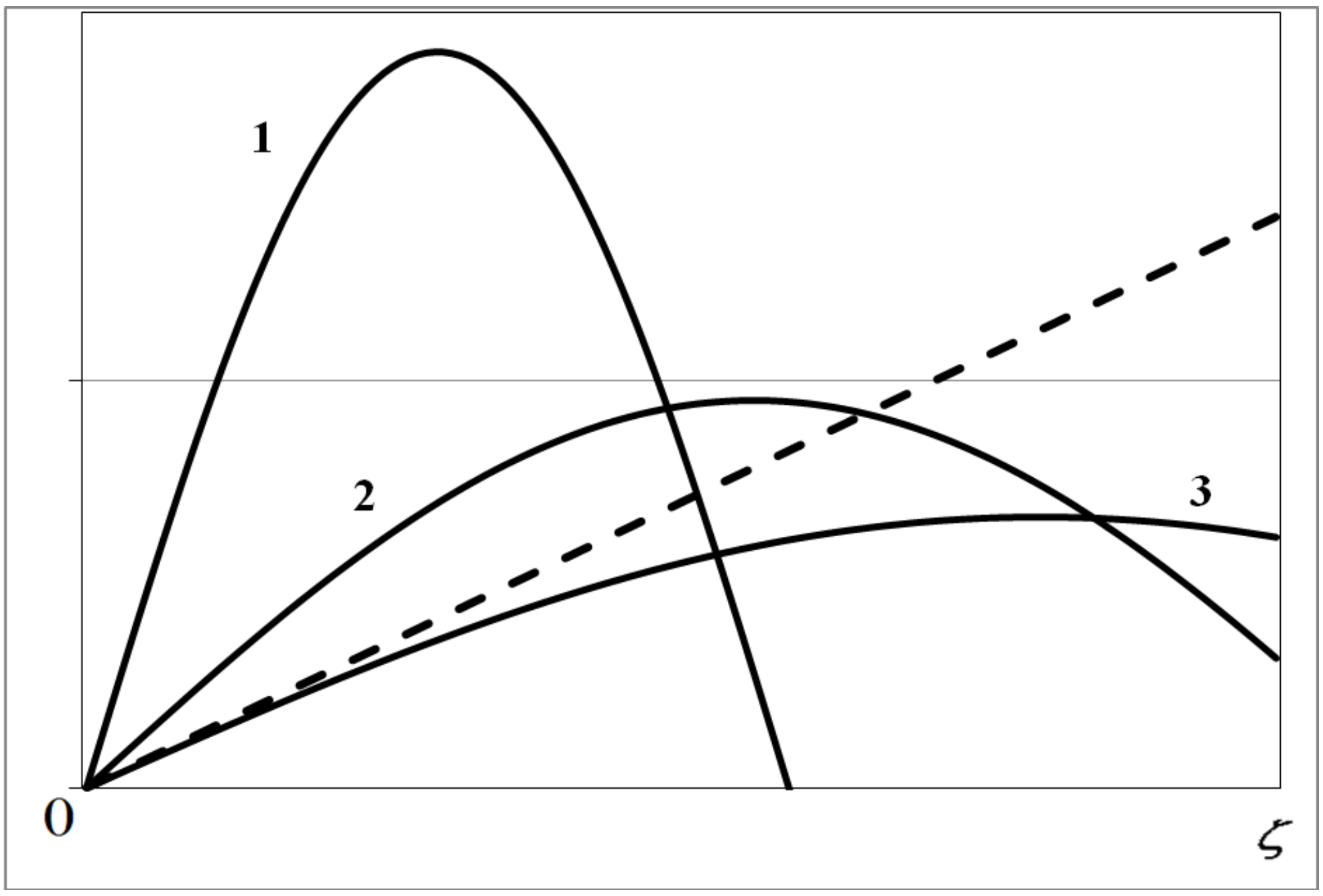

Fig.13. The graph of $f(\zeta)=\zeta$ (dashed line) and graphs of $f_{21}(\zeta)$ (continuous line) (1) and (2) for $f'_{21}(0)>1$; (3) for $f'_{21}(0)<1$.

We can rewrite inequality (144) in the following form:

$$\zeta > f_{21}(\zeta)=0{,}5c(a_d-a_s)\sin(\zeta(a_d-a_s)). \tag{145}$$

The illustration of three possible types of graphs of function $f_{21}(\zeta)$ (right hand side) and linear function (left hand side) of Eq. (145) is given in Fig.13. If $f'_{21}(0)\le 1$ inequality (145) is valid for all $\zeta>0$. Thus, the necessary condition for existence of root of Equation (137) is $f'_{21}(0)>1$ (cases 1 and 2 in Fig.13), i.e. when the coefficients of Eq. (137) satisfy condition:

$$0{,}5c(a_d-a_s)^2=\alpha_0^2\Lambda_1\Lambda_2(a_d-a_s)^2Y^2>1. \tag{146}$$

If condition (146) holds, the point of touch $\zeta>0$ of graphs $f_{11}(\zeta)$ and $f_{12}(\zeta)$ can exist (cases 1 and 2 in Fig.12), and Eq. (137) can have at least one common real positive root. Otherwise, the point of touch does not exist and Eqs. (137) do not have the real positive root (case 3 in Figs.12 and 13). From Eqs. (80) and (146) it follows the inequalities:

$$\left(\alpha_0^2\Lambda_1\Lambda_2(a_d-a_s)^2\right)^{-1}<Y^2<(\Lambda_1-\Lambda_2)^{-2} \tag{147}$$

$$\Lambda_1^2-\left(2+\alpha_0^2(a_d-a_s)^2\right)\Lambda_1\Lambda_2+\Lambda_2^2<0, \tag{148}$$

$$1-\left(2+\alpha_0^2(a_d-a_s)^2\right)\exp\left(-\alpha_0(a_d-a_s)\right)+\exp\left(-2\alpha_0(a_d-a_s)\right)<0. \tag{149}$$

It is easy to verify that function $f(x) = 1 - \left(2 + x^2\right)\exp(-x) + \exp(-2x) > 0$ for all $x > 0$. Thus, inequality (149) and, as a consequence, inequality (146) is not valid, and Eqs. (137) and (90) do not have real positive roots. Lemma 1 is proved.

**Appendix B.** The proof of Lemma 2.

It is easy to verify that the proof of Lemma 1 given in Appendix A can be applied to the proof of the statement of Lemma 2 for Eq. (113). Equation (90) differs from Equation (113) by the sign of coefficients $Y > 0$ in contrast with $Y_1 < 0$. But, all equations and inequalities in Appendix A (Eqs. (138), (139), (141), (146), (147)) use $Y^2$ and are valid for any values of $Y \neq 0$. Using these equations for $Y_1 < 0$ instead of $Y > 0$ we obtain the same results for Eq. (113). The statement of Lemma 2 can be proved also for Eq. (114) by analogy with Eq.(113). Lemma 2 is proved.

## Appendix C. The biological and model parameters of locust population and the data of observations.

In this section we provide the biological constants and parameters of the model for solitarious and gregarious locusts (Table 2). The area of invasion (dark gray), recession (gray) and gregarization (light gray) of desert locust considered in the numerical experiments is shown in Fig. 14. The time series of number of square degrees ($1^\circ \times 1^\circ$) on the map with observed swarms of gregarious locusts at the different periods over the last 30 years are shown in the Fig.15 a, b, c (dotted line). We used here the data from the works [32], [36], [37], [39], [40], [42], [43], numerous laboratory experiments and monthly reports of FAO.

| | $a_s$ | $\mu_s$ | $\omega_s$ | $m_s$ | $a_g$ | $\mu_g$ | $\omega_g$ | $m_g$ | $a_d$ | $\alpha_0$ | $\tau$ |
|---|---|---|---|---|---|---|---|---|---|---|---|
| Min | 40 | 60 | 0.5 | 1 | 30 | 40 | 0.5 | 1 | 100 | $10^{-8}$ | 10 |
| Expected | 60 | 70 | 0.5 | 3 | 46 | 60 | 0.5 | 2 | 200 | $5\times10^{-4}$ | 14 |
| Max | 240 | 100 | 0.5 | 6 | 120 | 70 | 0.5 | 5 | 300 | $10^{-1}$ | 30 |

Table 2. Biological constants of locust population: $a_s$, $a_g$ (age at maturation of solitarious and gregarious), $a_d$ (maximum age at death), $\tau$ (incubation time) are measured in days; $\omega_s$, $\omega_g$ (sex ratio), $m_s$, $m_g$ (number of eggpods/female), $\mu_s$, $\mu_g$ (number of eggs/eggpod) are dimensionless, $\alpha_0$ (natural death rate) is measured in fraction of all locusts per day.

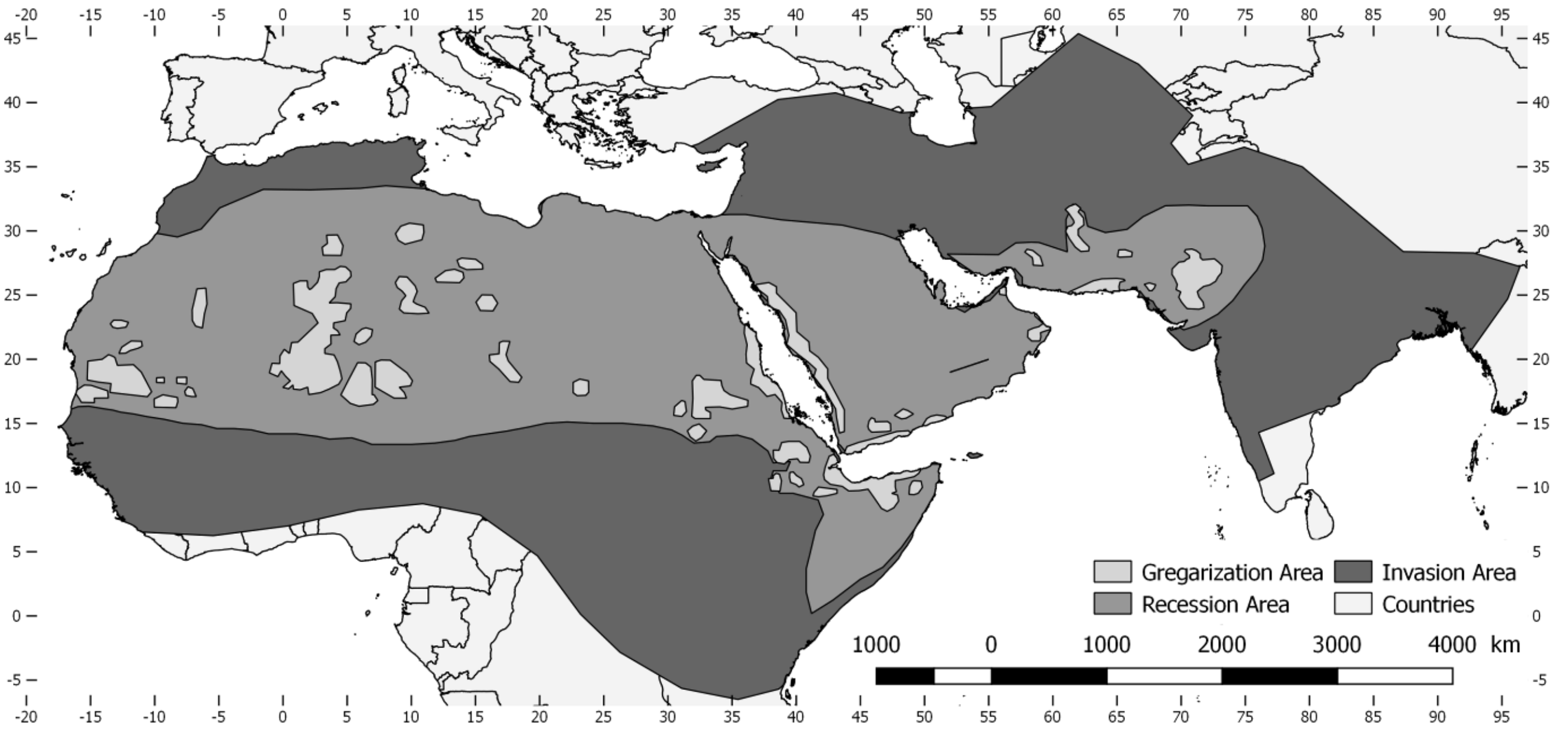

Fig.14. The distribution area of desert locust
in Northern Africa, Middle East and Western Asia (from Sword et al 2010).

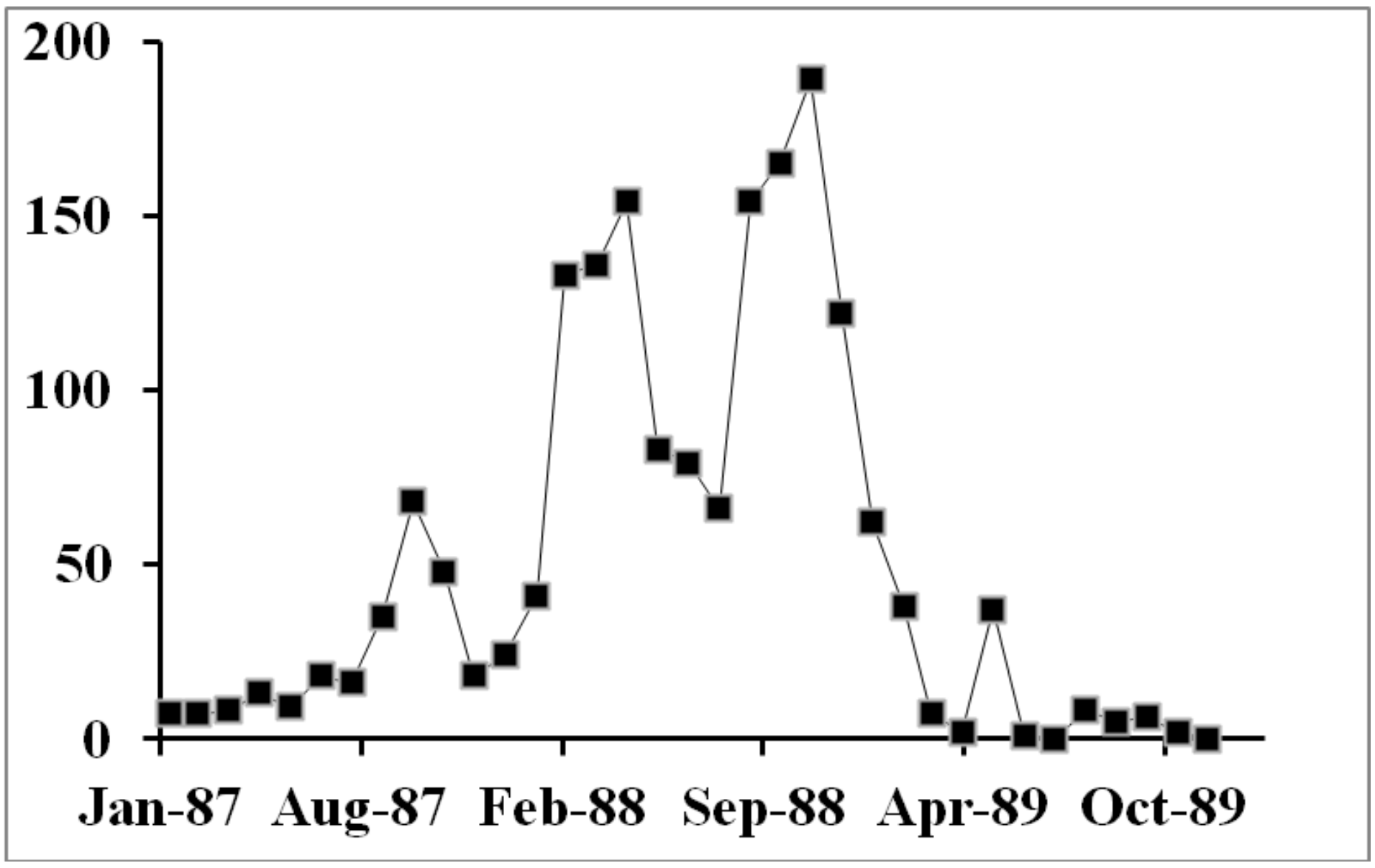

Fig.15.a.

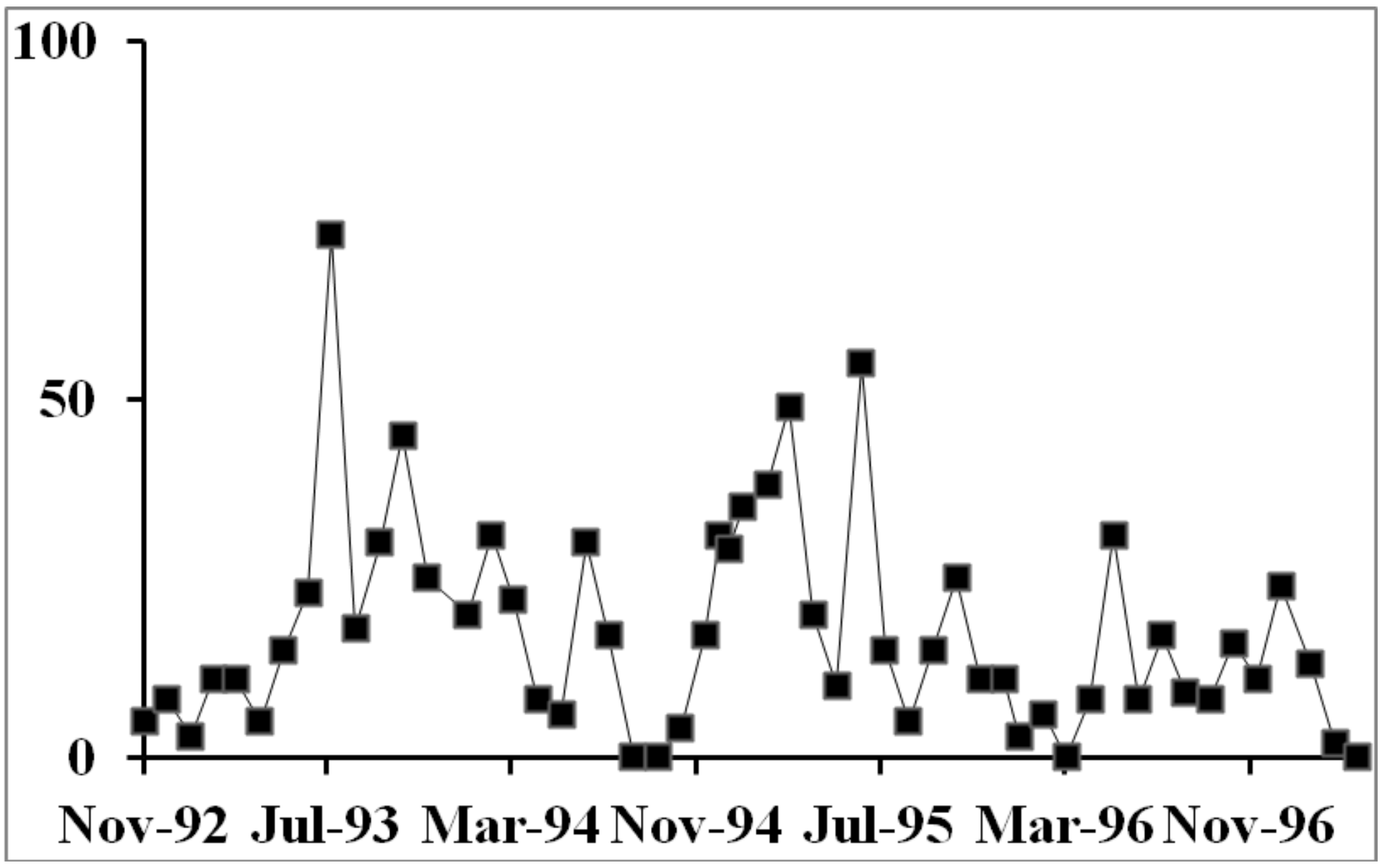

Fig.15.b.

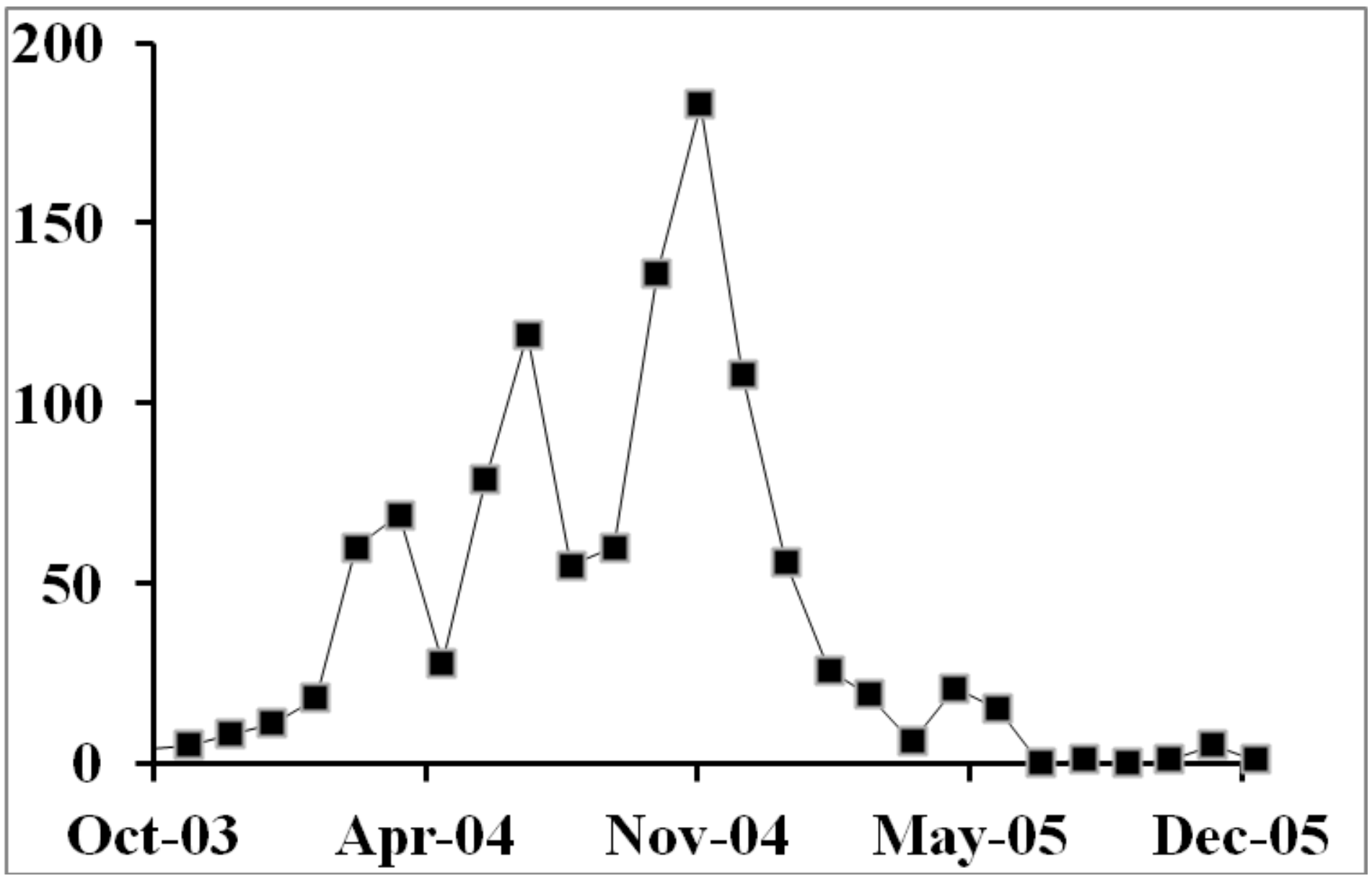

Fig.15.c.

Fig.15. The number of square degrees ($1^\circ \times 1^\circ$) with observed swarms of gregarious locusts in the area shown in Fig.14 at the periods: (a) January 1988 – December 1989; (b) November 1992 – March 1997; (c) October 2003 – December 2005.